\documentclass[aps,prd,superscriptaddress,floatfix,twocolumn,nofootinbib,preprintnumbers]{revtex4}

\usepackage{amsmath,amssymb,amsfonts,dcolumn,slashed,dsfont,multirow}
\usepackage{graphicx}
\usepackage{psfrag}
\usepackage{bold-extra}

\newcommand{\mpi}{M_{\pi}}

\newcommand{\Fpi}{F_\pi}
\newcommand{\gA}{g_A}
\newcommand{\beq}{\begin{equation}}
\newcommand{\eeq}{\end{equation}}
\newcommand{\mc}{m_\chi}
\newcommand{\mN}{m_N}

\newcommand{\muN}{\mu_N}

\newcommand{\mpp}{m_p}
\newcommand{\mn}{m_n}
\newcommand{\muu}{m_u}
\newcommand{\md}{m_d}
\newcommand{\qq}{\mathbf{q}}
\newcommand{\pp}{\mathbf{p}}
\newcommand{\PP}{\mathbf{P}}
\newcommand{\kk}{\mathbf{k}}
\newcommand{\KK}{\mathbf{K}}
\newcommand{\vv}{\mathbf{v}}
\newcommand{\vvp}{\mathbf{v}^\perp}
\newcommand{\N}{\mathcal{N}}
\newcommand{\M}{\mathcal{M}}
\newcommand{\F}{\mathcal{F}}
\newcommand{\spin}{\mathbf{S}}

\newcommand{\sig}{\boldsymbol{\sigma}}
\newcommand{\ttau}{\boldsymbol{\tau}}
\newcommand{\diff}{\text{d}}
\newcommand{\eps}{\epsilon}
\newcommand{\Order}{\mathcal{O}}
\newcommand{\Op}{\mathcal{O}}
\newcommand{\erf}{\text{erf}}
\newcommand{\kms}{\,\text{km}/\text{s}}

\renewcommand{\vec}[1]{\mathbf{#1}}

\providecommand{\MeV}{\,\text{MeV}}
\providecommand{\GeV}{\,\text{GeV}}

\allowdisplaybreaks[1]

\begin{document}

\preprint{INT-PUB-18-059}

\title{Nuclear structure factors for general spin-independent WIMP--nucleus scattering}

\author{Martin Hoferichter}
\email[E-mail:~]{mhofer@uw.edu}
\affiliation{Institute for Nuclear Theory, University of Washington, Seattle, WA 98195-1550, USA}
\affiliation{Kavli Institute for Theoretical Physics, University of California, Santa Barbara, CA 93106, USA}

\author{Philipp Klos}
\email[E-mail:~]{pklos@theorie.ikp.physik.tu-darmstadt.de}
\affiliation{Institut f\"ur Kernphysik, 
Technische Universit\"at Darmstadt, 
64289 Darmstadt, Germany}
\affiliation{ExtreMe Matter Institute EMMI, 
GSI Helmholtzzentrum f\"ur Schwerionenforschung GmbH, 
64291 Darmstadt, Germany}

\author{Javier Men\'endez}
\email[E-mail:~]{menendez@cns.s.u-tokyo.ac.jp}
\affiliation{Center for Nuclear Study, The University of Tokyo, 113-0033 Tokyo, Japan}

\author{Achim Schwenk}
\email[E-mail:~]{schwenk@physik.tu-darmstadt.de}
\affiliation{Institut f\"ur Kernphysik, 
Technische Universit\"at Darmstadt, 
64289 Darmstadt, Germany}
\affiliation{ExtreMe Matter Institute EMMI, 
GSI Helmholtzzentrum f\"ur Schwerionenforschung GmbH, 
64291 Darmstadt, Germany}
\affiliation{Max-Planck-Institut f\"ur Kernphysik, Saupfercheckweg 1, 
69117 Heidelberg, Germany}

\begin{abstract}
We present nuclear structure factors that describe the generalized
spin-independent coupling of weakly interacting massive particles
(WIMPs) to nuclei. Our results are based on state-of-the-art
nuclear structure calculations using the large-scale nuclear shell
model. Starting from quark- and gluon-level operators, we consider all
possible coherently enhanced couplings of spin-$1/2$ and spin-$0$ WIMPs to
one and two nucleons up to third order in chiral effective field
theory. This includes a comprehensive discussion of the structure
factors corresponding to the leading two-nucleon currents covering, for
the first time, the contribution of spin-$2$ operators. We provide
results for the most relevant nuclear targets considered in present and
planned dark matter direct detection experiments: fluorine, silicon,
argon, and germanium, complementing our previous work on xenon. All
results are also publicly available in a \textsc{Python} notebook. 
\end{abstract}

\maketitle

\section{Introduction}
\label{sec:intro}

Astrophysical observations have established that more than three quarters of the matter content of the universe are composed of dark matter. The nature of dark matter, however, remains elusive, and its very existence is one of the most compelling pieces of evidence for physics beyond the Standard Model of particle physics. A key step to unveil the composition of dark matter would be its direct detection in the laboratory~\cite{Baudis:2016qwx}. Such endeavor is led by international collaborations that use atomic nuclei as targets, in their aim to detect the nuclear recoils resulting from the scattering of dark matter particles~\cite{Angloher:2015ewa,Amole:2016pye,Armengaud:2016cvl,Akerib:2016vxi,Aprile:2017iyp,Agnese:2017jvy,Amaudruz:2017ekt,Cui:2017nnn,Agnese:2017njq,Agnes:2018ves,XMASS:2018bid,Aprile:2018dbl}. In spite of the very high sensitivities achieved by reduced backgrounds combined with extended exposures, there has been no conclusive evidence to date for the direct detection of dark matter. Next generations of experiments plan to push the frontier for dark matter direct detection by several orders of magnitude~\cite{Aprile:2015uzo,Akerib:2015cja,PandaXxT,Aalbers:2016jon}, until the background from coherent neutrino--nucleus scattering~\cite{Akimov:2017ade} becomes dominant.

Direct detection experiments are motivated by extensions of the Standard Model that propose dark matter candidates interacting with quarks and gluons, the Standard Model fields that ultimately form atomic nuclei. Prominent candidates are weakly interacting massive particles (WIMPs)~\cite{Bertone}. Because the WIMPs forming dark matter would be non-relativistic (NR), their scattering off atomic nuclei would transfer energies and momenta much smaller than the nuclear or nucleon masses. As a consequence nucleons and nuclei, instead of quarks and gluons, become the relevant degrees of freedom. The interpretation of the present experimental limits and future values of WIMP--nucleus cross sections, therefore, naturally depends on the nuclear physics aspects of the scattering. This information is encoded in the so-called nuclear structure factors~\cite{Engel:1992bf}. While for some simple cases a phenomenological prescription can be a good approximation to the structure factor~\cite{Lewin:1995rx,Vietze:2014vsa}, in general a good description of the nucleus, obtained with a dedicated nuclear many-body calculation, is needed.

In the absence of experimental data on the spin~\cite{Baudis:2013bba,McCabe:2015eia}, momentum-transfer~\cite{Rogers:2016jrx,Fieguth:2018vob}, or isospin dependence, the character of the WIMP--nucleus interaction is unknown. Nevertheless standard analyses usually assume so-called spin-independent (SI) interactions, which receive the coherent contribution of all nucleons in the nucleus. However, additional interactions are possible, and could be dominant if the SI coherence is compensated by suppressed values of the corresponding WIMP--nucleon couplings. Some examples analyzed experimentally are the so-called spin-dependent (SD)~\cite{Amole:2016pye,Aprile:2016swn,Fu:2016ega,Akerib:2017kat,Aprile:2019dbj,Amole:2019fdf} or momentum-transfer-dependent interactions~\cite{Angloher:2016jsl}.

In order to organize different possible WIMP--nucleus interactions, two alternative schemes have been proposed recently. On the one hand, a non-relativistic effective field theory (NREFT)~\cite{Fan:2010gt,Fitzpatrick:2012ix,Anand:2013yka} based on the lowest-order operators that can describe the coupling of a WIMP to a nucleon. The NREFT approach proposes a set of one-body operators, considered in  recent experimental analyses~\cite{Schneck:2015eqa,Aprile:2017aas,Xia:2018qgs,Angloher:2018fcs}.
On the other hand, an organization based on chiral effective field theory (ChEFT)~\cite{Epelbaum:2008ga,Machleidt:2011zz,Hammer:2012id}, a low-energy effective theory of quantum chromodynamics (QCD) that preserves the QCD symmetries, in particular capturing the important role played by pions at low energies. The ChEFT approach can be mapped onto the single-nucleon couplings of NREFT~\cite{Hoferichter:2015ipa}, implying certain interdependencies for the latter. In addition, ChEFT predicts consistent couplings of WIMPs to two nucleons~\cite{Prezeau:2003sv,Cirigliano:2012pq,Menendez:2012tm,Klos:2013rwa,Cirigliano:2013zta,Hoferichter:2016nvd,Hoferichter:2017olk,Andreoli:2018etf}, reflecting that nucleons are strongly interacting in nuclei. Such contributions occur, e.g., when the WIMP couples to a virtual pion exchanged between the two nucleons, an effect recently constrained for the first time by direct-detection experiments~\cite{Aprile:2018cxk}, for the case of a scalar WIMP--quark interaction.
Similar couplings to two nucleons are important in electromagnetic and weak transitions of atomic nuclei~\cite{Gazit:2008ma,Menendez:2011qq,Bacca:2014tla,Pastore:2017uwc}. Furthermore, ChEFT provides a power counting that suggests a hierarchy, guided by QCD, for the expected importance of the different NREFT operators~\cite{Hoferichter:2015ipa}.
This hierarchy is only tentative, because the couplings describing the interaction of the WIMP to the Standard Model fields are not known. A related ChEFT approach limited to WIMP interactions with one nucleon has been proposed in Refs.~\cite{Bishara:2016hek,Bishara:2017pfq}.

In this work we follow Ref.~\cite{Hoferichter:2016nvd} and combine the ChEFT framework with large-scale shell model nuclear many-body calculations to calculate the leading nuclear structure factors that exhibit the coherent contribution of several nucleons in the nucleus. We call these generalized SI interactions. We consider both WIMP couplings to one and two nucleons, with special emphasis on the two-nucleon contributions related to the diagonal and non-diagonal parts of the energy-momentum tensor. 
We note that at this point our structure factors are not fully consistent, in the sense that our many-body calculations are based on phenomenological interactions instead of ChEFT. Such consistent studies are presently only available for few-nucleon systems~\cite{Korber:2017ery}. In addition, nuclear response functions based on nuclear states obtained using ChEFT interactions are available for light nuclei~\cite{Gazda:2016mrp}. However, such very light isotopes are not used in leading direct detection experiments. While Ref.~\cite{Hoferichter:2016nvd} was limited to a xenon target, here we study all the stable isotopes of fluorine, silicon, argon, and germanium, the nuclear targets used and considered in present and future direct detection experiments. In addition, we provide a \textsc{Python} notebook to facilitate the use of our structure factors for both theorists and experimentalists.

The rest of the article is organized as follows. In Sec.~\ref{sec:formalism} we describe the formalism and discuss which terms we include in the cross section,
focusing on a spin-$1/2$ WIMP. Section~\ref{sec:shell_model} introduces the nuclear structure many-body calculations that describe the nuclear targets used in direct-detection experiments. The results for our calculated structure factors are given in Sec.~\ref{sec:structure_factors}. We distinguish between WIMP couplings to one nucleon, discussed in Sec.~\ref{sec:1b}, and the coupling to two nucleons, the subject of Sec.~\ref{sec:2b}. The most important new aspects of our implementation of two-body effects are highlighted in Sec.~\ref{sec:two_body_pheno}. 
We conclude with a summary of the main findings of this work in Sec.~\ref{sec:summary}.
Details of the matching to NREFT, the nucleon matrix elements, matching for a spin-$0$ WIMP, and the calculation of two-body interactions are provided in Appendices~\ref{sec:NREFT}--\ref{sec:app2bc}. 
All our results are also available in the form of a \textsc{Python} notebook, with a brief User's guide in Appendix~\ref{sec:user_guide}. 

\section{Formalism}
\label{sec:formalism}

Based on Ref.~\cite{Hoferichter:2016nvd} we consider the following cross section for the generalized SI WIMP--nucleus scattering:
\begin{align}
\label{structure_factors}
\frac{\diff \sigma}{\diff q^2}&=\frac{1}{4\pi v^2}\bigg|\sum_{I=\pm}\Big(c_I^M-\frac{q^2}{m_N^2} \, \dot c_I^M\Big)\F_I^M(q^2)+c_\pi\F_\pi(q^2)\notag\\
&+c_\text{b}\F_\text{b}(q^2)+\frac{q^2}{2\mN^2}\sum_{I=\pm}c_I^{\Phi''}\F_I^{\Phi''}(q^2)\bigg|^2\notag\\
&+\frac{1}{4\pi v^2}\sum_{i=5,8,11}\bigg|\sum_{I=\pm}\xi_i(q,v^\perp_T)c_I^{M,i}\F_I^M(q^2)\bigg|^2,
\end{align}
where $q=|\qq|$ is the three-momentum transfer between the WIMP and the nucleus, and $\mc$, $\mN$, and $m_\N$ denote the WIMP, nucleon, and nuclear masses, respectively.
The relative velocity of the WIMP is $v=|\vv|$,
while $(v^\perp_T)^2=v^2-q^2/(4\mu_\N^2)$~\cite{Fitzpatrick:2012ix} is the velocity of the WIMP with respect to the nucleus, with reduced mass $\mu_\N = m_\N \mc/(m_\N+\mc)$. Each term in Eq.~\eqref{structure_factors} contains a structure factor $\F$ and a coupling $c$. The structure factors $\F$ encode the nuclear structure aspects of the scattering, and are the main subject of this work. The couplings $c$ are a convolution of the Wilson coefficients, which describe the fundamental interaction of the WIMPs with the quarks and gluons, and the hadronic matrix elements. 
They therefore depend on the particular scenario for physics beyond the Standard Model considered. Finally, the kinematic factors
\beq
\xi_5=\frac{\muN q v^\perp_T}{2\mc\mN},\qquad \xi_8=v^\perp_T,\qquad \xi_{11}=-\frac{q}{2\mc},
\eeq
with the reduced mass $\muN = \mN \mc/(\mN+\mc)$, appear in the contribution from the subleading NREFT operators $\Op_{5,8,11}$, defined in Appendix~\ref{sec:NREFT}. 

\subsection{Effective Lagrangian and couplings}
\label{sec:lagrangian}

To be definite, we consider the case of a spin-$1/2$, Standard-Model singlet $\chi$ throughout the main text, but to demonstrate that the decomposition in Eq.~\eqref{structure_factors} applies in full generality
we also provide the matching relations for a spin-$0$ WIMP, see Appendix~\ref{sec:spin0}.
For spin-$1/2$, the relevant terms in the effective Lagrangian~\cite{Goodman:2010ku,Drees:1993bu,Belanger:2008sj} are
\begin{align}
\label{Lagr}
\mathcal{L}_{\chi}&=\mathcal{L}_{\chi}^{(6)}+\mathcal{L}_{\chi}^{(7)}+\mathcal{L}_{\chi}^{(8)},\\
\mathcal{L}_{\chi}^{(6)}&=\frac{1}{\Lambda^2}\sum_{q}\Big[C_q^{VV}\bar\chi\gamma^\mu\chi \,\bar q\gamma_\mu q
+C_q^{AA}\bar\chi\gamma^\mu\gamma_5\chi\, \bar q\gamma_\mu\gamma_5 q\notag\\
&+C_q^{AV}\bar\chi\gamma^\mu\gamma_5\chi \,\bar q\gamma_\mu q\notag\\
&+C_q^{TT}\bar\chi\sigma^{\mu\nu}\chi \,\bar q\sigma_{\mu\nu} q
+\tilde C_q^{TT}\bar\chi\sigma^{\mu\nu}i\gamma_5\chi\, \bar q\sigma_{\mu\nu}q\Big],\notag\\
\mathcal{L}_{\chi}^{(7)}
&=\frac{1}{\Lambda^3}\bigg\{\sum_{q} \Big(C_{q}^{SS}+\frac{8\pi}{9}C'^S_{g}\Big)\bar \chi \chi \,m_q\bar q q\notag\\
&+ \sum_{q} \Big(C_{q}^{PS}+\frac{8\pi}{9}\tilde C'^S_{g}\Big)\bar \chi i\gamma_5\chi \,m_q\bar q q\notag\\
&-\frac{8\pi}{9}C'^S_{g}\bar \chi\chi\, \theta^\mu_\mu -\frac{8\pi}{9}\tilde C'^S_{g}\bar \chi i\gamma_5\chi\, \theta^\mu_\mu\bigg\},\notag\\
\mathcal{L}_{\chi}^{(8)}&=\frac{1}{\Lambda^4}\bigg\{\sum_qC_q^{(2)}\bar \chi \gamma_\mu i\partial_\nu \chi \,\bar \theta^{\mu\nu}_q
+C_g^{(2)} \bar \chi \gamma_\mu i\partial_\nu \chi\,\bar \theta^{\mu\nu}_g\bigg\},\notag
\end{align}
where we have only listed operators that can lead to coherently enhanced responses or feature in the standard SD interaction,
and $C_q^{VV}=C_q^{TT}=\tilde C_q^{TT}=0$ for a Majorana particle.
In $\mathcal{L}_{\chi}^{(7)}$ we already integrated out the heavy quarks~\cite{Shifman:1978zn}, whose effect is absorbed in
\begin{align}
C'^S_{g}&=C^S_{g}-\frac{1}{12\pi}\sum_{Q=c,b,t}C^{SS}_{Q},\notag\\
\tilde C'^S_{g}&=\tilde C^S_{g}-\frac{1}{12\pi}\sum_{Q=c,b,t}C^{PS}_{Q},
\end{align}
while elsewhere the sum runs, in principle, over all quark flavors $q$. $C^S_{g}$ and $\tilde C^S_{g}$ are the original coefficients of the gluon operator $\bar\chi\chi\,\alpha_s G_{\mu\nu}^aG^{\mu\nu}_a$ 
and $\bar\chi i\gamma_5\chi\,\alpha_s G_{\mu\nu}^aG^{\mu\nu}_a$, respectively,
rewritten in terms of the trace of the energy-momentum tensor:
\beq
\theta^\mu_\mu=\sum_q m_q\bar q q - \frac{9}{8\pi} \alpha_sG_{\mu\nu}^aG^{\mu\nu}_a+\Order(\alpha_s^2).
\eeq 
Moreover, we 
introduced its traceless components $\bar \theta^{\mu\nu}=\bar \theta^{\mu\nu}_q+\bar \theta^{\mu\nu}_g$ in the context of the dimension-$8$ spin-$2$ contribution:
\begin{align}
\label{spin_2_definition}
 \bar \theta^{\mu\nu}_q&=\frac{1}{2}\bar q\Big(\gamma^{\{\mu}i D_-^{\nu\}}-\frac{m_q}{2}g^{\mu\nu}\Big)q,\notag\\
 \bar \theta^{\mu\nu}_g&=\frac{g^{\mu\nu}}{4}G^a_{\lambda\sigma}G_a^{\lambda\sigma}-G^{\mu\lambda}_aG^\nu_{a\lambda},
\end{align}
with covariant derivative $D_-^\mu=\overrightarrow{D}^\mu-\overleftarrow{D}^\mu$ and symmetrizer $A^{\{\mu}B^{\nu\}}=(A^\mu B^\nu+A^\nu B^\mu)/2$.

In detail, the coefficients in Eq.~\eqref{structure_factors} are given by (nucleons $N$ encompass protons and neutrons, $N=p$ or $n$)
\begin{align}
\label{c_coeff}
 c_\pm^M&=\frac{\zeta}{2}\Big[f_p\pm f_n+f_1^{V,p}\pm f_1^{V,n}+\frac{3}{4}\big(f_p^{(2)}\pm f_n^{(2)}\big)\Big],\notag\\
\dot c_\pm^M&=\frac{\zeta m_N^2}{2}\bigg[\dot f_p\pm \dot f_n+\dot f_1^{V,p}\pm \dot f_1^{V,n}\notag\\
&+\frac{1}{4\mN^2}\big(f_2^{V,p}\pm f_2^{V,n}\big)
+\frac{1}{2\mc\mN}\big(f_1^{T,p}\pm f_1^{T,n}\big)\notag\\
&+\frac{1}{\mc\mN}\big(f_2^{T,p}\pm f_2^{T,n}\big)\bigg],\notag\\
c_\pi&=\zeta \Big(f_\pi+2f_\pi^\theta-\frac{1}{2}f_\pi^{(2)}\Big),\qquad
c_\text{b}=\zeta\Big(f_\pi^\theta+\frac{1}{4}f_\pi^{(2)}\Big),\notag\\
c_\pm^{\Phi''}&=\frac{\zeta}{2}\bigg[f_2^{V,p}\pm f_2^{V,n}+\frac{1}{2}\Big(1+\frac{\muN}{\mc}\Big)\big(f_1^{V,p}\pm f_1^{V,n}\big)\notag\\
&+\frac{\mN}{\mc}\Big(1+\frac{\muN}{\mN}\Big)\big(f^{T,p}_1\pm f^{T,n}_1\big)\bigg],\notag\\
c_\pm^{M,5}&=\frac{\zeta}{2}\bigg[\Big(1+\frac{\mN}{2\mc}\Big)\big(f_1^{V,p}\pm f_1^{V,n}\big)\notag\\
&+\Big(1+\frac{2\mc}{\mN}\Big)\big(f^{T,p}_1\pm f^{T,n}_1\big)\notag
+4\frac{\mc}{\muN}\big(f^{T,p}_2\pm f^{T,n}_2\big)\bigg],\notag\\
c_\pm^{M,8}&=\frac{\zeta}{2}\big(\tilde f_1^{V,p}\pm \tilde f_1^{V,n}\big),\notag\\
c_\pm^{M,11}&=\frac{\zeta}{2}\bigg[\tilde f_p\pm \tilde f_n-2\frac{\mc}{\mN}\big(\tilde f^{T,p}_1\pm \tilde f^{T,n}_1\big)\notag\\
&-4\frac{\mc}{\mN}\big(\tilde f^{T,p}_2\pm \tilde f^{T,n}_2\big)\bigg],
\end{align}
where $\zeta=1 (2)$ for a Dirac (Majorana) particle.\footnote{The matching relations thus include the Majorana symmetry factor $2$. In the Dirac case, the relative sign between scalar and vector coefficients changes for the cross section of $\bar\chi$--nucleon scattering~\cite{Belanger:2008sj}.}

The various couplings
refer to combinations of Wilson coefficients from Eq.~\eqref{Lagr} and nucleon matrix elements in the scalar ($f_N$, $\dot f_N$, $\tilde f_N$), 
vector ($f_1^{V,N}$, $\dot f_1^{V,N}$, $f_2^{V,N}$, $\tilde f_1^{V,N}$), tensor ($f^{T,N}$, $\tilde f^{T,N}$), and spin-$2$ ($f_N^{(2)}$) channels.
The explicit expressions, together with the matching onto NREFT, are collected in Appendix~\ref{sec:NREFT}. 

\subsection{Pion matrix elements}

Here, we highlight the couplings to the pion in Eq.~\eqref{c_coeff}:
\begin{align}
\label{hadronic_couplings_pion}
 f_\pi&=\frac{\mpi}{\Lambda^3}\sum_{q=u,d} \Big(C_{q}^{SS}+\frac{8\pi}{9}C'^S_{g}\Big)f_q^\pi,\notag\\
f_\pi^\theta&=-\frac{\mpi}{\Lambda^3}\frac{8\pi}{9}C'^S_{g},\notag\\
f_\pi^{(2)}&=\frac{\mc \mpi}{\Lambda^4}\bigg(\sum_q C_q^{(2)} f^{(2)}_{q,\pi}+ C_g^{(2)} f^{(2)}_{g,\pi}\bigg).
\end{align}
In this case, the relevant matrix elements are the following: first, the scalar couplings to $u$- and $d$-quarks are given by
\begin{align}
f_u^\pi&=\frac{\muu}{\muu+\md}=\frac{1}{2}\big(1-\xi_{ud}\big)=0.32(2),\notag\\
f_d^\pi&=\frac{\md}{\muu+\md}=\frac{1}{2}\big(1+\xi_{ud}\big)=0.68(2),
\end{align}
where $\xi_{ud}=(\md-\muu)/(\md+\muu)=0.37(3)$ with $\muu/\md=0.46(3)$ from Ref.~\cite{Aoki:2016frl} (and $f_s^\pi \approx 0$).
Second, the coupling to the trace anomaly is only sensitive to the sum $f_u^\pi+f_d^\pi=1$, so that no new coupling appears in $f_\pi^\theta$. 
The spin-$2$ couplings are related to moments of pion parton distribution functions (PDFs) $q_\pi(x)$:
\beq
\label{spin_2_coupling_pion}
f^{(2)}_{q,\pi}=\int_0^1\diff x\, x \big(q_\pi(x)+\bar q_\pi(x)\big),
\eeq
and similarly for the gluon, subject to the constraint
\beq
\label{sum_rule_pion}
\sum_q f^{(2)}_{q,\pi} + f^{(2)}_{g,\pi}=1.
\eeq
The pion PDFs have often been studied assuming a single valence and sea distribution with moments $\langle x \rangle_\text{v}^\pi$ and $\langle x \rangle_\text{s}^\pi$~\cite{Badier:1983mj,Sutton:1991ay,Wijesooriya:2005ir}, in such a way that
\beq
 f_{u,\pi}^{(2)}=f_{d,\pi}^{(2)}=\langle x \rangle_\text{v}^\pi+2\langle x \rangle_\text{s}^\pi,\qquad f_{s,\pi}^{(2)}=2\langle x \rangle_\text{s}^\pi,
\eeq
and by means of the sum rule in Eq.~\eqref{sum_rule_pion}
\begin{align}
 f_{u,\pi}^{(2)}&=f_{d,\pi}^{(2)}=\frac{1}{3}\Big(1-f^{(2)}_{g,\pi}+\langle x \rangle_\text{v}^\pi\Big),\notag\\
 f_{s,\pi}^{(2)}&=\frac{1}{3}\Big(1-f^{(2)}_{g,\pi}-2\langle x \rangle_\text{v}^\pi\Big).
\end{align}
With $f^{(2)}_{g,\pi}=0.47(15)$~\cite{Badier:1983mj} and $\langle x \rangle_\text{v}^\pi=0.217(11)$~\cite{Wijesooriya:2005ir} this gives
\beq
\label{f2_pion}
f_{u,\pi}^{(2)}=f_{d,\pi}^{(2)}=0.26(5),\qquad f_{s,\pi}^{(2)}=0.01(5),
\eeq
consistent with a recent calculation in lattice QCD $\langle x \rangle_\text{v}^\pi=0.214(15)(^{+12}_{-9})$~\cite{Abdel-Rehim:2015owa}.\footnote{Note that all spin-$2$ couplings are scale-dependent quantities, so that, at a scale of $\mu=2\GeV$ in $\overline{\text{MS}}$, the value from Ref.~\cite{Wijesooriya:2005ir} actually becomes $\langle x \rangle_\text{v}^\pi=0.256(13)$~\cite{Abdel-Rehim:2015owa}. This has been taken into account in Eq.~\eqref{f2_pion}.}
A more recent, global analysis finds at $\mu=2\GeV$~\cite{Barry:2018ort}
\begin{align}
\label{f2_pion_JAM}
f_{u,\pi}^{(2)}&=f_{d,\pi}^{(2)}=0.298(8),\qquad f_{s,\pi}^{(2)}=0.055(4),\notag\\
f^{(2)}_{g,\pi}&=0.341(19),
\end{align}
in agreement with Eq.~\eqref{f2_pion}, but considerably more precise. 
For completeness, we reproduce the analogous expressions for the nucleon matrix elements in Appendix~\ref{sec:NREFT}.

\subsection{Dipole operators}

For a Dirac WIMP, a possible extension beyond Eq.~\eqref{Lagr} concerns dark matter candidates with a non-vanishing dipole moment, corresponding to the effective dimension-$5$ Lagrangian~\cite{Barger:2010gv,Banks:2010eh}
\beq
\label{dipole_L}
\mathcal{L}_{\chi}^{(5)}=\frac{C_F}{\Lambda} \bar \chi \sigma^{\mu\nu}\chi\, F_{\mu\nu} + \frac{\tilde C_F}{\Lambda} \bar \chi \sigma^{\mu\nu}\chi\, \tilde F_{\mu\nu},
\eeq
where due to $\sigma^{\mu\nu}\gamma_5=\frac{i}{2}\eps^{\mu\nu\alpha\beta}\sigma_{\alpha\beta}$ the second term involving the dual field strength tensor $\tilde F^{\mu\nu}=\frac{1}{2}\eps^{\mu\nu\alpha\beta}F_{\alpha\beta}$ is equivalent to $-\bar \chi \sigma^{\mu\nu}i\gamma_5\chi F_{\mu\nu}$. These operators produce long-range tree-level interactions via the exchange of a photon,
which leads to the NR one-body amplitudes
\begin{align}
\label{dipole_NR}
\M_{1,\text{NR}}^{F}&=\frac{e C_F}{\Lambda}\bigg[\Big(\frac{1}{\mc}\Op_1-\frac{4}{q^2}\Op_5\Big)F_1^N(t)\notag\\
&+\frac{4}{\mN}\Big(\Op_4-\frac{1}{q^2}\Op_6\Big)\Big(F_1^N(t)+F_2^N(t)\Big)\bigg],\notag\\
\M_{1,\text{NR}}^{\tilde F}&=-\frac{4e \tilde C_F}{\Lambda}\frac{1}{q^2}\Op_{11}\Big(F_1^N(t)-\frac{q^2}{4\mN^2}F_2^N(t)\Big).
\end{align}
Here, $F_{1/2}^N(t)$ are the Dirac/Pauli form factors of the nucleon with the full dependence on the relativistic momentum transfer $t=-q^2$ (up to relativistic corrections). 
The corresponding extension of Eq.~\eqref{structure_factors} is straightforward, with the photon poles introducing new terms, besides additional contributions to some of the existing ones.
Formally, this amounts to ($q$-dependent) terms in the matching relations in Eq.~\eqref{c_coeff}:
\begin{align}
 \Delta c_\pm^M&=\frac{1}{2}\frac{e C_F}{\Lambda\mc},\notag\\
 \Delta \dot c_\pm^M&=\frac{1}{2}\frac{e C_F}{\Lambda\mc}\frac{\mN^2}{6}\big(\langle r_1^2\rangle^p\pm\langle r_1^2\rangle^n\big),\notag\\
 \Delta c_\pm^{M,5}&=-\frac{2e C_F(\mc+\mN)}{\Lambda}\bigg[\frac{1}{q^2}-\frac{1}{6}\big(\langle r_1^2\rangle^p\pm\langle r_1^2\rangle^n\big)\bigg],\notag\\
 \Delta c_\pm^{M,11}&=\frac{2e \tilde C_F\mc}{\Lambda}\bigg[\frac{1}{q^2}-\frac{1}{6}\big(\langle r_E^2\rangle^p\pm\langle r_E^2\rangle^n\big)\bigg],
 \label{Eq:matching_dipole}
\end{align}
where we have already taken $\zeta=1$ due to the absence of tensor currents for Majorana particles.
The Dirac-form-factor radii are related to the Sachs ones by
\beq
\langle r_1^2\rangle^N=\langle r_E^2\rangle^N-\frac{3\kappa_N}{2\mN^2}.
\eeq
In the spirit of the present study, Eq.~\eqref{Eq:matching_dipole} neglects the non-coherent contributions from $\Op_4$ and $\Op_6$.
Further, there are many more dimension-$7$ operators involving the (electroweak) 
field strength tensors~\cite{Crivellin:2014gpa,Brod:2017bsw}, and matching relations similar to Eqs.~\eqref{dipole_NR} and~\eqref{Eq:matching_dipole} could be extended accordingly.
 
In the form~\eqref{dipole_NR} the amplitudes automatically include radius corrections,
subsumed in the full electromagnetic form factors.
In addition, Eq.~\eqref{dipole_L} could in principle produce new two-body currents.
In the end, such terms take a similar form as the axial-vector--vector two-body currents identified in Ref.~\cite{Hoferichter:2015ipa},
and are only suppressed by two chiral orders compared to the leading pieces of Eq.~\eqref{dipole_NR}.
However, as argued in Ref.~\cite{Hoferichter:2016nvd}, after summation over spins their isospin structure $[\ttau_1\times\ttau_2]^3$ leaves only an isovector coherent enhancement suppressed by $(N-Z)/A$ with respect to the scalar two-body current. We will continue to neglect such effects in the remainder of this work.

\begin{table*}[t]
	\centering
	\renewcommand{\arraystretch}{1.3}
	\begin{tabular}{cccccc}
		\toprule
		Structure factor & QCD operators & Chiral scaling & NR operators & Overall scaling & Interference with $\Op_1$\\\colrule
		$\F^M$ & $\bar\chi\chi\, m_q\bar q q$ & $\Order(\mpi^2)=\Order(p^2)$ & $\Op_1$ & $\Order\big(\frac{\mpi^2}{\Lambda_\chi^2}A\big)$ & yes\\
		& $\bar\chi\chi\, \theta^\mu_\mu$ & $\Order(1)$ & $\Op_1$ & $\Order(A)$ & yes\\
		& $\bar\chi\gamma_\mu i\partial_\nu\chi\, \bar\theta^{\mu\nu}$ & $\Order(1)$ & $\Op_1$ & $\Order(A)$ & yes\\
		& $\bar\chi\gamma^\mu\chi\, \bar q\gamma_\mu q$ & $\Order(1)$ & $\Op_1$& $\Order(A)$ & yes\\
		&  & $\Order\big(\frac{q v^\perp}{\mc+\mN}\big)=\Order(p^4)$ & $\Op_5$ & $\Order(\frac{q v^\perp_T}{\mc+\mN}A)$ & no\\
		& $\bar\chi\sigma^{\mu\nu}\chi\, \bar q\sigma_{\mu\nu} q$ & $\Order\big(\frac{q^2}{\mN\mc}\big)=\Order(p^4)$ & $\Op_1$ & $\Order\big(\frac{q^2}{\mN\mc}A\big)$ & yes\\
		&  & $\Order\big(\frac{q v^\perp}{\mN}\big)=\Order(p^4)$ & $\Op_5$ & $\Order\big(\frac{q v^\perp_T}{\mN}A\big)$ & no\\
		& $\bar\chi\gamma^\mu\gamma_5\chi\, \bar q\gamma_\mu q$ & $\Order(v^\perp)=\Order(p^2)$ & $\Op_8$ & $\Order(v^\perp_T A)$ & no\\
		& $\bar\chi\sigma^{\mu\nu}i\gamma_5\chi\, \bar q \sigma_{\mu\nu} q$ & $\Order\big(\frac{q}{\mN}\big)=\Order(p^2)$ & $\Op_{11}$ & $\Order\big(\frac{q}{\mN}A\big)$ & no\\
		& $\bar\chi i\gamma_5\chi\, m_q\bar q q$ & $\Order\big(\frac{q\mpi^2}{\mc}\big)=\Order(p^4)$ & $\Op_{11}$ & $\Order\big(\frac{q\mpi^2}{\mc\Lambda_\chi^2}A\big)$ & no\\\colrule
		$\F_\pi$ & $\bar\chi\chi\, m_q\bar q q$ & $\Order(\mpi^3)=\Order(p^3)$ & -- & $\Order\big(\frac{\mpi^3}{\Lambda_\chi^3}A\big)$ & yes\\
		& $\bar\chi\chi\, \theta^\mu_\mu$ & $\Order(\mpi^3)=\Order(p^3)$ & -- & $\Order\big(\frac{\mpi^3}{\Lambda_\chi^3}A\big)$ & yes\\
		& $\bar\chi\gamma_\mu i\partial_\nu\chi\, \bar\theta^{\mu\nu}$ & $\Order(\mpi^3)=\Order(p^3)$ & -- & $\Order\big(\frac{\mpi^3}{\Lambda_\chi^3}A\big)$ & yes\\\colrule
		$\F_\text{b}$ & $\bar\chi\chi\, \theta^\mu_\mu$ & $\Order(\mpi^3)=\Order(p^3)$ & -- & $\Order\big(\frac{\mpi^3}{\Lambda_\chi^3}A\big)$ & yes\\
		& $\bar\chi\gamma_\mu i\partial_\nu\chi\, \bar\theta^{\mu\nu}$ & $\Order(\mpi^3)=\Order(p^3)$ & -- & $\Order\big(\frac{\mpi^3}{\Lambda_\chi^3}A\big)$ & yes\\\colrule
		$\F^{\Phi''}$ & $\bar\chi\gamma^\mu\chi\, \bar q\gamma_\mu q$ & $\Order(q v^\perp)=\Order(p^3)$ & $\Op_3$ & $\Order\big(\frac{q^2}{\mN\Lambda_\chi} \xi A\big)$ & yes\\
		& $\bar\chi\sigma^{\mu\nu}\chi\, \bar q\sigma_{\mu\nu} q$ & $\Order\big(\frac{q v^\perp}{\mc}\big)=\Order(p^4)$ & $\Op_3$ & $\Order\big(\frac{q^2}{\mN\mc}\xi A\big)$ & yes\\\colrule
		$S_{ij}$ & $\bar\chi \gamma^\mu\gamma_5\chi\, \bar q\gamma_\mu\gamma_5 q$ & $\Order(1)$ & $\Op_4$, $\Op_6$ & $\Order(1)$ & no\\
		& $\bar\chi \sigma^{\mu\nu}\chi\, \bar q\sigma_{\mu\nu} q$ & $\Order(1)$ & $\Op_4$ & $\Order(1)$ & no\\
		& $\bar\chi \gamma_5\chi\,\bar q\gamma_5 q$ & $\Order(1)$ & $\Op_6$ & $\Order(1)$ & no\\
		\botrule
	\end{tabular}
	\renewcommand{\arraystretch}{1.0}
	\caption{Nuclear structure factors and associated QCD and NR operators. Isospin indices are suppressed because the identification pertains to isoscalar and isovector combinations alike.
	The table summarizes all operators that can be coherently enhanced. 
	Likewise, the table only shows coherently enhanced contributions for operators with velocity $v^\perp$, which in nuclei produce both a term that behaves as $q/\mN$ and a remainder determined by the target velocity $v^\perp_T\sim 10^{-3}$ (see Appendix~\ref{sec:NREFT} for more details).
	For comparison the last rows for $S_{ij}$ show operators that lead to SD interactions.
	In the chiral counting we have $\mpi=\Order(p)$, $v^\perp=\Order(p^2)$, with relativistic corrections counted as $\partial/\mN=\Order(p^2)$ and $\partial/\mc=\Order(p^2)$.
	Two-body structure factors, $\F_\pi$ and $\F_\text{b}$, cannot be matched onto single-nucleon NREFT operators, and
	the quasi-coherence of $\F^{\Phi''}$ is characterized by $\xi\sim 0.2$.
	Finally, the last column indicates whether each entry interferes with the leading $\Op_1$ operator.
	}
	\label{tab:structure_factors}
\end{table*}

\subsection{Scaling of operators}
\label{sec:scaling}

The global picture that arises in this way is summarized in Table~\ref{tab:structure_factors}. For each nuclear structure factor we list: the relativistic operators that contribute, 
the corresponding NREFT operator if applicable, the chiral scaling as well as the overall scaling including coherence, and finally whether or not the respective contribution interferes 
with the leading $\Op_1$ operator.

Table~\ref{tab:structure_factors} reflects the strategy underlying the present work, in that it includes all contributions, either two-body currents or subleading one-body operators, that appear up to third order in the chiral expansion and receive some form of coherent enhancement. Selected coherent chiral fourth-order contributions are shown as well, mainly because
in many cases this is the order when some of the coherent NREFT operators first enter. For comparison, the table also includes the leading relativistic operators that produce the NR expansion related to the standard SD interactions, even if these are not coherent.  

\section{Nuclear structure calculations}
\label{sec:shell_model}

The calculation of the nuclear structure factors requires a many-body approach that describes the ground states of the target nuclei considered. As in previous works~\cite{Menendez:2012tm,Klos:2013rwa,Baudis:2013bba,Vietze:2014vsa,Hoferichter:2016nvd} we use the nuclear shell model, one of the most successful many-body approaches in medium-mass and heavy nuclei~\cite{Caurier:2004gf}. For all calculations we have used the shell model code ANTOINE~\cite{Caurier:1999,Caurier:2004gf}.

The shell model is based on the solution of the quantum many-body problem in a reduced configuration space where the Schr\"odinger equation for the nuclear ground and low-energy excited states can be solved exactly. We highlight two important aspects. First, the configuration space used in the calculation needs to capture the nuclear structure properties relevant for the process of interest. The limitation of using a restricted configuration space stems from the difficulty to solve the nuclear many-body problem in a nontruncated space for heavier nuclei. For nuclear targets used in direct detection experiments, calculations of structure factors without such truncations exist up to $^4$He~\cite{Korber:2017ery,Gazda:2016mrp,Andreoli:2018etf} and could be performed in the near future up to $^{40}$Ca. Second, calculations must use an effective interaction appropriate for such a configuration space.

Until very recently, nuclear shell-model calculations relied on phenomenological effective interactions. In spite of being derived from nucleon--nucleon ($NN$) scattering data, these interactions have to be adjusted phenomenologically, mostly the part that describes the single-particle aspects of the nuclear interaction, referred to as monopole part, in order to achieve a better agreement with the nuclear structure of heavier nuclei. Progress in nuclear theory has improved this picture by including in the starting point of the derivation of effective interactions, in addition to $NN$ interactions, also nuclear forces between three nucleons, 3$N$ forces. Three-nucleon forces are the analog of two-body currents in the coupling to external probes, such as WIMPs, to nucleons. Consistent $NN$ and 3$N$ interactions can be derived in the ChEFT framework~\cite{Hammer:2012id,Machleidt:2011zz,Epelbaum:2008ga}. Starting from these ChEFT $NN$ and 3$N$ forces it is possible to derive effective interactions that, without further phenomenological adjustments, reproduce nuclear spectroscopy rather well in nuclei with nucleon number $A$ comparable to direct-detection nuclear targets~\cite{Hebeler:2015hla,Hergert:2015awm,Hagen:2013nca}. First studies have just started to extend these techniques to study electromagnetic and weak transitions in medium-mass nuclei~\cite{Hagen:2015yea,Parzuchowski:2017wcq,Morris:2017vxi}.
In addition, the effective-theory character of ChEFT, combined with the consistency between nuclear forces and currents, provides a framework to quantify nuclear-structure uncertainties~\cite{Epelbaum:2014efa,Furnstahl:2015rha,Carlsson:2015vda}. In this work, however, we follow the standard approach and use phenomenological effective interactions. This prevents the assessment of reliable nuclear-structure uncertainties, which will become possible in the future.

The configuration spaces and effective interactions we have used are described below. All single-particle orbitals belong to a three-dimensional harmonic oscillator basis $nl_j$, where $n$ is the principal quantum number, and $l,j$ denote the orbital and total angular momentum.

\begin{figure}[t]
	\begin{center}
		\includegraphics[width=0.48\textwidth,clip=]{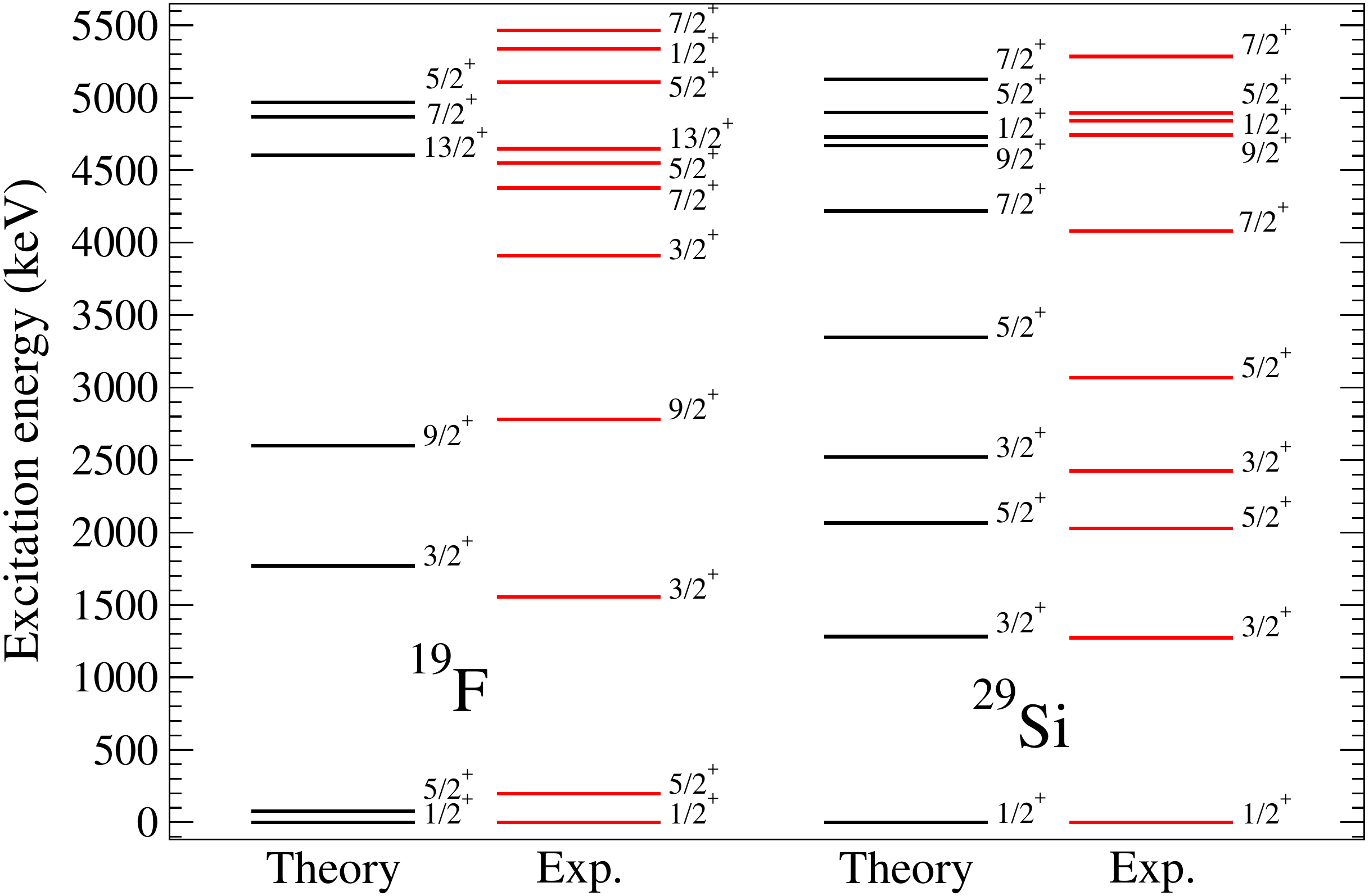}
	\end{center}
	\caption{Calculated $^{19}$F and $^{29}$Si spectra compared to experiment.\label{fig:f19si29_spectrum}}
\end{figure}

The lightest target we have studied has only one stable isotope, $^{19}$F, which we already considered in Ref.~\cite{Klos:2013rwa}. Here we use the same USDB effective interaction and the $0d_{5/2}$, $1s_{1/2}$, and $0d_{3/2}$ single-particle orbitals, for both neutrons and protons. This configuration space is known as the $sd$ shell. Reference~\cite{Klos:2013rwa} used the same configuration space and effective interaction for $^{29}$Si, and here we extend the study to the other two stable silicon isotopes, $^{28,30}$Si.

\begin{figure}[t]
	\begin{center}
		\includegraphics[width=0.48\textwidth,clip=]{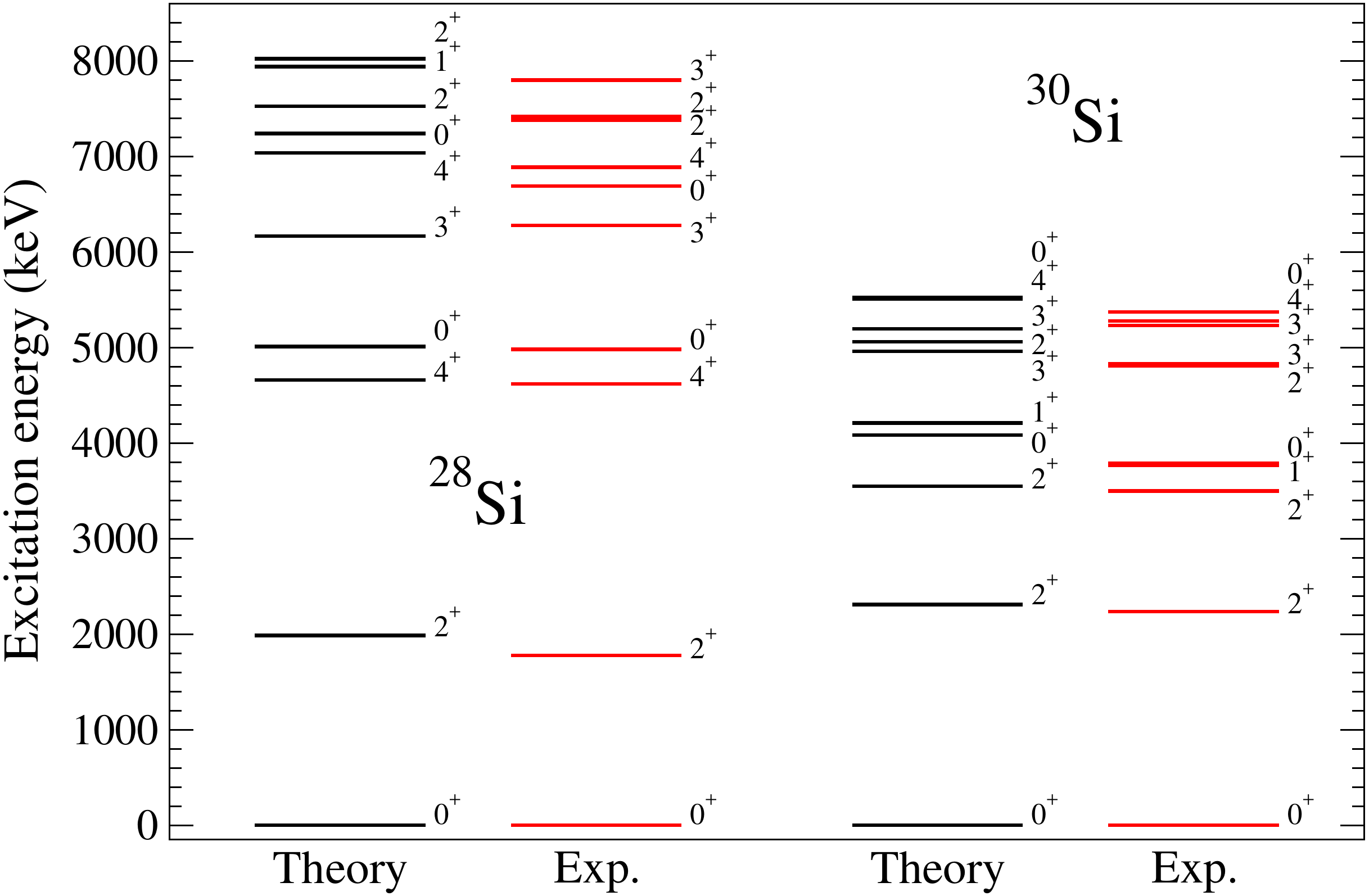}
	\end{center}
	\caption{Calculated $^{28,30}$Si spectra compared to experiment.\label{fig:si2830_spectrum}}
\end{figure}

\begin{figure}[t]
	\begin{center}
		\includegraphics[width=0.48\textwidth,clip=]{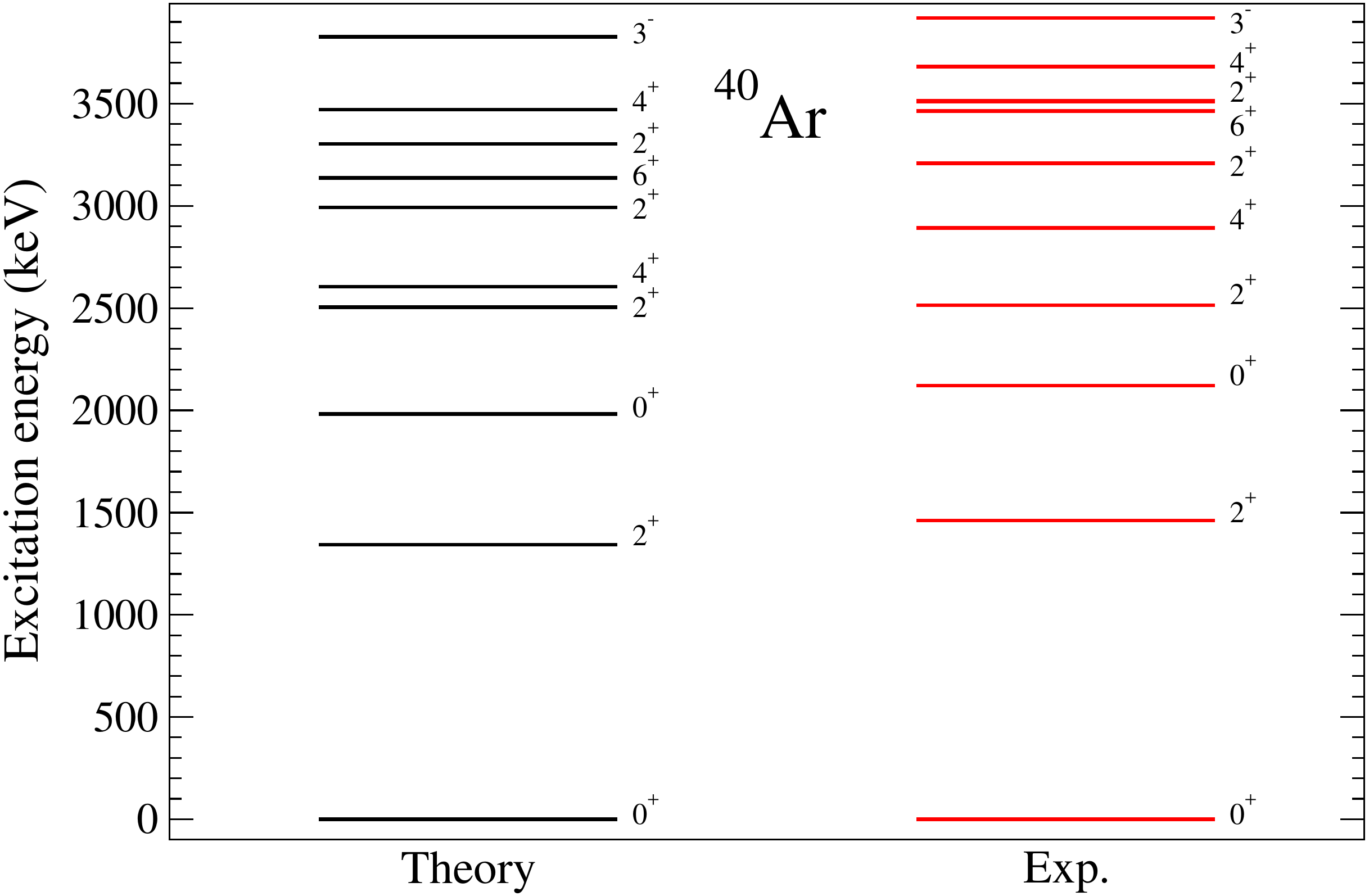}
	\end{center}
	\caption{Calculated $^{40}$Ar spectrum compared to experiment.\label{fig:ar_spectrum}}
\end{figure}

For argon, we study only $^{40}$Ar. The configuration space we consider is significantly larger, and comprises seven single-particle orbitals for neutrons and protons, the three $sd$-shell orbitals plus the $0f_{7/2}$, $1p_{3/2}$, $0f_{5/2}$, and  $1p_{3/2}$, where the latter comprise the so-called $pf$ shell. We use the SDPF.SM effective interaction, which describes well the electromagnetic properties of ground states and the coexistence of spherical and deformed states in this mass region~\cite{Caurier:2007ee,Ruiz:2015yra}. In order to make the diagonalizations in the configuration space feasible, we need to truncate our many-body calculations, by keeping the $0d_{5/2}$ orbital filled with nucleons, and restricting the number of excitations from $sd$-shell to $pf$-shell orbitals to 8. These are similar truncations to those in Ref.~\cite{Caurier:2007ee}, and limit the dimension of the diagonalization to $5\cdot10^8$. We have also preformed calculations of the stable isotopes $^{36,38}$Ar which are of the same quality as those for $^{40}$Ar. Nonetheless we have not included them in our study, because the natural abundance of these isotopes is very minor with less than $0.3\%$.

Finally, there are five stable stable germanium isotopes $^{70,72,73,74,76}$Ge. Consistently with our study of $^{73}$Ge in Ref.~\cite{Klos:2013rwa} we use the RG effective interaction~\cite{Menendez:2009xa} in a configuration space consisting on the $1p_{3/2}$, $0f_{5/2}$, $1p_{3/2}$, and $g_{9/2}$ single-particle orbitals. We have performed calculations with alternative effective interactions in the same configuration space~\cite{Caurier:2007wq,Menendez:2008jp,Honma:2009zz,jj44b}, and while the excitation spectra may be somewhat different to the ones predicted by the RG interaction, the impact on the nuclear structure factors is very small at the momentum transfers relevant to direct detection searches.

\begin{figure}[t]
	\begin{center}
		\includegraphics[width=0.48\textwidth,clip=]{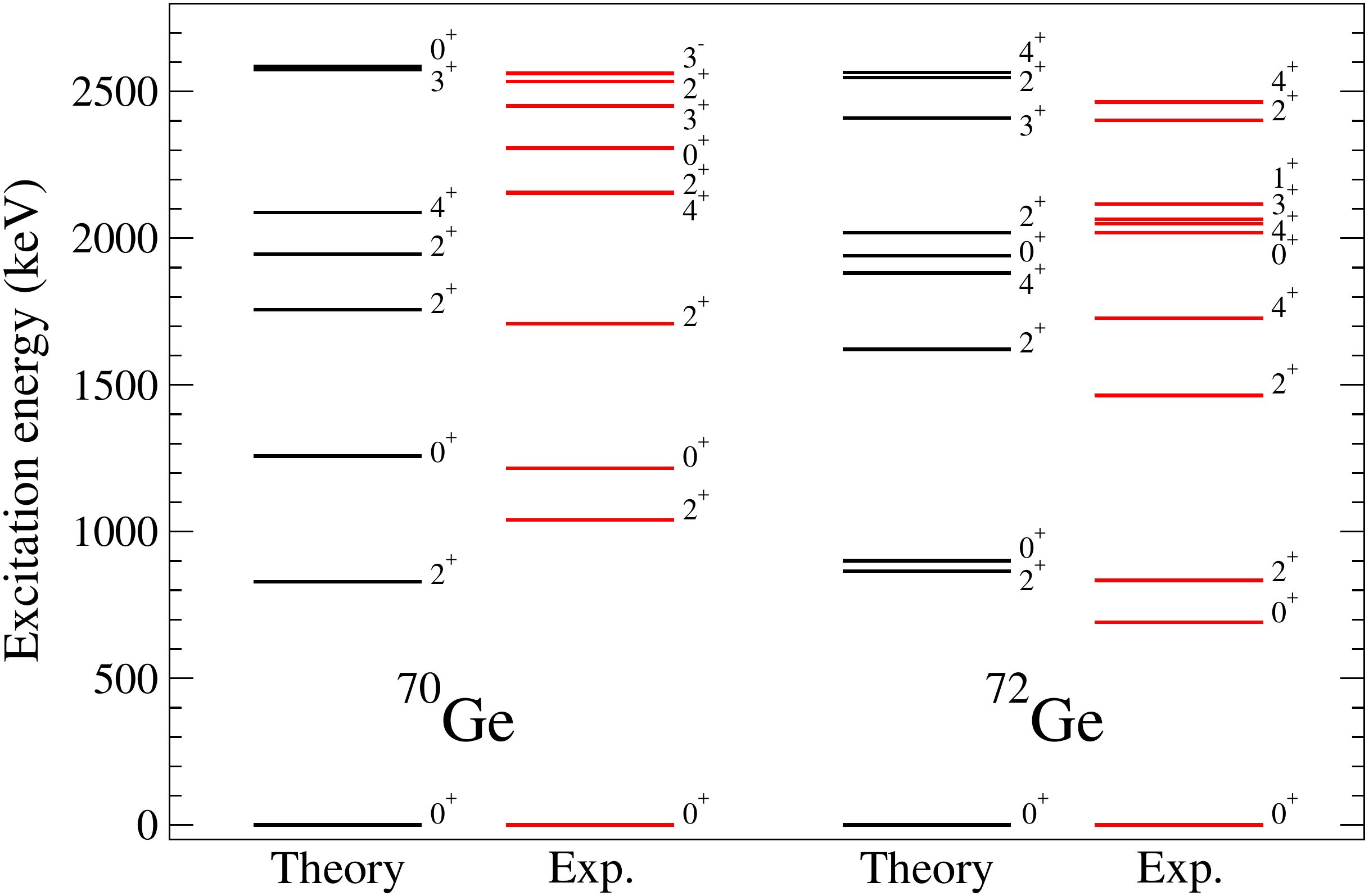}
	\end{center}
	\caption{Calculated $^{70,72}$Ge spectra compared to experiment.\label{fig:ge7072_spectrum}}
\end{figure}

\begin{figure}[t]
	\begin{center}
		\includegraphics[width=0.48\textwidth,clip=]{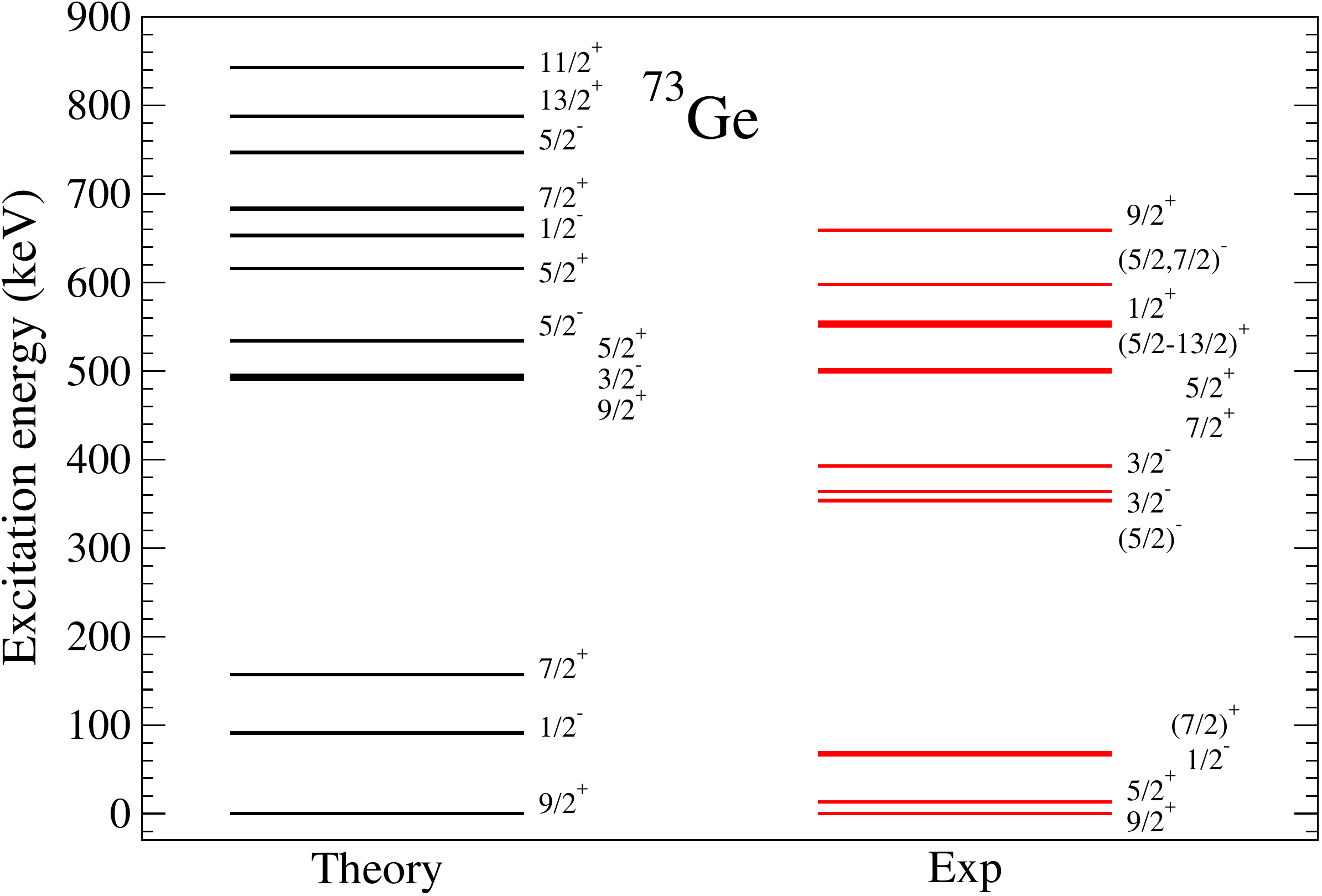}
	\end{center}
	\caption{Calculated $^{73}$Ge spectrum compared to experiment.\label{fig:ge73_spectrum}}
\end{figure}

\begin{figure}[t]
	\begin{center}
		\includegraphics[width=0.48\textwidth,clip=]{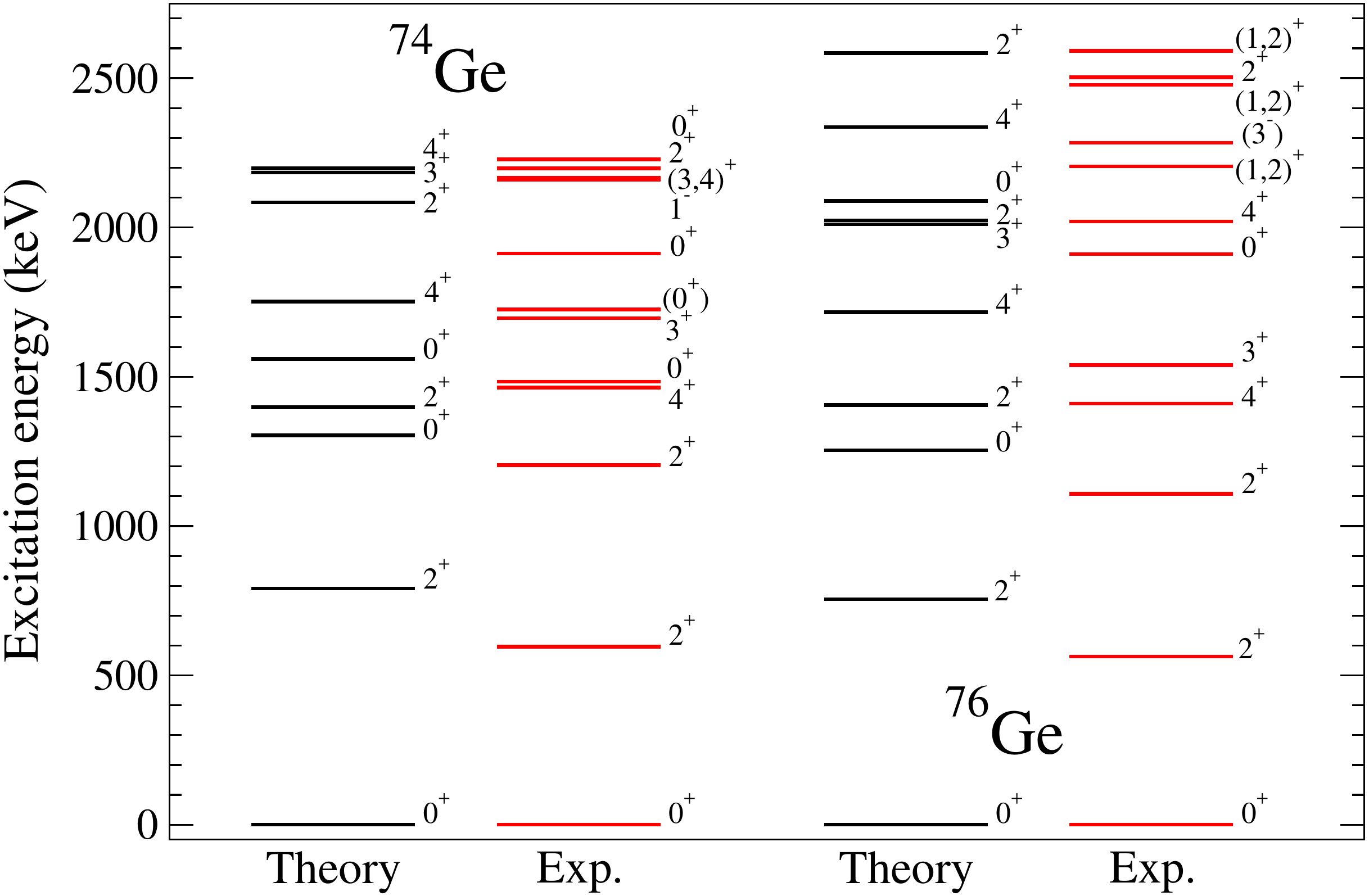}
	\end{center}
	\caption{Calculated $^{74,76}$Ge spectra compared to experiment.\label{fig:ge7476_spectrum}}
\end{figure}

Figures~\ref{fig:f19si29_spectrum}--\ref{fig:ge7476_spectrum} compare the low-energy excitation spectra of the stable isotopes of fluorine, silicon, argon, and germanium with our theoretical predictions. In all cases, our calculations are in very good agreement with experiment, especially for nuclei with even number of nucleons. In some cases, especially for the odd-mass nucleus $^{73}$Ge, some experimental states are not well reproduced. This is very likely due to the limitation of the configuration space used in our calculation, because the description of $^{73}$Ge is of similar quality with the other effective interactions we have studied.

\begin{table*}[t]
	\centering
	\renewcommand{\arraystretch}{1.3}
	\begin{tabular}{c|c|cc|cc|cc|cccc}
		\toprule
	Nucleus	& State / 
		& \multicolumn{2}{c|}{$\langle r^2 \rangle_\text{ch}^{1/2}$ [fm]} 
		& \multicolumn{2}{c|}{$Q$ [$e\,$fm$^2$]} 
		& \multicolumn{2}{c|}{$\mu$ [n.m.]}
		& \multicolumn{2}{c}{B(E2) [$e^2\,$fm$^4$]$\quad$}
		& \multicolumn{2}{c}{B(M1) [n.m.$^2$]} \\
		& Transition & Th & Exp & Th & Exp & Th & Exp & Th & Exp & Th & Exp \\ \colrule
		$^{19}$F & $1/2^+_\text{gs}$  & $2.83$ & $2.898(2)$ & -- & -- & $2.9$ & $2.628868(8)$ & & \\
		& $5/2^+_1\rightarrow1/2^+_\text{gs}$ & &  &  & & & & $19$ & $20.9(2)$ & -- & -- \\
		& $3/2^+_1\rightarrow5/2^+_1$ & &  &  & & & & & & $3.4$ & $4.1(2.5)$ \\
		& $9/2^+_1\rightarrow5/2^+_1$ & &  &  & & & & $19$ & $25(3)$ & -- & -- \\
		\colrule
		$^{28}$Si & $0^+_\text{gs}$   & $3.19$ & $3.122(2)$ & -- & -- & -- & -- & & \\
		& $2^+_1$ & & & $+19$ & $+16(3)$ & $1.1$ & $1.12(18)$ & & \\
		& $2^+_1\rightarrow0^+_\text{gs}$  &  &  &  & & & & $67$ & $70(3)$ & -- & --\\
		& $4^+_1\rightarrow2^+_1$ &  &  &  & & & & $110$ & $87(10)$ & -- & --\\
		& $0^+_2\rightarrow2^+_1$ &  &  &  & & & & $82$ & $50(3)$ & -- & --\\
		& $3^+_1\rightarrow2^+_1$ &  &  &  & & & & $6\times10^{-4}$ & $0.007(2)$ & $2\times10^{-4}$ & $48(4)\times10^{-5}$\\
		& $0^+_3\rightarrow2^+_1$ &  &  &  & & & & $0.6$ & $1.4(1)$ & -- & --\\
		\colrule		
		$^{29}$Si & $1/2^+_\text{gs}$ & $3.20$ & $3.118(5)$ & -- & -- & $-0.49$ & $-0.55529(3)$ & & \\
		& $3/2^+_1\rightarrow1/2^+_\text{gs}$ & &  &  & & & & $31$ & $22(2)$ & $0.011$ & $0.063(2)$ \\
		& $3/2^+_2\rightarrow1/2^+_\text{gs}$ & &  &  & & & & $33$ & $29(12)$ & $0.21$ & $0.116(7)$ \\
		& $5/2^+_2\rightarrow3/2^+_1$ & &  &  & & & & $64$ & $47(7)$ & $0.20$ & $0.15(1)$ \\
		& $5/2^+_2\rightarrow5/2^+_1$ & &  &  & & & & $2.5$ & $4(4)$ & $0.18$ & $0.18(3)$ \\
		& $7/2^+_1\rightarrow3/2^+_1$ & &  &  & & & & $50$ & $40(12)$ & -- & -- \\
		& $7/2^+_1\rightarrow5/2^+_1$ & &  &  & & & & $1.8$ & $1.0(6)$ & $0.013$ & $0.04(1)$ \\
		\colrule		
		$^{30}$Si & $0^+_\text{gs}$   & $3.21$ & $3.134(4)$ & -- & -- & -- & -- & & \\
		& $2^+_1$ & & & $+2.1$ & $-5(6)$ & $0.70$ & $0.76(18)$ & & \\
		& $2^+_1\rightarrow0^+_\text{gs}$  &  &  &  & & & & $48$ & $47(6)$ & -- & --\\
		& $2^+_2\rightarrow0^+_{\text{gs}}$ &  &  &  & & & & $14$ & $33(9)$ & -- & --\\
		& $2^+_2\rightarrow2^+_1$ &  &  &  & & & & $81$ & $50(30)$ & $0.18$ & $0.16(5)$\\
		& $1^+_1\rightarrow0^+_{\text{gs}}$ &  &  &  & & & & -- & -- & $0.007$ & $0.009(3)$\\		
		& $1^+_1\rightarrow2^+_1$ &  &  &  & & & & $2.3$ & $8(6)$ & $0.29$ & $0.16(4)$\\
		\colrule
		\botrule
	\end{tabular}
	\renewcommand{\arraystretch}{1.0}
	\caption{Root-mean-square charge radii ($\langle r^2 \rangle_\text{ch}^{1/2}$) of ground states (gs), electric quadrupole ($Q$) and magnetic dipole ($\mu$, in units of nuclear magnetons, n.m.) moments of ground and lowest-excited states, and nuclear matrix elements of selected electric quadrupole [B(E2)] and magnetic dipole [B(M1)] transitions between low-lying states of stable fluorine and silicon isotopes. Theoretical charge radii use calculated orbital occupancies combined with the harmonic oscillator length $b$ from Eq.~\eqref{oscillator_length}. Electromagnetic moments and transitions are obtained with effective neutron and proton electric charges $e_n=0.45$ and $e_p=1.36$, and effective orbital ($l$) and spin ($s$) $g$-factors $g_{p}^l=1.16$, $g_{p}^s=5.15$ for protons and $g_{n}^l=-0.09$, $g_{n}^s=-3.55$ for neutrons~\cite{usdb_obs}. Quadrupole moments $Q>0$ ($Q<0$) indicate nuclei with prolate (oblate) deformation. Theoretical results are compared to experimental data from Refs.~\cite{Angeli:2013epw,nndc,Raghavan:1989zz}.
	}
	\label{tab:EM_properties_f_si}
\end{table*}

\begin{table*}[t]
	\centering
	\renewcommand{\arraystretch}{1.3}
	\begin{tabular}{c|c|cc|cc|cc|cccc}
		\toprule
	Nucleus	& State /
		& \multicolumn{2}{c|}{$\langle r^2 \rangle_\text{ch}^{1/2}$ [fm]} 
		& \multicolumn{2}{c|}{$Q$ [$e\,$fm$^2$]} 
		& \multicolumn{2}{c|}{$\mu$ [n.m.]}
		& \multicolumn{2}{c}{B(E2) [$e^2$fm$^4$]$\quad$}
		& \multicolumn{2}{c}{B(M1) [n.m.$^2$]} \\
		& Transition & Th & Exp & Th & Exp & Th & Exp & Th & Exp & Th & Exp \\ \colrule
		$^{40}$Ar & $0^+_\text{gs}$   & $3.43$ & $3.427(3)$ & -- & -- & -- & -- & & \\
		& $2^+_1$ & & & $+2.6$ & $+1(4)$ & $-0.54$ & $-0.04(6)$ & & \\
		& $2^+_1\rightarrow0^+_\text{gs}$  &  &  &  & & & & $50$ & $73(3)$ & -- & --\\
		& $0^+_2\rightarrow2^+_1$  &  &  & & & & & $29$ & $43(7)$ & -- & -- \\
		& $2^+_2\rightarrow0^+_\text{gs}$  & & & & & & & $0.7$ & $10(2)$ & -- & --\\
		& $2^+_2\rightarrow2^+_1$ &  &  &  & & & & $55$ & $150(50)$ & $0.016$ & $0.07(1)$ \\
		& $4^+_1\rightarrow2^+_1$ &  &  &  & & & & $36$ & $43(8)$ & -- & --\\
		& $6^+_1\rightarrow4^+_1$ &  &  &  & & & & $16$ & $13.6(5)$ & -- & --\\\colrule
		$^{70}$Ge & $0^+_\text{gs}$   & $4.05$ & $4.0414(12)$ & -- & -- & -- & -- & \\
		& $2^+_1$ & & & $+23$ & $+4(3)$ & $0.96$ & $0.91(5)$ & & \\
		& $2^+_1\rightarrow0^+_\text{gs}$  &  &  &  & & & & $240$ & $360(7)$ & -- & --\\
		& $0^+_2\rightarrow2^+_1$  &  &  & & & & & $36$ & $820(120)$ & -- & -- \\
		& $2^+_2\rightarrow0^+_\text{gs}$  & & & & & & & $8.0$ & $9(1)$ & -- & --\\
		& $2^+_2\rightarrow2^+_1$ &  &  &  & & & & $16$ & $1100(190)$ & $0.022$ &
		$0.003(2)$ \\
		& $2^+_2\rightarrow0^+_2$          & & & & & & & $270$ & $270(50)$ & -- & --\\
		& $4^+_1\rightarrow2^+_1$ &  &  &  & & & & $370$ & $430(90)$ & -- & --\\\colrule
		$^{72}$Ge & $0^+_\text{gs}$   & $4.07$ & $4.0576(12)$ & -- & -- & -- & -- & \\
		& $2^+_1$ & & & $+16$ & $-13(6)$ & $0.55$ & $0.77(5)$ & & \\
		& $2^+_1\rightarrow0^+_\text{gs}$  &  &  &  & & & & $260$ & $418(7)$ & -- & --\\ & $2^+_1\rightarrow0^+_2$  &  &  & & & & & $60$ & $317(5)$ & -- & -- \\
		& $2^+_2\rightarrow0^+_\text{gs}$  & & & & & & & $29$ & $2.3(4)$ & -- & --\\
		& $2^+_2\rightarrow0^+_2$          & & & & & & & $15$ & $0.5(1)$ & -- & --\\
		& $2^+_2\rightarrow2^+_1$ &  &  &  & & & & $360$ & $1100(180)$ & $0.023$ &
		$29(9)\times10^{-5}$ \\
		& $4^+_1\rightarrow2^+_1$ &  &  &  & & & & $430$ & $660(90)$ & -- & --\\\colrule
		$^{73}$Ge & $9/2^+_\text{gs}$ & $4.07$ & $4.0632(14)$ & $-15$ & $-17(3)$ & $-1.0$ & $-0.8794677(2)$ &  & \\
		& $5/2^+_1$ & & & $+20$ & $+70(8)$ & $-0.90$ & $-1.08(3)$ &  & \\
		& $7/2^+_1\rightarrow9/2^+_\text{gs}$ & &  &  & & & & $260$ & $114(7)$ & $0.002$ & $14(9)\times10^{-4}$ \\
		& $7/2^+_2\rightarrow9/2^+_\text{gs}$ & &  &  & & & & $26$ & $740(140)$ & $0.014$ & $57(4)\times10^{-4}$ \\
		& $13/2^+_1\rightarrow9/2^+_\text{gs}$ & &  &  & & & & $250$ & $540(40)$ & -- & -- \\ \colrule
		$^{74}$Ge & $0^+_\text{gs}$   & $4.08$ & $4.0742(12)$ & -- & -- & -- & -- &  & \\
		& $2^+_1$ & & & $+0.7$ & $-19(2)$ & $0.49$ & $0.87(4)$ & & \\
		& $2^+_1\rightarrow0^+_\text{gs}$  &  &  &  & & & & $310$ & $609(7)$ & -- & --\\
		& $2^+_2\rightarrow0^+_\text{gs}$  & & & & & & & $3.2$ & $13(2)$ & -- & --\\
		& $2^+_2\rightarrow2^+_1$ &  &  &  & & & & $470$ & $790(110)$ & $4\times10^{-5}$ &
		$18(3)\times10^{-4}$ \\
		& $4^+_1\rightarrow2^+_1$ &  &  &  & & & & $450$ & $760(60)$ & -- & --\\\colrule
		$^{76}$Ge & $0^+_\text{gs}$   & $4.09$ & $4.0811(12)$ & -- & -- & -- & -- &  & \\
		& $2^+_1$ & & & $-14$ & $-19(6)$ & $0.42$ & $0.84(5)$ & & \\
		& $2^+_1\rightarrow0^+_\text{gs}$  &  &  &  & & & & $300$ & $550(19)$ & -- & --\\
		& $2^+_2\rightarrow0^+_\text{gs}$  & & & & & & & $1.2$ & $17(4)$ & -- & --\\
		& $2^+_2\rightarrow2^+_1$ &  &  &  & & & & $400$ & $800(170)$ & $0.006$ &
		$14(7)\times10^{-4}$ \\
		& $4^+_1\rightarrow2^+_1$ &  &  &  & & & & $430$ & $730(170)$ & -- & --\\\colrule
		\botrule
	\end{tabular}
	\renewcommand{\arraystretch}{1.0}
	\caption{Same as Table~\ref{tab:EM_properties_f_si}, for stable argon and germanium isotopes. Here electromagnetic moments and transitions are obtained with effective neutron and proton electric charges $e_n=0.5$, and $e_p=1.5$, and bare $g$-factors~\cite{Caurier:2004gf}. Quadrupole moments $Q>0$ ($Q<0$) indicate nuclei with prolate (oblate) deformation. Theoretical results are compared to experimental data from Refs.~\cite{Angeli:2013epw,nndc,Raghavan:1989zz}.
	}
	\label{tab:EM_properties_ar_ge}
\end{table*}

\begin{table}[t]
	\centering
	\renewcommand{\arraystretch}{1.3}
	\begin{tabular}{cc|ccccccc}
		\toprule
& & 	
$^{128}$Xe & $^{129}$Xe & $^{130}$Xe & $^{131}$Xe  \\ \colrule
$\langle r^2 \rangle_\text{ch}^{1/2}\;$ & Th & $4.75$ & $4.75$ & $4.76$ & $4.77$  \\
$\left[\text{fm}\right]\;$ & Exp & $4.7774(50)$ & $4.7775(50)$ & $4.7818(49)$ & $4.7808(49)$  \\ \colrule
& & 	
 $^{132}$Xe & $^{134}$Xe & $^{136}$Xe & \\ \colrule
$\langle r^2 \rangle_\text{ch}^{1/2}\;$ & Th &  $4.77$ & $4.78$ & $4.79$ & \\
$\left[\text{fm}\right]\;$ & Exp &  $4.7808(49)$ & $4.7899(47)$ & $4.7964(47)$ & \\ 
		\botrule
	\end{tabular}
	\renewcommand{\arraystretch}{1.0}
	\caption{Theoretical root-mean-square charge radii of stable xenon isotopes compared to experimental data from Ref.~\cite{Angeli:2013epw}.
	}
	\label{tab:radii_xe}
\end{table}

To complement the assessment of the quality of the nuclear structure calculations,
Tables~\ref{tab:EM_properties_f_si} and \ref{tab:EM_properties_ar_ge} compare
theoretical and experimental electromagnetic observables for all these nuclear targets.
The comparison includes nuclear charge radii,
electromagnetic moments for ground and lowest-excited states,
and nuclear matrix elements for selected electromagnetic transitions
between low-lying states.
Details on the shell model calculation of nuclear moments and matrix elements
can be found, e.g., in Ref.~\cite{Caurier:2004gf}.
Charge radii~\cite{Bertozzi:1972jff}, the properties most relevant for coherent nuclear structure factors,
are very well reproduced, better than $3\%$ in all cases, and better than $1\%$ in the heavier argon and germanium.
For fluorine and silicon, Table~\ref{tab:EM_properties_f_si}
shows an excellent agreement between theory and experiment, as the majority of the predictions reproduce data within experimental uncertainties. For argon and germanium, Table~\ref{tab:EM_properties_ar_ge} 
also shows a reasonable agreement of the nuclear structure
calculations with experiment. Nuclear matrix elements within the same
isotope can vary over two orders of magnitude, and the theoretical
results reproduce well the corresponding hierarchy. In germanium, some theoretical electric moments and transitions underestimate experiment moderately. This suggests that the configuration space used in the calculation is not sufficient to fully account for the most collective states, consistently with the findings from the comparison of the $^{73}$Ge excitation spectrum. We do not expect, however, any significant effect in the ground states involved in elastic WIMP--nucleus scattering.

In Sec.~\ref{sec:structure_factors} and the \textsc{Python} notebook we also cover nuclear structure factors for xenon, which we studied in the context of coherent SI scattering in Refs.~\cite{Vietze:2014vsa,Hoferichter:2016nvd}, and more generally in Refs.~\cite{Klos:2013rwa,Menendez:2012tm}.
The quality of the nuclear structure calculations of stable xenon isotopes
is similar to that of argon or germanium,
the heaviest nuclear targets considered in this section.
Calculations for xenon have been compared to experimental data in Refs.~\cite{Vietze:2014vsa,Klos:2013rwa}
for excitation spectra, and in
 Refs.~\cite{Menendez:2012tm,Sieja:2009zz}
for electromagnetic properties.
Given that the charge radii are closely connected to the nuclear structure factors considered here, 
Table~\ref{tab:radii_xe} compares theoretical results with experiment.
In all isotopes the calculations reproduce measured radii to better than 1\%.

\section{Structure factors}
\label{sec:structure_factors}

\begin{figure*}[t]
	\begin{center}
		\includegraphics[width=0.9\textwidth,clip=]{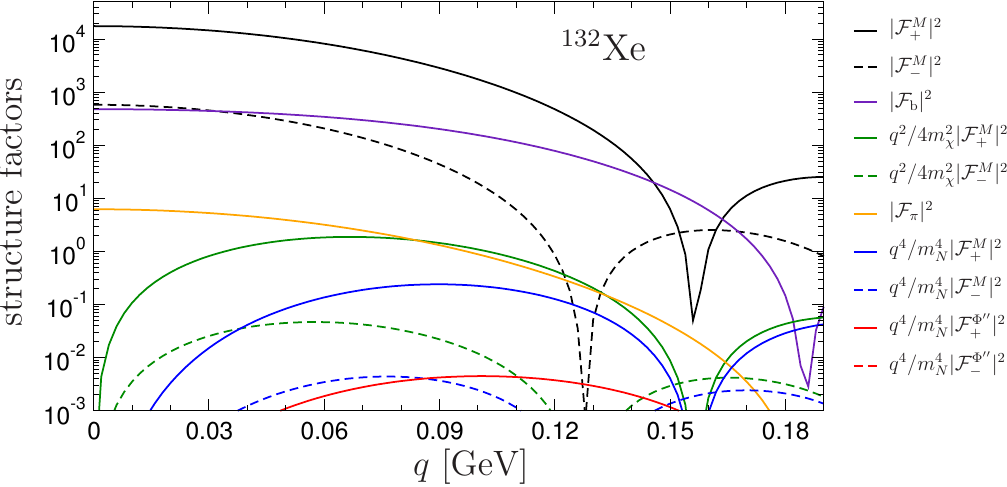}
	\end{center}
	\caption{Structure factors for $^{132}$Xe from 1b and 2b contributions without interference terms: 1b $\Op_1$ (black), $\Op_{11}$ (green), radius corrections (blue), and $\Op_{3}$ (red) contributions are shown together with 2b $\F_b$ (indigo) and $\F_\pi$ (orange) structure factors. Solid lines show isoscalar and 2b contributions, while dashed lines indicate isovector couplings. For $\Op_{11}$ $m_\chi=2$~GeV is assumed.
 \label{fig:xe132_overview}}
\end{figure*}

Apart from the dependence on $q$, $\mN$, and $\mc$ that is predicted by ChEFT, the generalized SI WIMP--nucleus cross section in Eq.~\eqref{structure_factors} depends on six  independent structure factors $\F$. Four of them correspond to the coupling of the WIMP to one nucleon, which can be the same for protons and neutrons (as in the two isoscalar structure factors) or opposite (as in the two isovector ones). In addition, two independent structure factors characterize the simultaneous coupling of WIMPs to two nucleons. In this section we evaluate these six structure factors for the nuclear targets considered in present and future direct detection experiments. An overview of the various contributions, excluding interference terms, is provided in Fig.~\ref{fig:xe132_overview}. In the following, we present in detail our results for the one-body (1b) and two-body (2b) structure factors.

\subsection{One-body structure factors}
\label{sec:1b}

As discussed in Refs.~\cite{Fitzpatrick:2012ix,Anand:2013yka} there are two different nuclear responses describing the coupling of a WIMP to a single nucleon, $\F^M$ and $\F^{\Phi''}$, which receive coherent contributions from several nucleons in the nucleus. In addition to the standard SI scattering, the nuclear response $\F^M$ describes a subleading contribution that corresponds to the NREFT operators $\Op_{5,8,11}$,
but by far the most important among them is $\Op_{11}$. In addition, the so-called radius correction to the standard SI structure factor is also coherent~\cite{Hoferichter:2016nvd}. 
Dropping the contributions from $\Op_{5,8}$, the scattering cross section including one-nucleon couplings simplifies to
\begin{align}
\label{structure_factors_1b}
\frac{\diff \sigma}{\diff q^2}&=\frac{1}{4\pi v^2}\bigg|\sum_{I=\pm}\Big(c_I^M-\frac{q^2}{m_N^2} \, \dot c_I^M\Big)\F_I^M(q^2)\notag \\
&\qquad+\frac{q^2}{2\mN^2}\sum_{I=\pm}c_I^{\Phi''}\F_I^{\Phi''}(q^2)\bigg|^2\notag \\
&+\frac{1}{4\pi v^2}\bigg|\sum_{I=\pm}\frac{q}{2\mc}\tilde c_I^M\F_I^M(q^2)\bigg|^2,
\end{align}
where $\tilde c_I^M$ can be identified with $c_I^{M,11}$ from Eq.~\eqref{structure_factors}. 
Even though there are only four independent isoscalar contributions (plus four isovector ones), in the most general case where all contributions in Eq.~\eqref{structure_factors_1b} compete, 
the interference of all of them generates a plethora of individual terms that could be considered.

\begin{figure}[t]
	\begin{center}
		\includegraphics[width=0.46\textwidth,clip=]{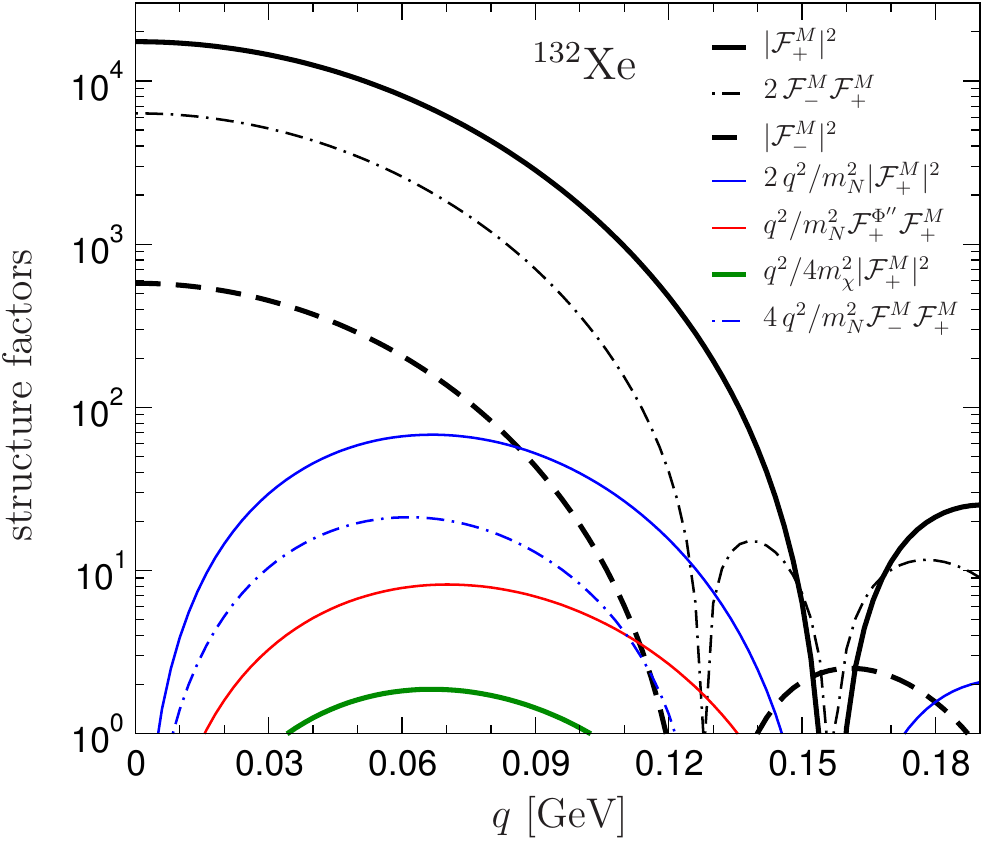}
	\end{center}
	\caption{Structure factors for $^{132}$Xe, 1b contributions only. Thick lines correspond to individual terms in Eq.~\eqref{structure_factors_1b}, while interference terms are shown as thin lines. Description as in Fig.~\ref{fig:xe132_overview}, with dotted-dashed lines representing interference terms involving isovector couplings.
	\label{fig:xe132_1b}}
\end{figure}

\begin{figure}[t]
	\begin{center}
		\includegraphics[width=0.48\textwidth,clip=]{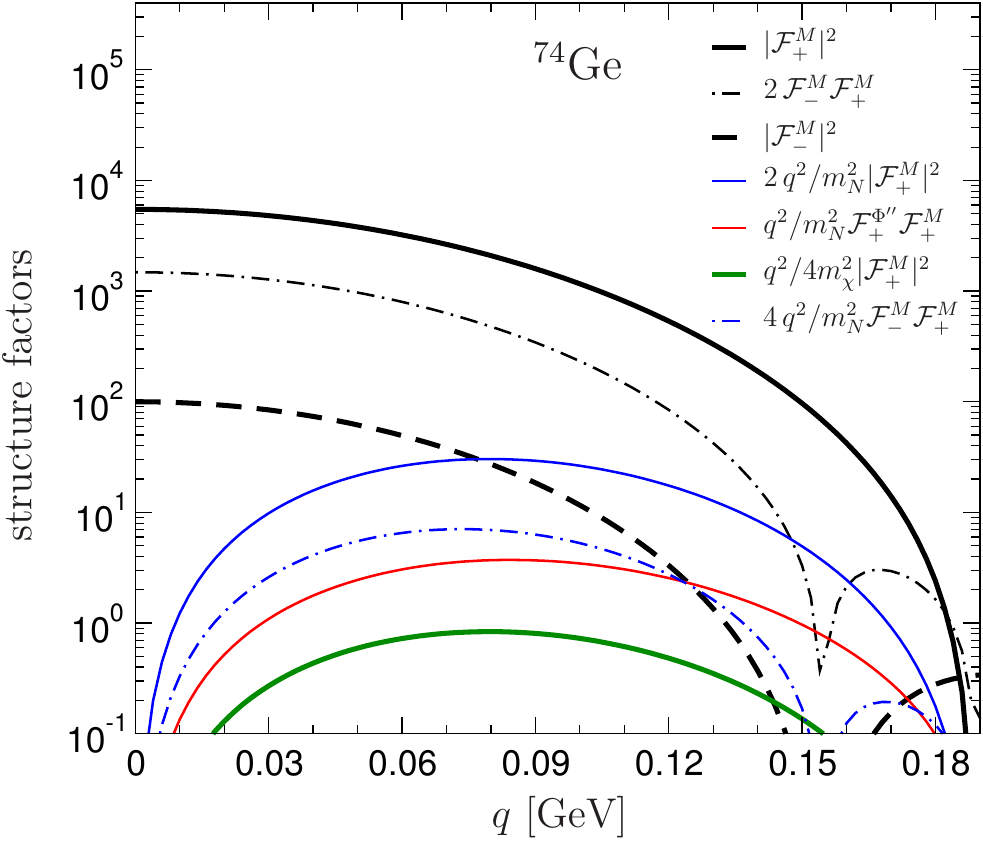}
	\end{center}
	\caption{Structure factors for $^{74}$Ge, 1b contributions only including interference terms. The description is as in Fig.~\ref{fig:xe132_1b}.\label{fig:ge74_1b}}
\end{figure}

\begin{figure}[t]
	\begin{center}
		\includegraphics[width=0.48\textwidth,clip=]{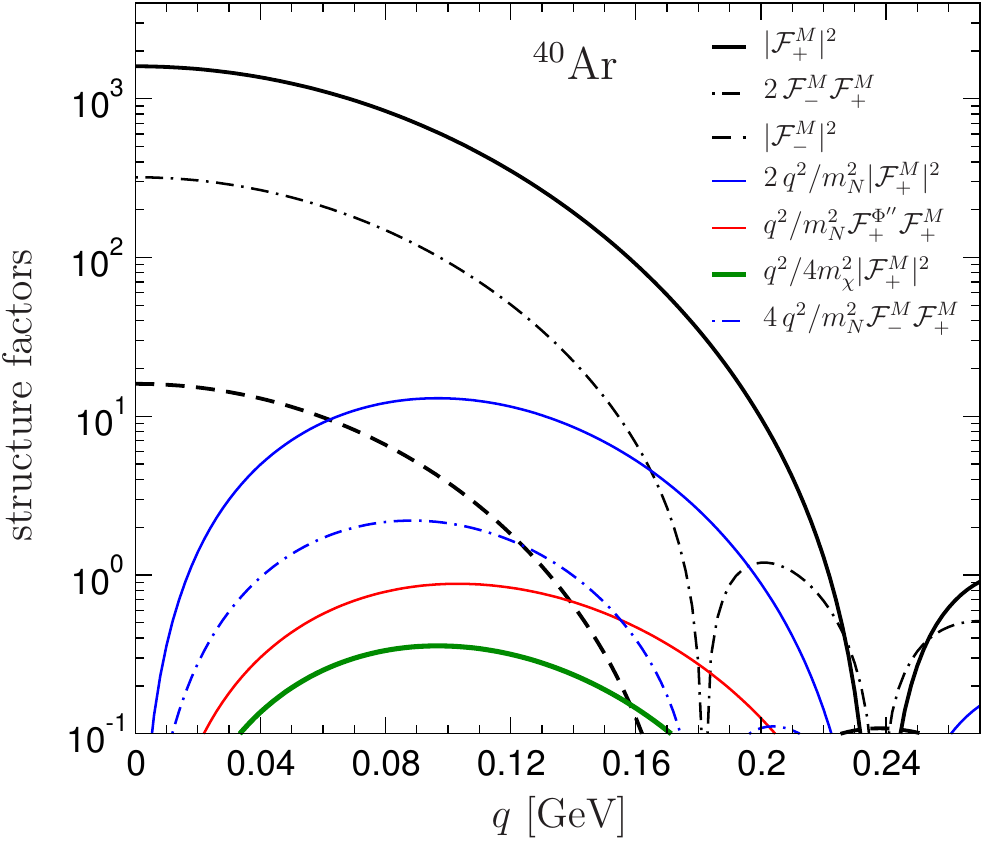}
	\end{center}
\caption{Structure factors for $^{40}$Ar, 1b contributions only including interference terms. The description is as in Fig.~\ref{fig:xe132_1b}.\label{fig:ar40_1b}}
\end{figure}

\begin{figure}[t]
	\begin{center}
		\includegraphics[width=0.48\textwidth,clip=]{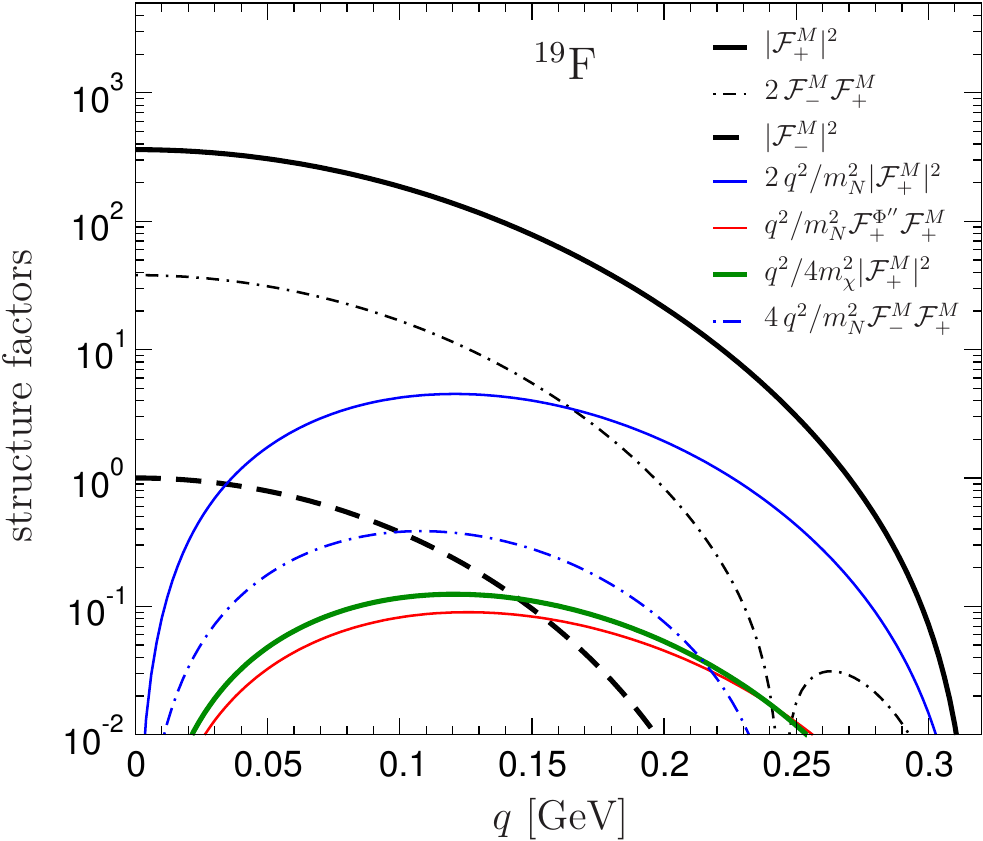}
	\end{center}
\caption{Structure factors for $^{19}$F, 1b contributions only including interference terms. The description is as in Fig.~\ref{fig:xe132_1b}.\label{fig:f19_1b}}
\end{figure}

Figures~\ref{fig:xe132_1b},~\ref{fig:ge74_1b},~\ref{fig:ar40_1b}, and~\ref{fig:f19_1b} show the leading contributions to the cross section for $^{132}$Xe, $^{74}$Ge, $^{40}$Ar, and $^{19}$F, respectively. 
The results for xenon use the results of Ref.~\cite{Hoferichter:2016nvd}. Figures~\ref{fig:xe132_1b},~\ref{fig:ge74_1b},~\ref{fig:ar40_1b}, and ~\ref{fig:f19_1b} assume that all couplings $c$ are equal to $1$, and for the NREFT $\Op_{11}$ term $\mc=2$~GeV, which implies that for heavier WIMPs the importance of this term will always be smaller than in the figures.

Figures~\ref{fig:xe132_1b},~\ref{fig:ge74_1b},~\ref{fig:ar40_1b}, and~\ref{fig:f19_1b} highlight that the standard SI contribution (solid black lines) is indeed expected to be dominant. Moreover, leading corrections to generalized SI scattering come from the interference of the standard SI and other terms, such as its isovector counterpart (dotted-dashed black), the isoscalar radius corrections (solid blue), or the isoscalar $\F^{\Phi''}$ term (solid red). The only exceptions are, first, the purely isovector SI structure factor (dashed black), suppressed by one or two orders of magnitude. Second, the $\Op_{11}$ contributions (solid green), suppressed by around four orders of magnitude in all cases (the suppression will be larger for heavy WIMPs $\mc>2\GeV$).

The variable importance of these contributions is set by the nuclear structure of the corresponding nuclear targets. Isovector contributions are relatively more important in neutron-rich xenon than in the $N \approx Z$ fluorine. On the other hand, the $\F^{\Phi''}$ contributions are relatively more important in the heavier xenon and germanium, because these targets have more nucleons in single-particle orbitals with aligned spin and orbital angular momentum. In contrast, the $\F^{\Phi''}$ contributions are more suppressed in lighter targets such as argon and especially fluorine, which tend to have nucleons more equally distributed in orbitals with spin parallel and antiparallel to the orbital angular momentum.

\begin{figure}[t]
	\begin{center}
		\includegraphics[width=0.46\textwidth,clip=]{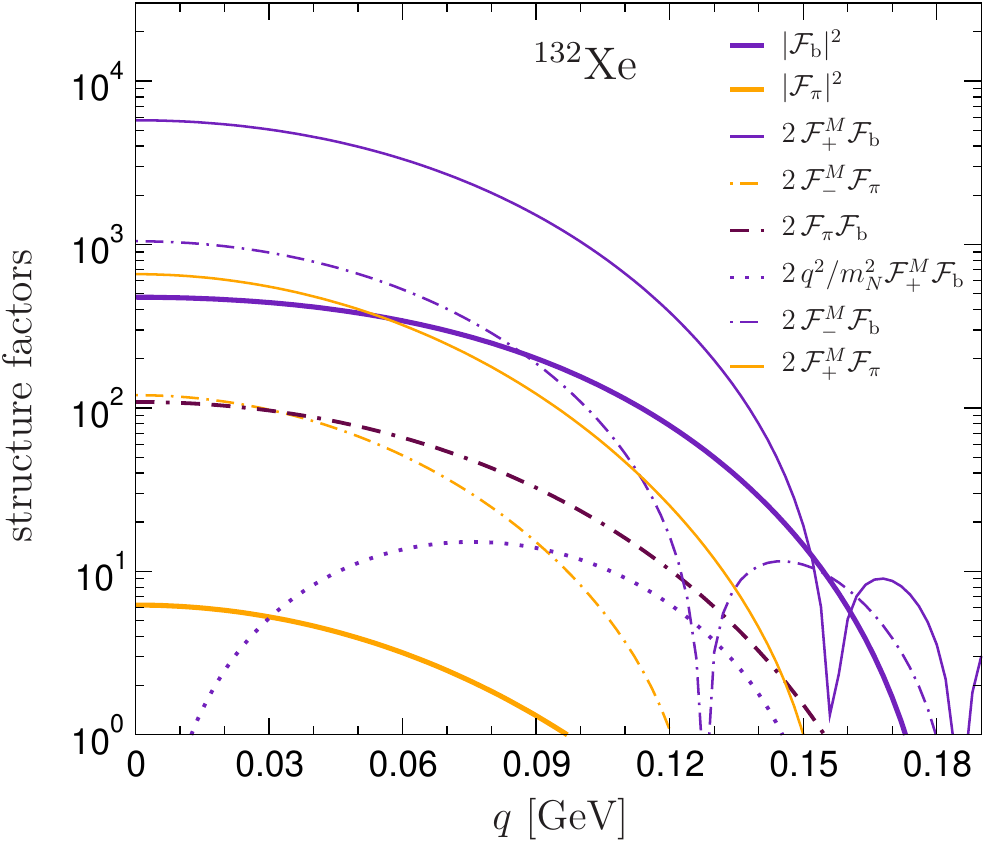}
	\end{center}
	\caption{Structure factors for $^{132}$Xe from 2b contributions and 1b--2b interferences. Thick lines correspond to individual 2b terms in Eq.~\eqref{structure_factors}, while interference terms are shown as thin lines. The description is as in Fig.~\ref{fig:xe132_1b}, with contributions involving $\F_b$ ($\F_\pi$) structure factors in indigo (orange), except for the  $\F_b$--$\F_\pi$ interference, represented by the maroon dotted-double-dashed line.
		\label{fig:xe132_1b2b}}
\end{figure}

\begin{figure}[t]
	\begin{center}
		\includegraphics[width=0.48\textwidth,clip=]{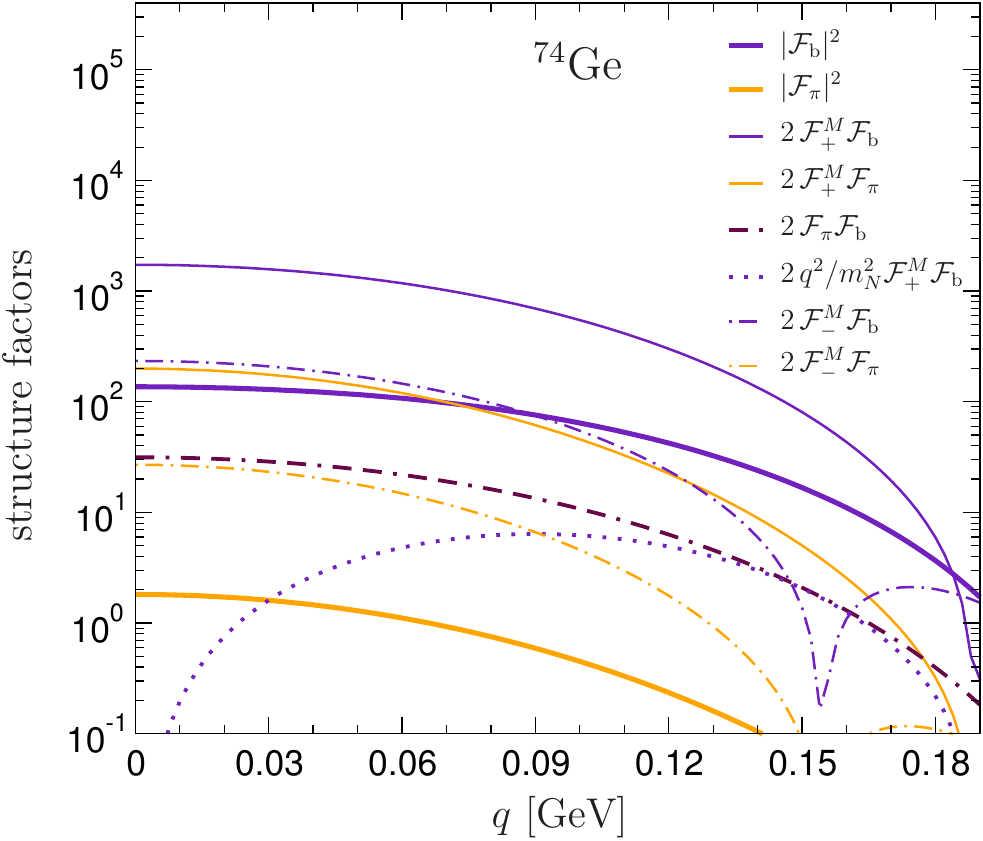}
	\end{center}
	\caption{Structure factors for $^{74}$Ge from contributions including 2b terms only. The description is as in Fig.~\ref{fig:xe132_1b2b}.\label{fig:ge74_2b}}
\end{figure}

\begin{figure}[t]
	\begin{center}
		\includegraphics[width=0.48\textwidth,clip=]{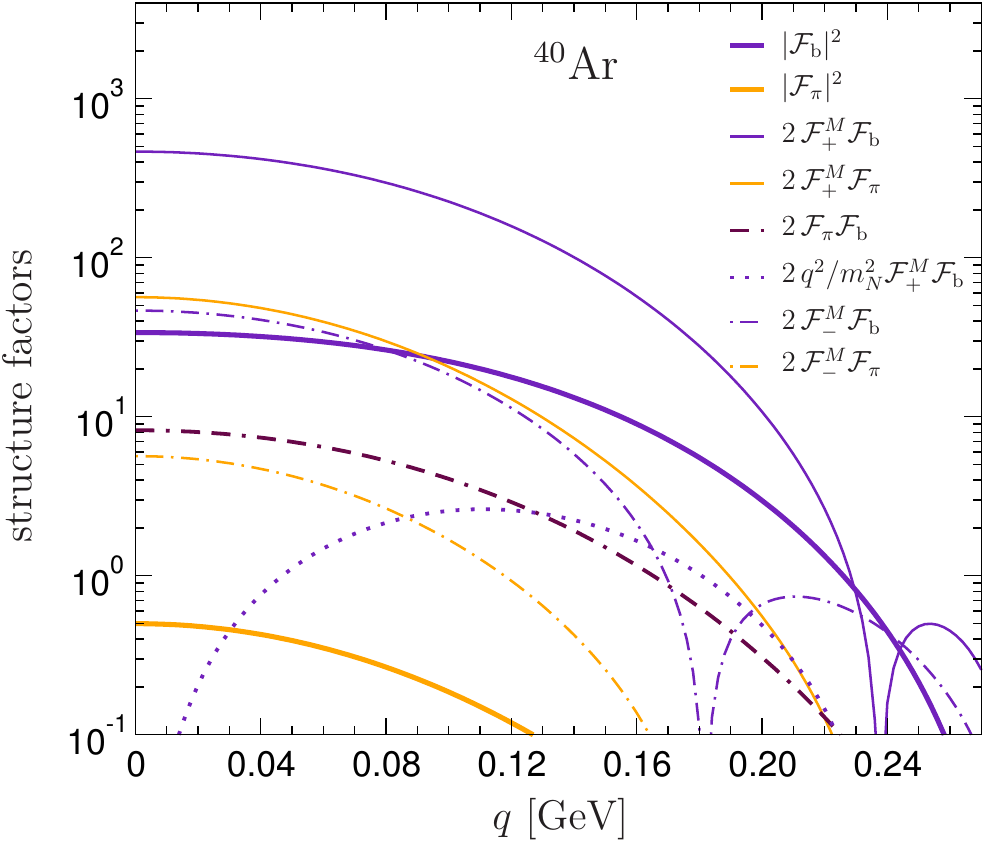}
	\end{center}
\caption{Structure factors for $^{40}$Ar from contributions including 2b terms only. The description is as in Fig.~\ref{fig:xe132_1b2b}.\label{fig:ar40_2b}}
\end{figure}

\begin{figure}[t]
	\begin{center}
		\includegraphics[width=0.48\textwidth,clip=]{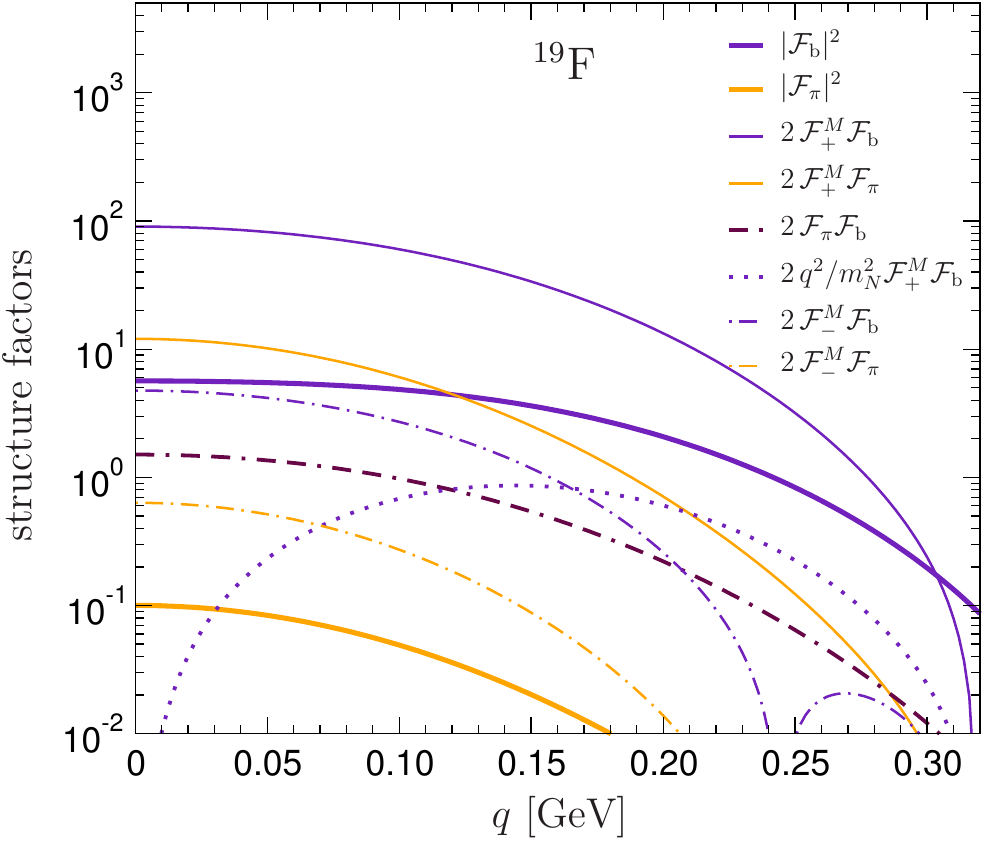}
	\end{center}
\caption{Structure factors for $^{19}$F from contributions including 2b terms only. The description is as in Fig.~\ref{fig:xe132_1b2b}.\label{fig:f19_2b}}
\end{figure}

\subsection{Two-body structure factors}
\label{sec:2b}

The two-body amplitudes of the three scalar channels, scalar--scalar ($SS$), trace anomaly $\theta^\mu_\mu$ (``$\theta$'' in short), and spin-$2$ (referred to by ``$(2)$''), read
\begin{align}
\M_{2,\text{NR}}^{SS}=&-\bigg(\frac{\gA}{2\Fpi}\bigg)^2f_\pi\mpi \frac{\ttau_1\cdot\ttau_2\,\sig_1\cdot\vec{q}_1\,\sig_2\cdot\vec{q}_2}{\big(q_1^2+\mpi^2\big)\big(q_2^2+\mpi^2\big)},
\label{Eq:MSS_2b}\\
\M_{2,\text{NR}}^{\theta}=&\frac{4\mpi^2-2\vec{q}_1\cdot \vec{q}_2}{\mpi^2}\frac{f_\pi^\theta}{f_\pi}\M_{2,\text{NR}}^{SS}\notag\\
&-\frac{f_\pi^\theta}{\mpi}2(C_S+C_T \sig_1\cdot\sig_2),
\label{Eq:Mtheta_2b}\\
\M_{2,\text{NR}}^{(2)}=&-\frac{\vec{q}_1\cdot\vec{q}_2}{2\mpi^2}\frac{f_\pi^{(2)}}{f_\pi}
\M_{2,\text{NR}}^{SS}\notag\\
&-\frac{f_\pi^{(2)}}{2\mpi}(C_S+C_T \sig_1\cdot\sig_2),
\label{Eq:Mspin2_2b}
\end{align}
where the couplings $f_\pi$, $f_\pi^\theta$, $f_\pi^{(2)}$ are defined in Eq.~\eqref{hadronic_couplings_pion}. $\sig_i$ and $\ttau_i$ refer to spin and isospin operators for nucleon $i$, $\Fpi$ is the pion decay constant, and $\gA$ the axial charge of the nucleon. Throughout, we use PDG values~\cite{Tanabashi:2018oca}, except for the particle masses, for which we use isospin averages of $\mN=938.92\MeV$ and $\mpi=138\MeV$.

While Refs.~\cite{Cirigliano:2012pq,Hoferichter:2015ipa,Hoferichter:2016nvd} introduced coherent two-body currents, the present work includes for the first time the contact-term contributions to the $\theta^\mu_\mu$ response involving $C_S$ and $C_T$ and the entire spin-$2$ contribution. In particular the consistent inclusion of the contact operators is a crucial improvement in this work (the result 
for $\theta^\mu_\mu$ was already used in Ref.~\cite{Hoferichter:2017olk}), see Sec.~\ref{sec:two_body_pheno} for an extended discussion.

By including the relativistic corrections of subleading one-body terms, both in the $\theta^\mu_\mu$ and spin-$2$ channels, it is possible to write the three physical responses in terms of just two 
new structure factors: see Sec.~\ref{sec:two_body_pheno} and Appendix~\ref{sec:app2bc} for more details and the precise definition of contact operators.
The two structure factors in the naive (non-interacting) shell model read
\begin{align}
\label{F_pi_b_shell_model}
\F_\pi(q^2)=&\frac{1}{2}\sum_{\text{occ}}\langle N_1N_2|(1-P_{12})|\frac{1}{f_\pi}\M_{2,\text{NR}}^{SS}|N_1N_2\rangle,\notag\\
\F_\text{b}(q^2)=&\frac{1}{2}\sum_{\text{occ}}\langle N_1N_2|(1-P_{12})|\M_\text{b}|N_1N_2\rangle\notag\\
&
-\frac{q^2}{\mpi^2} \F_\pi(q^2),
\end{align}
where the second structure factor is normalized as $\F_\text{b}(0)=-2E_\text{b}/\mpi$ with the binding energy of the nucleus $E_\text{b}<0$.
$\F_\pi(q^2)$ corresponds to the scalar--scalar two-body current,
and by defining $\F_\text{b}(q^2)$ as above,
the three physical channels ($SS$, $\theta$, $(2)$) are described in terms of these two structure factors because
\begin{align}
\label{F_theta_2}
	\F^\theta_\pi(q^2) &= 2\F_\pi(q^2) + \F_\text{b}(q^2),\notag\\
	\F^{(2)}_\pi(q^2) &= -\frac{1}{2}\F_\pi(q^2) + \frac{1}{4} \F_\text{b}(q^2).
\end{align}

Figures~\ref{fig:xe132_1b2b},~\ref{fig:ge74_2b},~\ref{fig:ar40_2b}, and~\ref{fig:f19_2b} show the structure factors for $^{132}$Xe, $^{74}$Ge, $^{40}$Ar, and $^{19}$F that include contributions from the coupling to two nucleons. Scalar couplings are described by the $\F_\pi$ contributions (thick solid orange line), which can interfere with the SI contribution (thin solid orange) and its isovector counterpart (dotted-dashed orange) and with an independent two-nucleon coupling (maroon dotted-double-dashed).

The two-nucleon coupling to the trace anomaly receives two contributions.
According to Eq.~\eqref{F_theta_2}, the first one can be described by $\F_\pi$
and the second one by the structure factor $\F_b$.
Figures~\ref{fig:xe132_1b2b},~\ref{fig:ge74_2b},~\ref{fig:ar40_2b}, and~\ref{fig:f19_2b} show the $\F_\pi$, $\F_b$ structure factors, and their interferences with one-nucleon couplings. In particular, besides the terms described above, the figures show the full $\F_b$ structure factor (thick solid indigo line), its interference with the SI term (thin solid indigo), and its isovector counterpart (dotted-dashed indigo), and finally the interference with the radius correction term (dotted indigo).

\begin{figure}[t]
	\begin{center}
		\includegraphics[width=0.46\textwidth,clip=]{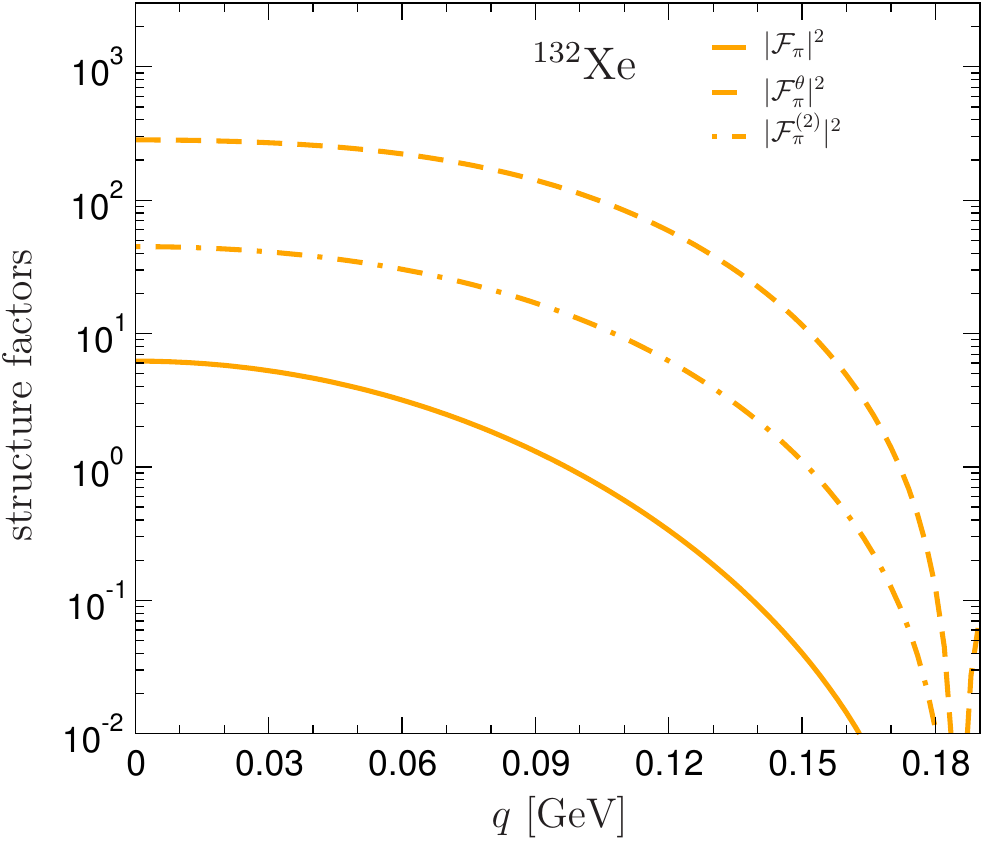}
	\end{center}
	\caption{Structure factors for $^{132}$Xe for the physical combinations of the response functions $\F_\pi$ and $\F_\text{b}$ corresponding to the $SS$ (solid line), $\theta$ (dashed), and spin-$2$ (dotted-dashed) channels as in Eq.~\eqref{F_theta_2}.
		\label{fig:xe132_phys}}
\end{figure}

\begin{figure}[t]
	\begin{center}
		\includegraphics[width=0.46\textwidth,clip=]{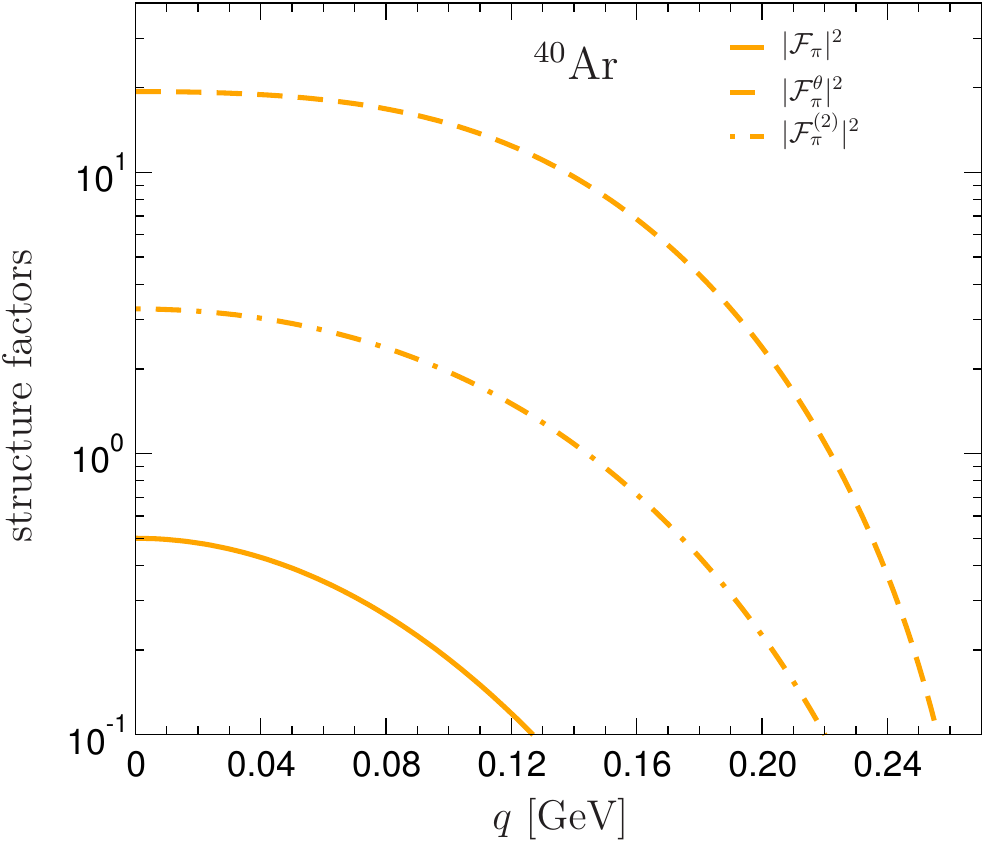}
	\end{center}
	\caption{Structure factors for $^{40}$Ar for the physical combinations of the response functions $\F_\pi$ and $\F_\text{b}$. The description is as in Fig.~\ref{fig:xe132_phys}.
		\label{fig:ar40_phys}}
\end{figure}

Similarly, spin-$2$ two-nucleon couplings contribute via $\F_\pi$ and $\F_b$ terms according to Eq.~\eqref{F_theta_2}.
Figures~\ref{fig:xe132_phys} and~\ref{fig:ar40_phys} compare the contribution of the physical combination of the $\F_\pi$ and $\F_b$ structure factors that originate in the scalar, trace anomaly, and spin-$2$ 
two-nucleon couplings. Taking all coefficients $c$ to unity, the dominant effect is given by the trace anomaly (dashed line), followed by the spin-$2$ term (dotted-dashed), and the scalar contribution (solid). However,  we stress that this hierarchy only reflects the nuclear structure aspects and does not need to be followed by particular models with definite Wilson coefficients.
The nucleon matrix elements and especially the BSM couplings can alter the importance of the different structure factors in Figs.~\ref{fig:xe132_phys} and~\ref{fig:ar40_phys}.

\section{Contact operators in two-body structure factors}
\label{sec:two_body_pheno}

The results for the two-body currents presented in Sec.~\ref{sec:2b} are based on the ChEFT formalism developed in Ref.~\cite{Hoferichter:2016nvd}, see Appendix~\ref{sec:app2bc} for more details. 
In particular, the chiral power counting follows the proposal from Ref.~\cite{Weinberg:1990rz,Weinberg:1991um}, in which the scaling of operators is estimated by dimensional analysis.
In this way, the leading contribution for a scalar current stems from pion-exchange diagrams, since $(N^\dagger N)^2$ contact operators require an additional insertion of a scalar source
that is counted in the same way as the quark mass matrix and therefore only appears at subleading order, see, e.g., Ref.~\cite{Epelbaum:2002gb}.
In alternative formulations~\cite{Kaplan:1998tg,Kaplan:1998we} where contact operators are promoted to lower orders, the pion-exchange diagrams would be accompanied by additional contact-term contributions at the same order, as also suggested by renormalization group arguments for external currents~\cite{Valderrama:2014vra}. A conclusive test of the importance of these contact operators would require detailed studies 
of the scalar current in light nuclei along the lines of Ref.~\cite{Korber:2017ery}, using recent precision chiral potentials~\cite{Epelbaum:2014sza,Entem:2017gor,Reinert:2017usi}, 
contrasted to nuclear $\sigma$-terms from lattice QCD~\cite{Beane:2013kca,Chang:2017eiq}. Work along these lines is in progress.

In practice, the question arises as to how to deal with such potential contact operators in nuclear many-body calculations. In fact, even in the Weinberg power counting contact operators do occur at leading order in the coupling to the trace anomaly $\theta^\mu_\mu$ and a spin-$2$ source. These terms were neglected in Ref.~\cite{Hoferichter:2016nvd}, so that the resulting $\theta^\mu_\mu$ response was incomplete (their contributions were, however, included in Ref.~\cite{Hoferichter:2017olk}). Here, we show how in the Weinberg power counting the contact operators in these channels are canonically renormalized in terms of nuclear binding energies, a mechanism that no longer applies once additional contact terms are introduced. 

\subsection{Trace anomaly}

\begin{figure}[t]
\center
\includegraphics[width=\linewidth]{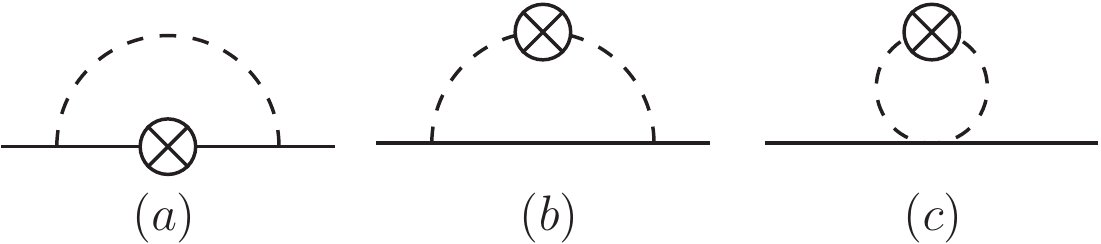}
\caption{One-loop diagrams for the coupling of $\theta^\mu_\mu$ to one nucleon (the diagram for the wave function renormalization is omitted). 
Solid and dashed lines refer to nucleons and pions, respectively, and crosses represent the coupling of the external current.}
\label{fig:diagrams_radius}
\end{figure}

The simplest example for the chiral realization of the energy-momentum tensor $\theta_{\mu\nu}$ in ChEFT is the tree-level 
pion matrix element:
\beq
\label{theta_munu_pion}
\langle\pi(p')|\theta_{\mu\nu}|\pi(p)\rangle=p_\mu p'_\nu+p'_\mu p_\nu+g_{\mu\nu}\big(\mpi^2-p\cdot p'),
\eeq
i.e., for an on-shell pion
\beq
\langle\pi(p')|\theta^\mu_\mu|\pi(p)\rangle=2\mpi^2+t, \qquad t=(p'-p)^2.
\eeq
One-loop corrections have been worked out in Ref.~\cite{Donoghue:1991qv}. Likewise the diagrams in Fig.~\ref{fig:diagrams_radius} for the nucleon give
\begin{align}
\label{theta_mumu_nucleon}
 \langle N(p')|\theta^\mu_\mu|N(p)\rangle&=\mN-\frac{3\gA^2\mpi}{32\pi\Fpi^2}\notag\\
 &\qquad\times\bigg[\frac{4\mpi^4-t^2}{4\mpi^2}I\Big(\frac{t}{\mpi^2}\Big)+t-\mpi^2\bigg]\notag\\
 &=\mN-\frac{13\gA^2\mpi}{128\pi\Fpi^2}t+\Order(t^2),
\end{align}
with loop integral
\beq
I(a)=\int_0^1\frac{\diff x}{\sqrt{1-x(1-x)a}}=\frac{1}{\sqrt{a}}\log\frac{2+\sqrt{a}}{2-\sqrt{a}}.
\eeq

\begin{figure}[t]
\center
\includegraphics[width=\linewidth]{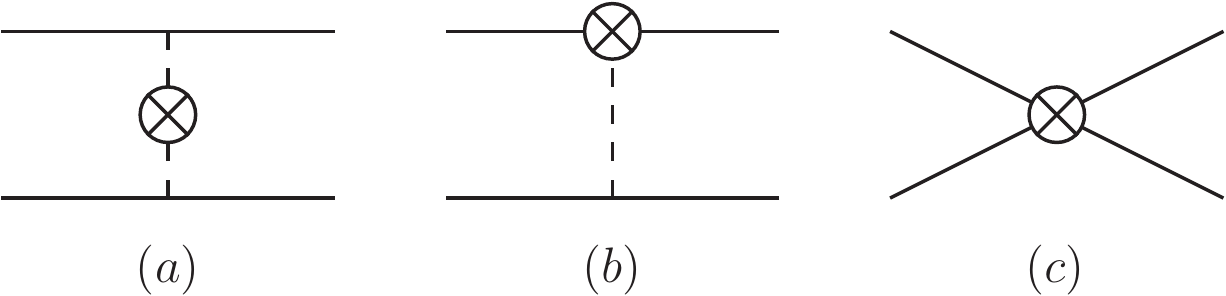}
\caption{Diagrams for the coupling of $\theta^\mu_\mu$ to two nucleons. The notation is as in Fig.~\ref{fig:diagrams_radius}.}
\label{fig:2b_diagrams}
\end{figure}

In the two-nucleon sector, see Fig.~\ref{fig:2b_diagrams}, the first term in Eq.~\eqref{Eq:Mtheta_2b} follows from the pion-exchange diagram $(a)$ by means of Eq.~\eqref{theta_munu_pion} (diagram $(b)$ only enters at higher orders). An additional contribution arises from
the contact-term Lagrangian:
\beq
\label{contacts}
{\mathcal L}_{2N}=-\frac{C_S}{2}(N^\dagger N)^2 -\frac{C_T}{2}(N^\dagger\sig N)^2,
\eeq
where the second term derives from the NR expansion of the spin vector $S^\mu=(0,\sig/2)$. The corresponding term in the trace 
\beq
\theta^\mu_\mu=-C_S(N^\dagger N)^2 -C_T(N^\dagger\sig N)^2,
\eeq
gives the second term in Eq.~\eqref{Eq:Mtheta_2b}, represented by diagram $(c)$ in Fig.~\ref{fig:2b_diagrams}.

In contrast to the scalar current, the pion-exchange piece in Eq.~\eqref{Eq:Mtheta_2b} behaves as a contact term for $q_i\to\infty$, due to the momentum dependence of the pion coupling, see Eq.~\eqref{theta_munu_pion}. As first noted in Ref.~\cite{Hoferichter:2017olk}, for vanishing momentum transfer these two pieces combine to the $NN$ potential $V_{NN}$. Together with
the kinetic-energy operator $T$, this suggests the renormalization prescription
\beq
\langle \Psi|T+V_{NN}|\Psi\rangle=E_\text{b},
\eeq
where $E_\text{b}<0$ is the binding energy of the nucleus, whose
wave function $|\Psi\rangle$ should be obtained from $NN$ interactions only.
In practice, the kinetic-energy operator formally enters at higher orders but arises naturally from relativistic corrections, while $3N$ forces correspond to higher-order corrections. 
In order to determine $C_S$ and $C_T$ we therefore use experimental binding energies, corrected for Coulomb interactions according to Ref.~\cite{Duflo:1995ep}.
Details of the renormalization both for $\theta^\mu_\mu$ and spin-$2$ terms are given in Appendix~\ref{sec:app2bc}.

\begin{table}[t]
	\centering
	\renewcommand{\arraystretch}{1.3}
	\begin{tabular}{ccccccc}
		\toprule
		 & $^{19}$F & $^{28}$Si & $^{29}$Si & $^{30}$Si & $^{40}$Ar & $^{70}$Ge\\
		$C_S\,[\text{GeV}^{-2}]$ & $-68.9$ & $-75.2$ & $-75.0$ & $-75.0$ & $-76.6$ & $-85.6$\\\colrule
		& $^{72}$Ge & $^{73}$Ge & $^{74}$Ge & $^{76}$Ge & $^{128}$Xe & $^{129}$Xe\\
		$C_S\,[\text{GeV}^{-2}]$ & $-85.9$ & $-85.9$ & $-86.2$ & $-86.4$ & $-98.0$ & $-98.2$\\\colrule
		  & $^{130}$Xe & $^{131}$Xe & $^{132}$Xe & $^{134}$Xe & $^{136}$Xe &\\
		$C_S\,[\text{GeV}^{-2}]$ & $-98.5$ & $-98.6$ & $-98.8$ & $-99.1$ & $-99.3$ & \\
		\botrule
	\end{tabular}
	\renewcommand{\arraystretch}{1.0}
	\caption{$C_S$ values, renormalized to the nuclear binding energy, for all isotopes considered in this work.
	}
	\label{tab:CS}
\end{table}

For the numerical analysis we ignore $C_T$ contributions, because the magnitude of $C_T$ is
expected to be small due to approximate $SU(4)$ symmetry of $NN$ interactions at low energies, with corrections that can be shown to be suppressed by $1/N_c^2$~\cite{Kaplan:1995yg,Kaplan:1996rk,Mehen:1999qs}. The resulting $C_S$ values in Table~\ref{tab:CS} are consistent with the expectation
from dimensional analysis~\cite{Epelbaum:2004fk}:
\beq
|C_S|=\frac{1}{16\pi}\big|\tilde C_{^1S_0}+3\tilde C_{^3S_1}\big|\sim \frac{1}{\Fpi^2}\sim 120\GeV^{-2},
\eeq
where we have used that $C_T=0$ implies $\tilde C_{^1S_0}=\tilde C_{^3S_1}$. Moreover, the values in Table~\ref{tab:CS} also agree with typical fits to the $NN$ system, e.g., at LO 
Ref.~\cite{Epelbaum:2014efa} finds $C_S=(-56.5\ldots -118.3)\GeV^{-2}$ for cutoffs in the range $R=(0.8\ldots 1.2)\,\text{fm}$. 

\subsection{Spin 2}

Despite only entering at dimension-$8$ in Eq.~\eqref{Lagr}, spin-$2$ contributions become relevant, for instance, in the context of heavy WIMPs, where significant cancellations
with spin-$0$ terms have been observed~\cite{Hill:2013hoa}, enhancing the importance of higher-order corrections.
The relevant operators are the traceless parts of the energy-momentum tensor, given in Eq.~\eqref{spin_2_definition}.
As can be seen from the WIMP part of ${\mathcal L}^{(8)}_\chi$ in Eq.~\eqref{Lagr}, the dominant contribution arises from the $\mu=\nu=0$ components, leading to the amplitudes $\M_{1,\text{NR}}^{(2)}$ in Eq.~\eqref{matching} and $\M_{2,\text{NR}}^{(2)}$ in Eq.~\eqref{Eq:Mspin2_2b}. An extension to subleading components is straightforward, since due to Lorentz invariance the pion-exchange contribution 
becomes proportional to $q_\mu q_\nu-\frac{g_{\mu\nu}}{4}q^2$. Then the full expression can be reconstructed from the $00$ component, which we have identified with $\F^{(2)}_\pi(q^2)$.

The chiral realizations of the matrix elements of $\bar\theta^{\mu\nu}_{q,g}$ have been studied in detail in the literature, both for the pion and the nucleon~\cite{Arndt:2001ye,Chen:2001nb,Detmold:2005pt,Diehl:2005rn,Ando:2006sk,Diehl:2006js,Wein:2014wma}.   
Here, we only retain the leading couplings related to moments of pion and nucleon PDFs, resulting in the one- and two-body contributions $\M_{1,\text{NR}}^{(2)}$ and $\M_{2,\text{NR}}^{(2)}$,
as well as the relativistic corrections in Eq.~\eqref{Eq:Mspin2_1b}.
Motivated by the EMC effect~\cite{Aubert:1983xm}, similar methods have been applied in the context of spin-$2$ couplings in multi-nucleon systems~\cite{Chen:2004zx,Chen:2016bde}. 
Therefore, measurements of nuclear PDFs could, in principle, provide independent cross checks on the resulting spin-$2$ structure factor. However, 
in practice this is not possible with currently employed parameterizations~\cite{Hirai:2007sx,deFlorian:2011fp,Kovarik:2015cma,Eskola:2016oht}: 
in the dark matter context, the main effect of the two-body corrections modifies the normalization at $q=0$ away from the fully coherent single-particle expectation $\F(0)=A$, 
see~\cite{Cirigliano:2012pq,Hoferichter:2016nvd} for the scalar channel. Presently, nuclear PDFs $q^A(x)$ are studied based on bound-proton PDFs $q^{p/A}(x)$ restricted onto the range $x\in [0,1]$, 
in such a way that the full PDF is reconstructed by $q^A(x) = Z q^{p/A}(x) + N q^{n/A}(x)$.
The moments of this $q^A(x)$ are therefore, by definition, normalized to the coherent limit $A$ and cannot be used to cross check the two-body effect that we derived from the 
spin-$2$ couplings of the pion.

\section{Summary}
\label{sec:summary}

We have presented a comprehensive analysis of the generalized SI scattering of spin-$1/2$ and spin-$0$ WIMPs off atomic nuclei. 
Our analysis considers all contributions that can receive the coherent enhancement from several nucleons in the nucleus, keeping terms up to third order in ChEFT.
This includes both the coupling of WIMPs to one nucleon as well as to two nucleons.
For two-body interactions we provide, for the first time, a full and consistent treatment of the contact operators that appear at the same order as pion-exchange diagrams,
arguing that these contributions can be renormalized to the nuclear binding energy. As a result, just two nuclear structure factors are enough to characterize two-body interactions via scalar operators, the trace of the energy-momentum tensor, and spin-$2$ operators.

Taking into account all these contributions,
we give all one-body and two-body nuclear structure factors relevant for the coherent WIMP scattering off fluorine, silicon, argon, germanium and xenon, covering the targets of the most advanced direct detection searches.
For that purpose, we perform large-scale nuclear shell model calculations with configuration spaces and nuclear interactions that describe very well the structure of these nuclei.

Our analysis identifies the parameters that can, at least in principle, be separately constrained in direct detection experiments.
These parameters subsume both the BSM couplings of WIMPs with quarks and gluons and the hadronic matrix elements 
that embed these quark-level operators into hadrons. 
The corresponding matching relations are illustrated in detail for both spin-1/2 and spin-0 WIMPs.

The main results of our work, encoded in the nuclear structure factors and the relation between direct-detection experiments and BSM couplings, are
available as supplementary material in a \textsc{Python} notebook.
These results form the basis for a comprehensive study of WIMP--nucleus interactions based on ChEFT.
Future extensions concern non-coherent WIMP--nucleus interactions, for which 
more parameters and nuclear structure factors need to be considered.
Accordingly, if the coherent contributions studied in this paper are strongly suppressed,
the identification of the underlying quark-level interactions becomes even more challenging.
On the other hand, progress in ab initio nuclear theory paves the way towards fully consistent structure factors from many-body calculations based on ChEFT~\cite{Andreoli:2018etf,Hebeler:2015hla,Hergert:2015awm,Hagen:2013nca,Korber:2017ery,Gazda:2016mrp}. Such improved nuclear structure factors, including their momentum-dependence, will further help distinguish among possible BSM scenarios. 

\begin{acknowledgments}
We thank V.~C.~Antochi, P.~Barry, W.~Detmold, E.~Epelbaum, A.~Fieguth, H.-W.~Hammer, C.~Hasterok, D.~B.~Kaplan, K.~Kova\v r\'ik, H.~Krebs, T.~Marrodan, F.~Olness, and J.~de Vries for valuable discussions.  
This work was supported in part by the US DOE (Grant No.\ DE-FG02-00ER41132),
the National Science Foundation (Grant No.\ NSF PHY-1748958),
the ERC (Grant No.\ 307986 STRONGINT), the DFG through SFB 1245 (Projektnummer 279384907),
the Max-Planck Society, the Japanese Society for the Promotion of Science KAKENHI through grant 18K03639, 
MEXT as ``Priority Issue on Post-K computer'' (Elucidation of the fundamental laws and evolution of the universe), JICFuS,  and the CNS-RIKEN joint project for large-scale nuclear structure calculations. 
J.~M.~and A.~S.~thank the Institute for Nuclear Theory at the University of
Washington for its hospitality and the US DOE for partial support.
\end{acknowledgments}

\appendix

\section{Matching to NREFT and nucleon matrix elements}
\label{sec:NREFT}

For the matching onto NR single-nucleon operators we use the conventions
\beq
N(\pp)+\chi(\kk)\to N(\pp')+\chi(\kk'),
\eeq
with
\beq
\qq=\kk'-\kk=\pp-\pp',\qquad
\PP=\pp+\pp',\qquad \KK=\kk+\kk',
\eeq
and
\beq
\label{velocity}
\vvp=\frac{\KK}{2\mc}-\frac{\PP}{2\mN},
\eeq
as well as the operator basis
\begin{align}
\label{Op_Wick}
 \Op_1&=\mathds{1}, & \Op_2&=\big(\vvp\big)^2, \notag\\
 \Op_3&=i\spin_N\cdot (\qq\times\vvp), &\Op_4&=\spin_\chi\cdot \spin_N,\notag\\
 \Op_5&=i\spin_\chi\cdot \big(\qq\times\vvp\big), & \Op_6&=\spin_\chi\cdot \qq\,\spin_N\cdot \qq,\notag\\
  \Op_7&=\spin_N\cdot \vvp, & \Op_8&=\spin_\chi\cdot \vvp, \notag\\
  \Op_9&=i\spin_\chi\cdot \big(\spin_N\times \qq\big), & \Op_{10}&=i\spin_N\cdot \qq,\notag\\
  \Op_{11}&=i\spin_\chi\cdot \qq, & \Op_{12}&=\spin_\chi\cdot(\spin_N\times \vvp),
\end{align}
where $q=|\qq|$.

The expressions above refer to the WIMP--nucleon system.
In the nucleus, the operator $\vvp$ generates two kinds of contributions~\cite{Fitzpatrick:2012ix,Anand:2013yka}.
First, there are operators dependent on the WIMP velocity with respect to the nucleus, $\vvp_T$ [see the terms involving $\xi_i(q,v^\perp_T)$ in Eq.~\eqref{structure_factors}]. These terms are very suppressed in the scattering amplitude because $v^\perp_T=|\vvp_T|\sim10^{-3}$. On the other hand, $\vvp$ generates terms that contain the nucleon velocity operator. In this case, the operators are mildly suppressed by $q/\mN$, and include an additional derivative.
The latter terms are fully responsible for the $\F^{\Phi''}$ structure factor.

Back to the WIMP--nucleon level, the coherently enhanced terms listed in Table~\ref{tab:structure_factors}, complemented by the leading SD response, are derived from
\begin{align}
\label{matching}
 \M_{1,\text{NR}}^{SS}&= \Op_1 f_N(t),\notag\\
 \M_{1,\text{NR}}^{PS}&= -\tilde f_N(t) \frac{1}{\mc}\Op_{11},\notag\\
 \M_{1,\text{NR}}^{VV}&=\Op_1\Big(f_1^{V,N}(t)+\frac{t}{4\mN^2}f_2^{V,N}(t)\Big)\notag\\
 &+\frac{1}{\mN}\Op_3 \bigg[f_2^{V,N}(t)+\frac{1}{2}\Big(1+\frac{\muN}{\mc}\Big)f_1^{V,N}(t)\bigg]\notag\\ 
 &+ \frac{1}{2\mc}\Big(1+\frac{\muN}{\mN}\Big) \Op_5 f_1^{V,N}(t),\notag\\
\M_{1,\text{NR}}^{AV}&=2\Op_8 \tilde f_1^{V,N}(t),\notag\\
\M_{1,\text{NR}}^{AA}&=-4\Op_4 g_A^N(t)+\frac{1}{\mN^2}\Op_6 g_P^N(t),\notag\\
\M_{1,\text{NR}}^{PP}&=\frac{1}{\mc}\Op_6 h_5^N(t),\notag\\
\M_{1,\text{NR}}^{(2)}&=\frac{3}{4}\Op_1 f_N^{(2)},\notag\\
\M_{1,\text{NR}}^{TT}&=\frac{t}{2\mc\mN}\Big(f^{T,N}_1+2f^{T,N}_2\Big)\Op_1+f^{T,N}_2\frac{4}{\mN}\Op_5\notag\\
&+f^{T,N}_1\bigg[\frac{1}{\mc}\Big(1+\frac{\muN}{\mN}\Big)\Op_3+\frac{1}{\mN}\Big(1+\frac{\muN}{\mc}\Big)\Op_5\bigg],\notag\\
\M_{1,\text{NR}}^{T\tilde T}&=\Big(\tilde f^{T,N}_1+2\tilde f^{T,N}_2\Big)\frac{2}{\mN}\Op_{11},
\end{align}
where we have ignored the non-coherent terms in the NREFT expansion.
For the remaining couplings the momentum dependence is indicated by the
relativistic momentum transfer $t$, which reduces to $t=-q^2$ up to relativistic corrections.

Some of the amplitudes in Eq.~\eqref{matching} receive contributions that break Galilean invariance. For such terms, which only appear beyond $\Order(p^3)$ in the chiral expansion, 
we assume center-of-mass kinematics, for which the velocity in Eq.~\eqref{velocity} simplifies to
\beq
\vvp=\frac{\KK}{2\muN}=-\frac{\PP}{2\muN}.
\eeq
In principle, corrections to this identification would need to be considered when calculating the nuclear structure factors, similar to the boost correction in Ref.~\cite{Beane:2002wk}, but given that these contributions are already 
highly suppressed, we only keep the center-of-mass component. The appearance of such Galilean-invariance-breaking terms at subleading orders in the NR expansion has been
pointed out in Ref.~\cite{Bishara:2017pfq}. However, at variance with Ref.~\cite{Bishara:2017pfq}, we already find such contributions in the context of the Pauli form factor $F_1$ in the $VV$ channel. This is reflected by the 
corresponding coefficients of $\Op_3$ and $\Op_5$. We find similar discrepancies to Ref.~\cite{Bishara:2017pfq} in the NR expansion of the tensor current. 
Besides the $\Op_3$ and $\Op_5$ coefficients, we also disagree in that in our expressions the induced form factors $F_{2,T}$ and $F_{3,T}$ [see Eq.~\eqref{tensor_decomposition} below] combine
to the tensor magnetic moments, so 
that the less well determined individual form factors are not required at this order in the expansion.

Expressed in terms of nucleon matrix elements we have
\begin{align}
\label{hadronic_couplings}
f_N&=\frac{m_N}{\Lambda^3}\bigg(\sum_{q=u,d,s}C^{SS}_{q}f_q^N-12\pi f^N_QC'^S_{g}\bigg),\\
\tilde f_N&=\frac{m_N}{\Lambda^3}\bigg(\sum_{q=u,d,s}C^{PS}_{q}f_q^N-12\pi f^N_Q\tilde C'^S_{g}\bigg),\notag\\
\dot f_N&=\frac{1}{\Lambda^3}\bigg(C^{SS}_{u}\frac{1-\xi_{ud}}{2}\dot\sigma+C^{SS}_{d}\frac{1+\xi_{ud}}{2}\dot\sigma+C^{SS}_{s}\dot\sigma_s\bigg),\notag\\
 f_1^{V,p}&=\frac{1}{\Lambda^2}\Big(2C^{VV}_{u}+C^{VV}_{d}\Big),\notag\\
f_2^{V,p}&=\frac{1}{\Lambda^2}\bigg[\Big(2C^{VV}_{u}+C^{VV}_{d}\Big)\kappa_p+\Big(C^{VV}_{u}+2C^{VV}_{d}\Big)\kappa_n\notag\\
&+\Big(C^{VV}_{u}+C^{VV}_{d}+C^{VV}_{s}\Big)\kappa_N^s\bigg],\notag\\
\dot f_1^{V,p}&=\frac{1}{\Lambda^2}\bigg[\Big(2C^{VV}_{u}+C^{VV}_{d}\Big)\bigg(\frac{\langle r_E^2\rangle^p}{6}-\frac{\kappa_p}{4\mpp^2}\bigg)\notag\\
&+\Big(C^{VV}_{u}+2C^{VV}_{d}\Big)\bigg(\frac{\langle r_E^2\rangle^n}{6}-\frac{\kappa_n}{4\mn^2}\bigg)\notag\\
&+\Big(C^{VV}_{u}+C^{VV}_{d}+C^{VV}_{s}\Big)\bigg(\frac{\langle r_{E,s}^2\rangle^N}{6}-\frac{\kappa_N^s}{4\mN^2}\bigg)\bigg],\notag\\
f_N^{(2)}&=\frac{\mc \mN}{\Lambda^4}\bigg(\sum_q C_q^{(2)} f^{(2)}_{q,N}+ C_g^{(2)} f^{(2)}_{g,N}\bigg),\notag\\
f^{T,N}_1&=\frac{1}{\Lambda^2}\sum_q f_q^{T,N} C_q^{TT},\qquad \tilde f^{T,N}_1=\frac{1}{\Lambda^2}\sum_q f_q^{T,N} \tilde C_q^{TT},\notag\\
f^{T,N}_2&=\frac{1}{\Lambda^2}\sum_q \tilde \kappa_q^{T,N} C_q^{TT},\qquad \tilde f^{T,N}_2=\frac{1}{\Lambda^2}\sum_q \tilde \kappa_q^{T,N} \tilde C_q^{TT},\notag
\end{align}
and $f_1^{V,n}$, $f_2^{V,n}$, $\dot f_1^{V,n}$ are given by
the exchange $u \leftrightarrow d$ in  $f_1^{V,p}$, $f_2^{V,p}$, $\dot f_1^{V,p}$.
Finally, $\tilde f_1^{V,N}$ follows from $f_1^{V,N}$ by replacing $C_q^{VV}\to C_q^{AV}$. For the axial-vector and pseudoscalar matrix elements $g_A^N$, $g_P^N$, and $h_5^N$ 
we refer to Ref.~\cite{Hoferichter:2015ipa}, since in the present paper
the numerical analysis is restricted to the coherently enhanced contributions.

The scalar couplings $f_q^N$, scalar radii $\dot \sigma$ and $\dot \sigma_s$, as well as charge radii $\langle r_E^2\rangle^N$, $\langle r_{E,s}^2\rangle^N$ and magnetic moments $\kappa_N$, $\kappa_N^s$
are discussed in detail in Ref.~\cite{Hoferichter:2016nvd} (see also Ref.~\cite{Hill:2014yxa} for the heavy-quark couplings). 
For strangeness in the vector channel there is increasing evidence from lattice QCD that these couplings are extremely small, we take
$\kappa_N^s=0.006(4)$ and $\langle r_{E,s}^2\rangle^N=0.0012(9)\,\text{fm}^2$
from Ref.~\cite{Alexandrou:2018zdf}.
The scalar couplings to $u$- and $d$-quarks can be reconstructed~\cite{Crivellin:2013ipa} from the pion--nucleon $\sigma$-term and input for the proton--neutron mass difference~\cite{Gasser:1974wd,Borsanyi:2014jba,Gasser:2015dwa,Brantley:2016our}.
Here, the tension between phenomenology~\cite{Hoferichter:2015dsa,Hoferichter:2015hva,Hoferichter:2016ocj} 
and lattice QCD~\cite{Durr:2015dna,Yang:2015uis,Abdel-Rehim:2016won,Bali:2016lvx} already discussed in Ref.~\cite{Hoferichter:2016nvd} persists.
More recently, the phenomenological determination from data on pionic atoms~\cite{Gotta:2008zza,Strauch:2010vu,Hennebach:2014lsa,Baru:2010xn,Baru:2011bw} has been confirmed
by an independent extraction from low-energy pion--nucleon cross sections~\cite{RuizdeElvira:2017stg}, making
the resolution of the tension all the more pressing.   

The tensor form factors of the nucleon are defined according to~\cite{Weinberg:1958ut,Adler:1975he}
\begin{align}
\label{tensor_decomposition}
 &\langle N(p')|\bar q \sigma^{\mu\nu} q|N(p)\rangle\notag\\
 &=\bar u(p')\Big[\sigma^{\mu\nu} F_{1,T}^q(t) + \frac{i}{\mN}\big(\gamma^\mu q^\nu-\gamma^\nu q^\mu\big)F^q_{2,T}(t)\notag\\
 &\qquad+\frac{i}{\mN^2}\big(P^\mu q^\nu-P^\nu q^\mu\big)F^q_{3,T}(t)\Big]u(p),
\end{align}
where the tensor charges of the proton $f_q^{T,p}=F_{1,T}^{q,p}(0)$ have been recently calculated to high precision in lattice QCD (evaluated at scale $\mu=2\GeV$)~\cite{Gupta:2018lvp}:
\begin{align}
f_u^{T,p}&=0.784(28),\qquad f_s^{T,p}=-0.0027(16),\notag\\
f_d^{T,p}&=-0.204(11),
\end{align}
and the neutron ones follow from isospin symmetry according to
\beq
\label{isospin}
f_u^{T,n}=f_d^{T,p},\qquad f_d^{T,n}=f_u^{T,p},\qquad f_s^{T,n}=f_s^{T,p}.
\eeq
The other couplings $\tilde \kappa_q^{T,p}=F_{2,T}^{q,p}(0)+2F_{3,T}^{q,p}(0)$, related to the tensor magnetic moments $\kappa_q^{T,p}=-2\tilde \kappa_q^{T,p}$, are less well determined. 
Lattice QCD information~\cite{Gockeler:2006zu} is consistent with an estimate 
based on analyticity and unitarity of the form factors in analogy to Ref.~\cite{Cirigliano:2017tqn}, 
combining the pion tensor charge~\cite{Baum:2011rm} with the electromagnetic form factors of the nucleon~\cite{Hoferichter:2016duk}. The resulting values
$\tilde \kappa_u^{T,p}=-1.3(5)$, $\tilde \kappa_d^{T,p}=-0.7(3)$, $\tilde \kappa_s^{T,p}=0.00(1)$~\cite{Hoferichter:2018zwu} indicate that 
for $u$- and $d$-quarks
the induced terms in the tensor decomposition in Eq.~\eqref{tensor_decomposition} are actually dominant. As in the vector case, the strangeness content is very small.
In addition, the known pion tensor charge~\cite{Baum:2011rm} allows us to calculate the pion-exchange diagram also in this channel, but, similarly to the vector current, 
this contribution is suppressed by $(N-Z)/A$ due to its isospin structure.

Finally, the spin-$2$ couplings of the nucleon are given as moments of nucleon PDFs $q(x)$ in complete analogy to Eq.~\eqref{spin_2_coupling_pion}:
\beq
f^{(2)}_{q,N}=\int_0^1\diff x\, x \big(q(x)+\bar q(x)\big),
\eeq
subject to the same sum rule
\beq
\label{sum_rule_nucleon}
\sum_q f^{(2)}_{q,N} + f^{(2)}_{g,N}=1.
\eeq
Numerically, this gives for the proton (again at $\mu=2\GeV$)~\cite{Martin:2009iq,Buckley:2014ana}
\begin{align}
\label{nucleon_PDF}
 f^{(2)}_{u,p}&=0.346(6), & f^{(2)}_{c,p}&=0.0088(3),\notag\\
 f^{(2)}_{d,p}&=0.192(5), & f^{(2)}_{g,p}&=0.419(11),\notag\\
 f^{(2)}_{s,p}&=0.034(3), & &
\end{align}
while the neutron couplings follow as in Eq.~\eqref{isospin} by exchanging $u$ and $d$.

\section{Matching to NREFT for scalar dark matter}
\label{sec:spin0}

For a spin-$0$, Standard-Model singlet $\chi$ the analysis is based on the effective Lagrangian
\begin{align}
\label{Lagr_0}
\mathcal{L}_{\chi}&=\mathcal{L}_{\chi}^{(6)}+\mathcal{L}_{\chi}^{(7)}+\mathcal{L}_{\chi}^{(8)},\\
\mathcal{L}_{\chi}^{(6)}
&=\frac{1}{\Lambda^2}\bigg\{\sum_{q} \Big(C_{q}^{SS}+\frac{8\pi}{9}C'^S_{g}\Big)\chi^\dagger \chi \,m_q\bar q q\notag\\
&-\frac{8\pi}{9}C'^S_{g}\chi^\dagger\chi\, \theta^\mu_\mu
+\sum_q C_q^{VV}\chi^\dagger i\partial^\mu\chi \,\bar q\gamma_\mu q
\bigg\},\notag\\
\mathcal{L}_{\chi}^{(7)}&=\frac{1}{\Lambda^3}\sum_q C_q^{TT}i\partial^\mu\chi^\dagger \partial^\nu \chi \,\bar q \sigma_{\mu\nu} q,\notag\\
\mathcal{L}_{\chi}^{(8)}&=\frac{1}{\Lambda^4}\bigg\{\sum_q C_q^{(2)}\chi^\dagger \partial_\mu\partial_\nu \chi \,\bar \theta^{\mu\nu}_q
+C_g^{(2)} \chi^\dagger \partial_\mu \partial_\nu \chi\,\bar \theta^{\mu\nu}_g\bigg\},\notag
\end{align}
where the notation follows closely the spin-$1/2$ case in Eq.~\eqref{Lagr}, and for a real scalar we have $C_q^{VV}=C_q^{TT}=0$.
We have not included an axial term of the form $\chi^\dagger i\partial^\mu\chi \,\bar q\gamma_\mu\gamma_5 q$, because such a contribution reduces to
a combination of $\Op_7$ and $\Op_{10}$ NREFT operators, and is therefore even further suppressed than the standard SD interaction in the spin-$1/2$ case. Similarly, we have neglected
a tensor operator with $\bar q\sigma_{\mu\nu}i\gamma_5 q$.

Most single-nucleon amplitudes come out as in the spin-$1/2$ case, up to an additional factor of $\mc$ whenever a derivative is required in the effective operator, and
the corresponding factors of $\Lambda$. To have the reduced amplitudes in the same conventions as for spin-$1/2$, a factor $\mc$ needs to be removed which leads to 
\begin{align}
\label{matching_spin_0}
 \M_{1,\text{NR}}^{SS}&= \Op_1 \frac{\Lambda}{\mc}f_N(t),\\
 \M_{1,\text{NR}}^{VV}&=\Op_1\Big(f_1^{V,N}(t)+\frac{t}{4\mN^2}f_2^{V,N}(t)\Big)\notag\\
 &+\frac{1}{\mN}\Op_3 \bigg[f_2^{V,N}(t)+\frac{1}{2}\Big(1+\frac{\muN}{\mc}\Big)f_1^{V,N}(t)\bigg],\notag\\
\M_{1,\text{NR}}^{(2)}&=\frac{3}{4}\Op_1 f_N^{(2)},\notag\\
\M_{1,\text{NR}}^{TT}&=\frac{1}{\Lambda}f^{T,N}_1\bigg[2\Op_3+\frac{t}{2\mN}\Op_1\bigg]
+\frac{1}{\Lambda}f^{T,N}_2\frac{t}{\mN}\Op_1,\notag
\end{align}
where we have kept the same notation for the couplings as in Eq.~\eqref{hadronic_couplings}, with the understanding that the Wilson coefficients therein now refer to the operators 
defined in Eq.~\eqref{Lagr_0}.

In total, the analog of Eq.~\eqref{c_coeff} for spin-$0$ becomes
\begin{align}
\label{c_coeff_0}
 c_\pm^M&=\frac{\zeta}{2}\bigg[\frac{\Lambda}{\mc}\big(f_p\pm f_n\big)+f_1^{V,p}\pm f_1^{V,n}\notag\\
 &+\frac{3}{4}\big(f_p^{(2)}\pm f_n^{(2)}\big)\bigg],\notag\\
\dot c_\pm^M&=\frac{\zeta m_N^2}{2}\bigg[\frac{\Lambda}{\mc}\big(\dot f_p\pm \dot f_n\big)+\dot f_1^{V,p}\pm \dot f_1^{V,n}\notag\\
&+\frac{1}{4\mN^2}\big(f_2^{V,p}\pm f_2^{V,n}\big)+\frac{1}{2\mN\Lambda}\big(f^{T,p}_1\pm f^{T,n}_1\big)\notag\\
&+\frac{1}{\mN\Lambda}\big(f^{T,p}_2\pm f^{T,n}_2\big)\bigg],\notag\\
c_\pi&=\zeta \bigg[\frac{\Lambda}{\mc}\big(f_\pi+2f_\pi^\theta\big)-\frac{1}{2}f_\pi^{(2)}\bigg],\notag\\
c_\text{b}&=\zeta\bigg[\frac{\Lambda}{\mc}f_\pi^\theta+\frac{1}{4}f_\pi^{(2)}\bigg],\notag\\
c_\pm^{\Phi''}&=\frac{\zeta}{2}\bigg[f_2^{V,p}\pm f_2^{V,n}+
\frac{1}{2}\Big(1+\frac{\muN}{\mc}\Big)\big(f_1^{V,p}\pm f_1^{V,n}\big)\notag\\
&+\frac{2\mN}{\Lambda}\big(f^{T,p}_1\pm f^{T,n}_1\big)\bigg],
\end{align}
where now $\zeta=1 (2)$ for a complex (real) scalar. At this order in the ChEFT expansion we do not find contributions from $\Op_{5,8,11}$.

\section{Two-body structure factors}
\label{sec:app2bc}

The two-body $\theta^\mu_\mu$ and spin-$2$ amplitudes given in Eqs.~\eqref{Eq:Mtheta_2b} and~\eqref{Eq:Mspin2_2b}
can be rewritten according to
\begin{align}
	\M^\theta_{2,\text{NR}} =& -\frac{f_\pi^\theta}{\mpi}\Bigg[
\bigg(\frac{\gA}{2\Fpi}\bigg)^2 \frac{\ttau_1\cdot\ttau_2\,\sig_1\cdot\vec{q}_1\,\sig_2\cdot\vec{q}_2}{\big(q_1^2+\mpi^2\big)\big(q_2^2+\mpi^2\big)}\notag\\
&\qquad\times(2\mpi^2-q^2)\notag\\	
	&+\bigg(\frac{g_A}{2F_\pi}\bigg)^2 \ttau_1\cdot\ttau_2\, \sig_1\cdot\vec{q}_1\sig_2\cdot\vec{q}_2\notag\\
	&\qquad\times\bigg(\frac{1}{q_1^2+\mpi^2}+\frac{1}{q_2^2+\mpi^2}\bigg)\notag\\
	&+ 2( C_S + C_T \sig_1 \cdot \sig_2)\Bigg],\\
	\M^{(2)}_{2,\text{NR}} =& \frac{f_\pi^{(2)}}{4\mpi}\Bigg[
\bigg(\frac{\gA}{2\Fpi}\bigg)^2 \frac{\ttau_1\cdot\ttau_2\,\sig_1\cdot\vec{q}_1\,\sig_2\cdot\vec{q}_2}{\big(q_1^2+\mpi^2\big)\big(q_2^2+\mpi^2\big)}\notag\\
&\qquad\times(2\mpi^2+q^2)\notag\\	
	&-\bigg(\frac{g_A}{2F_\pi}\bigg)^2 \ttau_1\cdot\ttau_2\, \sig_1\cdot\vec{q}_1\sig_2\cdot\vec{q}_2\notag\\
	&\qquad\times\bigg(\frac{1}{q_1^2+\mpi^2}+\frac{1}{q_2^2+\mpi^2}\bigg)\notag\\
	&- 2( C_S + C_T \sig_1 \cdot \sig_2)\Bigg],
\end{align}
where we have used that $\vec{q}=-\vec{q}_1-\vec{q}_2$ and thus
\beq
  2\vec{q}_1 \cdot \vec{q}_2 = q^2-q_1^2-q_2^2.
\eeq
In both cases most of the first term excluding the $q^2$-dependent piece can be related to $\M_{2,\text{NR}}^{SS}$, leading to the $\F_\pi$ contribution in Eq.~\eqref{F_theta_2}. The $q^2$-dependent part can be 
absorbed into a redefinition of $\F_\text{b}$ in Eq.~\eqref{F_pi_b_shell_model} to avoid the introduction of another structure factor.
This redefinition does not change the normalization of $\F_\text{b}(0)$ given by the nuclear binding energy.
The remaining term involving pion propagators can be expressed in terms of 
\begin{align}
	\M_{NN}^{(i)} =
	\bigg(\frac{g_A}{2F_\pi}\bigg)^2 \ttau_1\cdot\ttau_2\, \frac{\sig_1\cdot\vec{q}_1\sig_2\cdot\vec{q}_2}{q_i^2+\mpi^2},\qquad i=1,2.
	\label{Eq:vpotnn}
\end{align}

On the other hand, the relativistic corrections to the one-body $\theta$ term and the spin-$2$ channel read
\begin{align}
	\Delta \M_{1,\text{NR}}^{\theta} &= -\frac{f_\pi^\theta}{\mpi} \frac{p_1^2+p_2^2+p_1'^2+p_2'^2}{4\mN} \notag\\
	&\times\big(\delta(\vec{p}_1-\vec{p}_1')+\delta(\vec{p}_2-\vec{p}_2')\big),
	\label{Eq:Mtheta_1b}\\
	\Delta \M_{1,\text{NR}}^{(2)} &= \frac{f_N^{(2)}}{16\mN^2}\bigg[(p_2^2+p_2'^2-3p_1^2-3p_1'^2)
	\delta(\vec{p}_1-\vec{p}'_1)\notag\\
	&+(p_1^2+p_1'^2-3p_2^2-3p_2'^2)\delta(\vec{p}_2-\vec{p}'_2)\bigg].
	\label{Eq:Mspin2_1b}
\end{align}
In the limit $\vec{q}\rightarrow0$ these amplitudes become proportional to the kinetic-energy operator $T$: 
\begin{align}
	\Delta \M_{1,\text{NR}}^{\theta}&=-\frac{2f_\pi^\theta}{\mpi}T,\notag\\
	\Delta \M_{1,\text{NR}}^{(2)}&=-\frac{f_N^{(2)}}{2\mN}T.
\end{align}

Summarizing all 1b and 2b contributions we obtain
\begin{align}
	\M_{1+2,\text{NR}}^{\theta}=&f^\theta_\pi\bigg[\bigg(2-\frac{q^2}{\mpi^2}\bigg)\frac{1}{f_\pi}\M_{2,\text{NR}}^{SS} \label{Eq:Mtheta1+2}\\
	&-\frac{1}{\mpi}\Big[\M_{NN}^{(1)}+\M_{NN}^{(2)}\notag\\
	&+ 2( C_S + C_T \sig_1 \cdot \sig_2)\Big]+\frac{1}{f^\theta_\pi}\Delta \M_{1,\text{NR}}^{\theta}\bigg],\notag\\
	\M_{1+2,\text{NR}}^{(2)}=&f^{(2)}_\pi\bigg[-\frac{2\mpi^2+q^2}{4\mpi^2}\frac{1}{f_\pi}\M_{2,\text{NR}}^{SS}\label{Eq:Mspin21+2}\\
	 &-\frac{1}{4\mpi}\Big[\M_{NN}^{(1)}+\M_{NN}^{(2)}\notag\\
	 &+ 2( C_S + C_T \sig_1 \cdot \sig_2)\Big]+\frac{1}{f_\pi^{(2)}}\Delta \M_{1,\text{NR}}^{(2)}\bigg].\notag
\end{align}
In the limit $\vec{q} \rightarrow 0$ we have $\vec{q}_1=-\vec{q}_2$, so that the momentum transfers in $\M_{NN}^{(1)}$ and $\M_{NN}^{(2)}$ become equal
and both can be identified with the pion-exchange part $V_\text{OPE}^{(0)}$
of the leading-order chiral $NN$ potential:
\beq
  \M_{NN}^{(i)} \rightarrow  V_\text{OPE}^{(0)}.
\eeq
Together with the contact terms in Eqs.~\eqref{Eq:Mtheta1+2} and~\eqref{Eq:Mspin21+2} we recover twice the complete leading-order chiral $NN$ potential $V_{NN}^\text{LO}$.

In the limit $\vec{q}\rightarrow 0$ both $\Delta \M_{1,\text{NR}}^{\theta}$ and $\Delta \M_{1,\text{NR}}^{(2)}$ become proportional to $T$, leading in both cases to the linear combination $T+V_{NN}^\text{LO}$, that can be renormalized to the nuclear binding energy at LO order in the
chiral expansion.
For the $\theta^\mu_\mu$ current in Eqs.~\eqref{Eq:Mtheta_1b} and \eqref{Eq:Mtheta_2b} this conclusion follows directly from Eq.~\eqref{Eq:Mtheta1+2}.
On the other hand, the spin-$2$ one-body contribution in Eq.~\eqref{Eq:Mspin2_1b} 
carries a coupling $f_N^{(2)}$ that, in general, may differ from $f_\pi^{(2)}$, the coupling of the two-body term in Eq.~\eqref{Eq:Mspin2_2b},
see Eqs.~\eqref{hadronic_couplings} and~\eqref{hadronic_couplings_pion}. However, due to the sum rules in Eqs.~\eqref{sum_rule_pion} and~\eqref{sum_rule_nucleon},
when all Wilson coefficients are set equal we have
$f_N^{(2)}/\mN=f_\pi^{(2)}/\mpi$, so that 
these two couplings cancel in the last term of Eq.~\eqref{Eq:Mspin21+2} to ensure renormalizability at this order in ChEFT. 
Since the comparison of Eqs.~\eqref{f2_pion_JAM} and~\eqref{nucleon_PDF} shows that the individual spin-$2$ couplings do not differ much 
(especially when considering an isospin average for the nucleon), we assume that for spin-$2$ the kinetic-energy operator aligns as required. 
Similarly, in Eq.~\eqref{Eq:Mspin2_2b} we also assume that the coefficients of the spin-$2$ $(N^\dagger N)^2$ contact operators can be determined
in the same way, at least up to higher-order effects. 

In practice, once the contribution at $q=0$ is renormalized to the nuclear binding energy at LO by adjusting the contact terms $C_S$ and $C_T$, 
the contributions from $\Delta \M_{1,\text{NR}}^{\theta}$ and $\Delta \M_{1,\text{NR}}^{(2)}$ to the full structure factors are small. 
In addition, while in principle the structure factors could differ for finite $q$, due to the different functional form of Eqs.~\eqref{Eq:Mtheta_1b} and~\eqref{Eq:Mspin2_1b} for $q\neq 0$,
we find that the results are practically the same using either expression.
In view of the uncertainties from higher orders in the chiral expansion, this shows that the definition of a single new structure factor is sufficient.
We choose 
\begin{align}
	\M_\text{b}=&-\frac{1}{\mpi}\Big[\M_{NN}^{(1)}+\M_{NN}^{(2)}+ 2( C_S + C_T \sig_1 \cdot \sig_2)\Big]\notag\\
	&+\frac{1}{f^\theta_\pi}\Delta \M_{1,\text{NR}}^{\theta},
\end{align}
which in the limit $\vec{q}\rightarrow 0$ reduces to
\beq
	\M_\text{b}\rightarrow -\frac{2}{\mpi}(T+V_{NN}^\text{LO}).
	\label{Eq:Mbq0}
\eeq
The corresponding response functions in the naive shell model are
\begin{align}
\label{F_2b_final}
\F_\pi(q^2)&=\frac{1}{2}\sum_{\text{occ}}\langle N_1N_2|(1-P_{12})|\frac{1}{f_\pi}\M_{2,\text{NR}}^{SS}|N_1N_2\rangle,\notag\\
\bar\F_\text{b}(q^2)&=\frac{1}{2}\sum_{\text{occ}}\langle N_1N_2|(1-P_{12})|\M_\text{b}|N_1N_2\rangle,
\end{align}
where due to Eq.~\eqref{Eq:Mbq0} we have $\bar\F_\text{b}(0)=-2E_\text{b}/\mpi$ with the binding energy of the nucleus $E_\text{b}<0$.
In Eq.~\eqref{F_pi_b_shell_model} we redefine
$\F_\text{b}(q^2)=\bar\F_\text{b}(q^2)-\frac{q^2}{\mpi^2} \F_\pi(q^2)$.

\begin{table}[t]
	\centering
	\renewcommand{\arraystretch}{1.3}
	\begin{tabular}{cccccc}
		\toprule
		Structure factor & F & Si & Ar & Ge & Xe\\\colrule
		$\F^M_\pm$ & 2 & 2 & 3 & 4 & 5\\
		$\F^{\Phi''}_\pm$ & 1 & 1 & 2 & 3 & 4\\
		$\F_\pi$ & 3 & 3 & 3 & 4 & 5\\
		$\F_\text{b}$ & 2 & 2 & 3 & 4 & 5\\
		\botrule
	\end{tabular}
	\renewcommand{\arraystretch}{1.0}
	\caption{Maximal power $m$ in Eq.~\eqref{fit_function} for each structure factor and nuclear target considered in this work.
	}
	\label{tab:oscillator_model}
\end{table}

We evaluate the two-body structure factors in Eq.~\eqref{F_2b_final} in a non-interacting shell model, as described in detail in Ref.~\cite{Hoferichter:2016nvd},
with the multidimensional integrations performed using the \textsc{CUBA} library~\cite{Hahn:2004fe}.
For orbitals in the configuration space of the nuclear shell-model calculations, the occupation numbers are taken from the full diagonalization discussed in Sec.~\ref{sec:shell_model}.
The final results are represented by a fit function
\beq
\label{fit_function}
\F(u) = e^{-\frac{u}{2}} \sum\limits_{i=0}^m c_i u^i.
\eeq
For one-body contributions this is an exact analytic form with known maximal power $m$,
and the overall normalization is set by
 $c_0=A$ and $c_0=Z-N$ for $\F^M_{+}$ and $\F^M_{-}$,
respectively. The expansion is organized in the variable $u=q^2b^2/2$, with 
harmonic-oscillator lengths
\beq
\label{oscillator_length}
b=\sqrt{\frac{\hbar}{\mN \omega}},\quad
\hbar\omega = (45 A^{-1/3}-25 A^{-2/3})\MeV,
\eeq
that depend on the mass number $A$ of the nucleus.
The functional form in Eq.~\eqref{fit_function} proves an efficient representation of the two-body structure factors as well, 
but the optimal maximal power needs to be determined empirically.
Table~\ref{tab:oscillator_model} provides the complete list.

\section{ChiralEFT4DM: \textsc{Python} Notebook}
\label{sec:user_guide}

We strongly encourage current and future analyses of direct detection experiments
to use the structure factors for the different 
interaction channels discussed in this manuscript. For convenience we offer an accompanying \textsc{Python} package in the form of a \textsc{Jupyter} notebook, which can be downloaded from \path{ https://theorie.ikp.physik.tu-darmstadt.de/strongint/ChiralEFT4DM.html}. The notebook calculates both nuclear structure factors and differential recoil spectra: 
\beq
	\frac{\diff R}{\diff E_R}=\frac{\rho}{2\pi \mc}\times\big|\mathcal{F}(q^2)\big|^2\times\int_{v_\text{min}(E_R)}^{\infty}
\frac{f(\vv)}{v}\diff^3v,
\eeq
including the general response
\begin{align}
	|\mathcal{F}(q^2)\big|^2 =& \bigg|\sum_{I=\pm}\Big(c_I^M-\frac{q^2}{m_N^2} \, \dot c_I^M\Big)\mathcal{F}_I^M(q^2)
	+c_\pi\mathcal{F}_\pi(q^2)\notag\\
&+c_\text{b}\mathcal{F}_\text{b}(q^2)+\frac{q^2}{2m_N^2}\sum_{I=\pm}c_I^{\Phi''}\mathcal{F}_I^{\Phi''}(q^2)\bigg|^2\notag\\
&+\sum_{i=5,8,11}\bigg|\sum_{I=\pm}\xi_i(q,v)c_I^{M,i}\mathcal{F}_I^M(q^2)\bigg|^2,
\end{align}
where $\rho$ and $f(v)$ denote the local dark matter density and velocity distribution, respectively.
The notebook gives results
for all possible coherently enhanced couplings of spin-$1/2$ and spin-$0$ WIMPs to one and
two nucleons up to third order in ChEFT, as discussed in the present manuscript. All stable isotopes of the most relevant nuclear targets including fluorine, silicon, argon,
germanium, and xenon are available. In addition, our package calculates
the responses based upon the fundamental couplings at the level of quarks and gluons
as incorporated in the respective Wilson coefficients.

The notebook aims to be self-explanatory and easy-to-use even for users
new to \textsc{Python}. When downloaded from the website the files are stored in
an archive. When unpacked to a common directory the notebook can be
loaded. In the first part of the notebook, users can specify a given response
and create data sets for both nuclear structure factors and differential recoil
spectra. In the second part of the notebook, users can set specific
values for the Wilson coefficients that describe the WIMP--quark/gluon
couplings. The notebook generates the corresponding nucleon and pion
matrix elements. Finally, the package yields the response including all
channels that contribute to the choice of Wilson coefficients.

For completeness, we also implemented a routine that calculates the rate corresponding to the standard halo model (SHM)~\cite{Lewin:1995rx}. 
In this way we clarify conventions and facilitate the use of improved astrophysical input in the future, 
as is increasingly becoming available with the Gaia mission~\cite{Brown:2018dum}, see, e.g., Refs.~\cite{Necib:2018iwb,Eilers:2018,Evans:2018bqy}.
The SHM is defined by $\rho=0.3\GeV/\text{cm}^3$ and 
\begin{align}
f(\vv)&=\frac{e^{-\frac{(\vv+\vv_E)^2}{v_0^2}}}{\pi^{3/2}v_0^3 k}\theta(v_\text{esc}-|\vv+\vv_E|),\\
k&=\erf(z)-\frac{2z}{\sqrt{\pi}}e^{-z^2},\qquad \erf(z)=\frac{2}{\sqrt{\pi}}\int_0^z\diff x\,e^{-x^2},\notag
\end{align}
where $z=v_\text{esc}/v_0$, $v_\text{esc}=544\kms$, $v_0=220\kms$,
and the Earth's velocity $v_E=232\kms$ drops out in the normalization. 
For the operators considered in the notebook one needs
\begin{align}
 g(v_\text{min})&=\int_{v_\text{min}}^{\infty}
\frac{f(\vv)}{v}\diff^3v\notag\\
&=\frac{1}{2v_E k}\bigg[\erf(z')-\erf(x_\text{min}-\eta)-\frac{2}{\sqrt{\pi}}z''e^{-z^2}\bigg],\notag\\
\tilde g(v_\text{min})&=\int_{v_\text{min}}^{\infty}
v f(\vv)\diff^3v\notag\\
&=\frac{v_0^2}{4v_E k}\big(1+2\eta^2\big)\Big(\erf(z')-\erf(x_\text{min}-\eta)\Big)\notag\\
&+\frac{v_0^2}{\sqrt{\pi}v_E k}\bigg[\frac{x_\text{min}+\eta}{2}e^{-(x_\text{min}-\eta)^2}-\frac{x_\text{min}-\eta}{2}e^{-z'^2}\notag\\
&-\frac{e^{-z^2}}{3}\Big((z''+x_\text{min})^3-x_\text{min}^3+\frac{3}{2}z''+3\eta\notag\\
&+6\eta(z-z')(x_\text{min}+z+\eta)\Big)\bigg],
\end{align}
where $x_\text{min}=v_\text{min}/v_0$, $\eta=v_E/v_0$, and 
\beq
z'=\min(z,x_\text{min}+\eta),\qquad z''=z'+\eta-x_\text{min}.
\eeq


\begin{thebibliography}{999}

\bibitem{Baudis:2016qwx} 
  L.~Baudis,
  J.\ Phys.\ G {\bf 43}, 044001 (2016).

\bibitem{Angloher:2015ewa} 
G.~Angloher {\it et al.} [CRESST Collaboration],
Eur.\ Phys.\ J.\ C {\bf 76}, 25 (2016)
[arXiv:1509.01515 [astro-ph.CO]].

\bibitem{Amole:2016pye} 
C.~Amole {\it et al.} [PICO Collaboration],
Phys.\ Rev.\ D {\bf 93}, 061101 (2016)
[arXiv:1601.03729 [astro-ph.CO]].

\bibitem{Armengaud:2016cvl} 
E.~Armengaud {\it et al.} [EDELWEISS Collaboration],
JCAP {\bf 1605}, 019 (2016)
[arXiv:1603.05120 [astro-ph.CO]].

\bibitem{Akerib:2016vxi} 
D.~S.~Akerib {\it et al.} [LUX Collaboration],
Phys.\ Rev.\ Lett.\  {\bf 118}, 021303 (2017)
[arXiv:1608.07648 [astro-ph.CO]].

\bibitem{Aprile:2017iyp} 
E.~Aprile {\it et al.} [XENON Collaboration],
Phys.\ Rev.\ Lett.\  {\bf 119}, 181301 (2017)
[arXiv:1705.06655 [astro-ph.CO]].

\bibitem{Agnese:2017jvy} 
R.~Agnese {\it et al.} [SuperCDMS Collaboration],
Phys.\ Rev.\ D {\bf 97}, 022002 (2018)
[arXiv:1707.01632 [astro-ph.CO]].

\bibitem{Amaudruz:2017ekt} 
P.~A.~Amaudruz {\it et al.} [DEAP-3600 Collaboration],
Phys.\ Rev.\ Lett.\  {\bf 121}, 071801 (2018)
[arXiv:1707.08042 [astro-ph.CO]].

\bibitem{Cui:2017nnn}
X.~Cui {\it et al.} [PandaX-II Collaboration],
Phys.\ Rev.\ Lett.\  {\bf 119}, 181302 (2017)
[arXiv:1708.06917 [astro-ph.CO]].

\bibitem{Agnese:2017njq} 
R.~Agnese {\it et al.} [SuperCDMS Collaboration],
Phys.\ Rev.\ Lett.\  {\bf 120},  061802 (2018)
[arXiv:1708.08869 [hep-ex]].
	
\bibitem{Agnes:2018ves} 
P.~Agnes {\it et al.} [DarkSide Collaboration],
Phys.\ Rev.\ Lett.\ {\bf 121}, 081307 (2018) 
[arXiv:1802.06994 [astro-ph.HE]].

\bibitem{XMASS:2018bid} 
  K.~Abe {\it et al.} [XMASS Collaboration],
  Phys.\ Lett.\ B {\bf 789}, 45 (2019)
  [arXiv:1804.02180 [astro-ph.CO]].

\bibitem{Aprile:2018dbl} 
  E.~Aprile {\it et al.} [XENON Collaboration],
  Phys.\ Rev.\ Lett.\ {\bf 121}, 111302 (2018) 
  [arXiv:1805.12562 [astro-ph.CO]].
		
\bibitem{Aprile:2015uzo} 
E.~Aprile {\it et al.} [XENON Collaboration],
JCAP {\bf 1604}, 027 (2016)
[arXiv:1512.07501 [physics.ins-det]].

\bibitem{Akerib:2015cja} 
D.~S.~Akerib {\it et al.} [LZ Collaboration],
arXiv:1509.02910 [physics.ins-det].
		
\bibitem{PandaXxT}
\path{https://static.pandax.sjtu.edu.cn/download/IAC-2016-pandax.pdf}.

\bibitem{Aalbers:2016jon} 
J.~Aalbers {\it et al.} [DARWIN Collaboration],
JCAP {\bf 1611}, 017 (2016)
[arXiv:1606.07001 [astro-ph.IM]].

\bibitem{Akimov:2017ade} 
D.~Akimov {\it et al.} [COHERENT Collaboration],
Science {\bf 357}, 1123 (2017)
[arXiv:1708.01294 [nucl-ex]].

\bibitem{Bertone} 
G.\ Bertone, {\it Particle Dark Matter: Observations, Models and Searches},
Cambridge University Press, Cambridge, England, 2010.
    
\bibitem{Engel:1992bf} 
  J.~Engel, S.~Pittel and P.~Vogel,
  Int.\ J.\ Mod.\ Phys.\ E {\bf 01}, 1 (1992).

\bibitem{Lewin:1995rx} 
J.~D.~Lewin and P.~F.~Smith,
Astropart.\ Phys.\  {\bf 6}, 87 (1996).

\bibitem{Vietze:2014vsa} 
L.~Vietze, P.~Klos, J.~Men\'endez, W.~C.~Haxton and A.~Schwenk,
Phys.\ Rev.\ D {\bf 91}, 043520 (2015)
[arXiv:1412.6091 [nucl-th]].

\bibitem{Baudis:2013bba} 
  L.~Baudis, G.~Kessler, P.~Klos, R.~F.~Lang, J.~Men\'endez, S.~Reichard and A.~Schwenk,
  Phys.\ Rev.\ D {\bf 88}, 115014 (2013)
  [arXiv:1309.0825 [astro-ph.CO]].
  
\bibitem{McCabe:2015eia} 
  C.~McCabe,
  JCAP {\bf 1605}, 033 (2016)
  [arXiv:1512.00460 [hep-ph]].

\bibitem{Rogers:2016jrx} 
H.~Rogers, D.~G.~Cerde\~no, P.~Cushman, F.~Livet and V.~Mandic,
Phys.\ Rev.\ D {\bf 95}, 082003 (2017)
[arXiv:1612.09038 [astro-ph.CO]].

\bibitem{Fieguth:2018vob} 
  A.~Fieguth, M.~Hoferichter, P.~Klos, J.~Men\'endez, A.~Schwenk and C.~Weinheimer,
  Phys.\ Rev.\ D {\bf 97}, 103532 (2018)
  [arXiv:1802.04294 [hep-ph]].

\bibitem{Aprile:2016swn} 
E.~Aprile {\it et al.} [XENON100 Collaboration],
Phys.\ Rev.\ D {\bf 94}, 122001 (2016)
[arXiv:1609.06154 [astro-ph.CO]].

\bibitem{Fu:2016ega} 
    C.~Fu {\it et al.} [PandaX-II Collaboration],
    Phys.\ Rev.\ Lett.\  {\bf 118}, 071301 (2017)
    Erratum: [Phys.\ Rev.\ Lett.\ {\bf 120}, 049902 (2018)]
    [arXiv:1611.06553 [hep-ex]].

\bibitem{Akerib:2017kat} 
D.~S.~Akerib {\it et al.} [LUX Collaboration],
Phys.\ Rev.\ Lett.\  {\bf 118}, 251302 (2017)
[arXiv:1705.03380 [astro-ph.CO]].

\bibitem{Aprile:2019dbj} 
  E.~Aprile {\it et al.} [XENON Collaboration],
  arXiv:1902.03234 [astro-ph.CO].
  
\bibitem{Amole:2019fdf} 
  C.~Amole {\it et al.} [PICO Collaboration],
  arXiv:1902.04031 [astro-ph.CO].

\bibitem{Angloher:2016jsl} 
G.~Angloher {\it et al.},
Phys.\ Rev.\ Lett.\  {\bf 117},  021303 (2016)
[arXiv:1601.04447 [astro-ph.CO]].

\bibitem{Fan:2010gt}
J.~Fan, M.~Reece and L.~T.~Wang,
JCAP {\bf 1011}, 042 (2010) 
[arXiv:1008.1591 [hep-ph]].
  
\bibitem{Fitzpatrick:2012ix} 
A.~L.~Fitzpatrick, W.~Haxton, E.~Katz, N.~Lubbers and Y.~Xu,
JCAP {\bf 1302}, 004 (2013)
[arXiv:1203.3542 [hep-ph]].
  
\bibitem{Anand:2013yka} 
N.~Anand, A.~L.~Fitzpatrick and W.~C.~Haxton,
Phys.\ Rev.\ C {\bf 89},  065501 (2014)
[arXiv:1308.6288 [hep-ph]].
  
\bibitem{Schneck:2015eqa} 
K.~Schneck {\it et al.} [SuperCDMS Collaboration],
Phys.\ Rev.\ D {\bf 91}, 092004 (2015)
[arXiv:1503.03379 [astro-ph.CO]].

\bibitem{Aprile:2017aas} 
E.~Aprile {\it et al.} [XENON Collaboration],
Phys.\ Rev.\ D {\bf 96}, 042004 (2017)
[arXiv:1705.02614 [astro-ph.CO]].

 \bibitem{Xia:2018qgs} 
 J.~Xia {\it et al.} [PandaX-II Collaboration],
 arXiv:1807.01936 [hep-ex].

\bibitem{Angloher:2018fcs} 
  G.~Angloher {\it et al.} [CRESST Collaboration],
  Eur.\ Phys.\ J.\ C {\bf 79}, 43 (2019)
  [arXiv:1809.03753 [hep-ph]].

\bibitem{Epelbaum:2008ga} 
  E.~Epelbaum, H.~W.~Hammer and U.-G.~Mei{\ss}ner,
  Rev.\ Mod.\ Phys.\  {\bf 81}, 1773 (2009)
  [arXiv:0811.1338 [nucl-th]].
  
\bibitem{Machleidt:2011zz} 
  R.~Machleidt and D.~R.~Entem,
  Phys.\ Rept.\  {\bf 503}, 1 (2011)
  [arXiv:1105.2919 [nucl-th]].

\bibitem{Hammer:2012id} 
  H.-W.~Hammer, A.~Nogga and A.~Schwenk,
  Rev.\ Mod.\ Phys.\ {\bf 85}, 197 (2013)
  [arXiv:1210.4273 [nucl-th]].

\bibitem{Hoferichter:2015ipa} 
M.~Hoferichter, P.~Klos and A.~Schwenk,
Phys.\ Lett.\ B {\bf 746}, 410 (2015)
[arXiv:1503.04811 [hep-ph]].

\bibitem{Prezeau:2003sv}
  G.~Pr\'ezeau, A.~Kurylov, M.~Kamionkowski and P.~Vogel,
  Phys.\ Rev.\ Lett.\  {\bf 91}, 231301 (2003) 
  [astro-ph/0309115].

\bibitem{Cirigliano:2012pq}
  V.~Cirigliano, M.~L.~Graesser and G.~Ovanesyan,
  JHEP {\bf 1210}, 025 (2012) 
  [arXiv:1205.2695 [hep-ph]].

\bibitem{Menendez:2012tm}
  J.~Men\'endez, D.~Gazit and A.~Schwenk,
  Phys.\ Rev.\ D {\bf 86}, 103511 (2012) 
  [arXiv:1208.1094 [astro-ph.CO]].

\bibitem{Klos:2013rwa} 
  P.~Klos, J.~Men\'endez, D.~Gazit and A.~Schwenk,
  Phys.\ Rev.\ D {\bf 88}, 083516 (2013)
  Erratum: [Phys.\ Rev.\ D {\bf 89}, 029901 (2014)]
  [arXiv:1304.7684 [nucl-th]].

\bibitem{Cirigliano:2013zta}
  V.~Cirigliano, M.~L.~Graesser, G.~Ovanesyan and I.~M.~Shoemaker,
  Phys.\ Lett.\ B {\bf 739}, 293 (2014) 
  [arXiv:1311.5886 [hep-ph]].

\bibitem{Hoferichter:2016nvd} 
M.~Hoferichter, P.~Klos, J.~Men\'endez and A.~Schwenk,
Phys.\ Rev.\ D {\bf 94}, 063505 (2016)
[arXiv:1605.08043 [hep-ph]].

\bibitem{Hoferichter:2017olk} 
  M.~Hoferichter, P.~Klos, J.~Men\'endez and A.~Schwenk,
  Phys.\ Rev.\ Lett.\  {\bf 119}, 181803 (2017)
  [arXiv:1708.02245 [hep-ph]].
   
\bibitem{Andreoli:2018etf}
  L.~Andreoli, V.~Cirigliano, S.~Gandolfi and F.~Pederiva,
  Phys.\ Rev.\ C {\bf 99} (2019) 025501
  [arXiv:1811.01843 [nucl-th]].
  
\bibitem{Aprile:2018cxk} 
  E.~Aprile {\it et al.},
  Phys.\ Rev.\ Lett.\ {\bf 122} (2019) 071301
  [arXiv:1811.12482 [hep-ph]].

\bibitem{Gazit:2008ma}
    D.~Gazit, S.~Quaglioni and P.~Navr\'atil,
    Phys.\ Rev.\ Lett.\  {\bf 103}, 102502 (2009) 
    Erratum: [Phys.\ Rev.\ Lett.\ {\bf 122}, 029901 (2019)]
    [arXiv:0812.4444 [nucl-th]].

\bibitem{Menendez:2011qq} 
  J.~Men\'endez, D.~Gazit and A.~Schwenk,
  Phys.\ Rev.\ Lett.\  {\bf 107}, 062501 (2011)
  [arXiv:1103.3622 [nucl-th]].

\bibitem{Bacca:2014tla} 
S.~Bacca and S.~Pastore,
J.\ Phys.\ G {\bf 41}, 123002 (2014)
[arXiv:1407.3490 [nucl-th]].

\bibitem{Pastore:2017uwc} 
S.~Pastore, A.~Baroni, J.~Carlson, S.~Gandolfi, S.~C.~Pieper, R.~Schiavilla and R.~B.~Wiringa,
Phys.\ Rev.\ C {\bf 97}, 022501 (2018)
[arXiv:1709.03592 [nucl-th]].

\bibitem{Bishara:2016hek} 
F.~Bishara, J.~Brod, B.~Grinstein and J.~Zupan,
JCAP {\bf 1702}, 009 (2017)
[arXiv:1611.00368 [hep-ph]].
		
\bibitem{Bishara:2017pfq} 
F.~Bishara, J.~Brod, B.~Grinstein and J.~Zupan,
JHEP {\bf 1711}, 059 (2017)
[arXiv:1707.06998 [hep-ph]].

\bibitem{Korber:2017ery} 
C.~K\"orber, A.~Nogga and J.~de Vries,
Phys.\ Rev.\ C {\bf 96}, 035805 (2017)
[arXiv:1704.01150 [hep-ph]].

\bibitem{Gazda:2016mrp} 
D.~Gazda, R.~Catena and C.~Forss\'en,
Phys.\ Rev.\ D {\bf 95}, 103011 (2017)
[arXiv:1612.09165 [hep-ph]].

\bibitem{Goodman:2010ku} 
  J.~Goodman, M.~Ibe, A.~Rajaraman, W.~Shepherd, T.~M.~P.~Tait and H.~B.~Yu,
  Phys.\ Rev.\ D {\bf 82}, 116010 (2010)
  [arXiv:1008.1783 [hep-ph]].
  
\bibitem{Drees:1993bu} 
  M.~Drees and M.~Nojiri,
  Phys.\ Rev.\ D {\bf 48}, 3483 (1993)
  [hep-ph/9307208].
  
\bibitem{Belanger:2008sj} 
  G.~B\'elanger, F.~Boudjema, A.~Pukhov and A.~Semenov,
  Comput.\ Phys.\ Commun.\  {\bf 180}, 747 (2009)
  [arXiv:0803.2360 [hep-ph]].
  
\bibitem{Shifman:1978zn}
  M.~A.~Shifman, A.~I.~Vainshtein and V.~I.~Zakharov,
  Phys.\ Lett.\ B {\bf 78}, 443 (1978).
  
\bibitem{Aoki:2016frl} 
  S.~Aoki {\it et al.},
  Eur.\ Phys.\ J.\ C {\bf 77}, 112 (2017)
  [arXiv:1607.00299 [hep-lat]].
  
\bibitem{Badier:1983mj} 
  J.~Badier {\it et al.} [NA3 Collaboration],
  Z.\ Phys.\ C {\bf 18}, 281 (1983).
  
\bibitem{Sutton:1991ay} 
  P.~J.~Sutton, A.~D.~Martin, R.~G.~Roberts and W.~J.~Stirling,
  Phys.\ Rev.\ D {\bf 45}, 2349 (1992).
  
\bibitem{Wijesooriya:2005ir} 
  K.~Wijesooriya, P.~E.~Reimer and R.~J.~Holt,
  Phys.\ Rev.\ C {\bf 72}, 065203 (2005)
  [nucl-ex/0509012].
  
\bibitem{Abdel-Rehim:2015owa} 
  A.~Abdel-Rehim {\it et al.},
  Phys.\ Rev.\ D {\bf 92}, 114513 (2015)
  Erratum: [Phys.\ Rev.\ D {\bf 93}, 039904 (2016)]
  [arXiv:1507.04936 [hep-lat]].
  
\bibitem{Barry:2018ort} 
  P.~C.~Barry, N.~Sato, W.~Melnitchouk and C.~R.~Ji,
 Phys.\ Rev.\ Lett.\  {\bf 121}, 152001 (2018)
  [arXiv:1804.01965 [hep-ph]].
  
\bibitem{Barger:2010gv} 
  V.~Barger, W.~Y.~Keung and D.~Marfatia,
  Phys.\ Lett.\ B {\bf 696}, 74 (2011)
  [arXiv:1007.4345 [hep-ph]].
  
\bibitem{Banks:2010eh} 
  T.~Banks, J.~F.~Fortin and S.~Thomas,
  arXiv:1007.5515 [hep-ph].
  
\bibitem{Crivellin:2014gpa} 
  A.~Crivellin and U.~Haisch,
  Phys.\ Rev.\ D {\bf 90}, 115011 (2014)
  [arXiv:1408.5046 [hep-ph]].
  
\bibitem{Brod:2017bsw} 
  J.~Brod, A.~Gootjes-Dreesbach, M.~Tammaro and J.~Zupan,
  JHEP {\bf 1810}, 065 (2018)
  [arXiv:1710.10218 [hep-ph]].

\bibitem{Caurier:2004gf} 
E.~Caurier, G.~Mart\'inez-Pinedo, F.~Nowacki, A.~Poves and A.~P.~Zuker,
Rev.\ Mod.\ Phys.\  {\bf 77}, 427 (2005)
[nucl-th/0402046].

\bibitem{Caurier:1999}
E.~Caurier and F.~Nowacki, 
Acta Phys.\ Pol.\ B {\bf 30}, 705 (1999).

\bibitem{Hebeler:2015hla} 
K.~Hebeler, J.~D.~Holt, J.~Men\'endez and A.~Schwenk,
Annu.\ Rev.\ Nucl.\ Part.\ Sci.\  {\bf 65}, 457 (2015)
[arXiv:1508.06893 [nucl-th]].

\bibitem{Hergert:2015awm} 
H.~Hergert, S.~K.~Bogner, T.~D.~Morris, A.~Schwenk and K.~Tsukiyama,
Phys.\ Rept.\  {\bf 621}, 165 (2016)
[arXiv:1512.06956 [nucl-th]].

\bibitem{Hagen:2013nca}
G.~Hagen, T.~Papenbrock, M.~Hjorth-Jensen and D.~J.~Dean,
Rept.\ Prog.\ Phys.\  {\bf 77}, 096302 (2014)
[arXiv:1312.7872 [nucl-th]].

\bibitem{Hagen:2015yea} 
  G.~Hagen {\it et al.},
  Nature Phys.\ {\bf 12}, 186 (2016)
  [arXiv:1509.07169 [nucl-th]].

\bibitem{Parzuchowski:2017wcq} 
N.~M.~Parzuchowski, S.~R.~Stroberg, P.~Navr\'atil, H.~Hergert and S.~K.~Bogner,
Phys.\ Rev.\ C {\bf 96}, 034324 (2017)
[arXiv:1705.05511 [nucl-th]].

\bibitem{Morris:2017vxi} 
T.~D.~Morris, J.~Simonis, S.~R.~Stroberg, C.~Stumpf, G.~Hagen, J.~D.~Holt, 
G.~R.~Jansen, T.~Papenbrock, R.~Roth and A.~Schwenk,
Phys.\ Rev.\ Lett.\  {\bf 120}, 152503 (2018)
[arXiv:1709.02786 [nucl-th]].

\bibitem{Epelbaum:2014efa} 
E.~Epelbaum, H.~Krebs and U.-G.~Mei{\ss}ner,
Eur.\ Phys.\ J.\ A {\bf 51}, 53 (2015)
[arXiv:1412.0142 [nucl-th]].

\bibitem{Furnstahl:2015rha} 
R.~J.~Furnstahl, N.~Klco, D.~R.~Phillips and S.~Wesolowski,
Phys.\ Rev.\ C {\bf 92}, 024005 (2015)
[arXiv:1506.01343 [nucl-th]].

\bibitem{Carlsson:2015vda} 
B.~D.~Carlsson, A.~Ekstr\"{o}m, C.~Forss\'{e}n, D.~F.~Str\"{o}mberg, G.~R.~Jansen, O.~Lilja, M.~Lindby, B.~A.~Mattsson and K.~A.~Wendt,
Phys.\ Rev.\ X {\bf 6}, 011019 (2016)
[arXiv:1506.02466 [nucl-th]].

\bibitem{Caurier:2007ee} 
E.~Caurier, J.~Men\'endez, F.~Nowacki and A.~Poves,
Phys.\ Rev.\ C {\bf 75}, 054317 (2007)
Erratum: [Phys.\ Rev.\ C {\bf 76}, 049901 (2007)]
[nucl-th/0702047].

\bibitem{Ruiz:2015yra}
R.~F.~Garcia Ruiz {\it et al.},
Phys.\ Rev.\ C {\bf 91}, 041304 (2015) 
[arXiv:1504.04474 [nucl-ex]].

\bibitem{Menendez:2009xa}
J.~Men\'endez, A.~Poves, E.~Caurier and F.~Nowacki,
Phys.\ Rev.\ C {\bf 80}, 048501 (2009) 
[arXiv:0906.0179 [nucl-th]].

\bibitem{Caurier:2007wq} 
E.~Caurier, J.~Men\'endez, F.~Nowacki and A.~Poves,
Phys.\ Rev.\ Lett.\  {\bf 100}, 052503 (2008)
[arXiv:0709.2137 [nucl-th]].

\bibitem{Menendez:2008jp} 
J.~Men\'endez, A.~Poves, E.~Caurier and F.~Nowacki,
Nucl.\ Phys.\ A {\bf 818}, 139 (2009)
[arXiv:0801.3760 [nucl-th]].

\bibitem{Honma:2009zz} 
M.~Honma, T.~Otsuka, T.~Mizusaki and M.~Hjorth-Jensen,
Phys.\ Rev.\ C {\bf 80}, 064323 (2009).
      
\bibitem{jj44b}
B.~A.~Brown, private communication.

\bibitem{usdb_obs} 
W.~A.~Richter, T.~Mkhize, and B.~A.~Brown,
Phys.\ Rev.\ C {\bf 78}, 064302 (2008).

\bibitem{Angeli:2013epw} 
I.~Angeli and K.~P.~Marinova,
Atom.\ Data Nucl.\ Data Tabl.\  {\bf 99}, 69 (2013).

\bibitem{nndc}
 \path{https://www.nndc.bnl.gov/ensdf}.

\bibitem{Raghavan:1989zz} 
P.~Raghavan,
Atom.\ Data Nucl.\ Data Tabl.\  {\bf 42}, 189 (1989).

\bibitem{Bertozzi:1972jff} 
W.~Bertozzi, J.~Friar, J.~Heisenberg and J.~W.~Negele,
Phys.\ Lett.\ B {\bf 41}, 408 (1972).

\bibitem{Sieja:2009zz} 
K.~Sieja, G.~Mart\'inez-Pinedo, L.~Coquard, and N.~Pietralla,
Phys.\ Rev.\ C {\bf 80}, 054311 (2009).

\bibitem{Tanabashi:2018oca} 
  M.~Tanabashi {\it et al.} [Particle Data Group],
  Phys.\ Rev.\ D {\bf 98}, 030001 (2018).
  
\bibitem{Weinberg:1990rz} 
  S.~Weinberg,
  Phys.\ Lett.\ B {\bf 251}, 288 (1990).
  
\bibitem{Weinberg:1991um} 
  S.~Weinberg,
  Nucl.\ Phys.\ B {\bf 363}, 3 (1991).
  
\bibitem{Epelbaum:2002gb} 
  E.~Epelbaum, U.-G.~Mei{\ss}ner and W.~Gl\"ockle,
  Nucl.\ Phys.\ A {\bf 714}, 535 (2003)
  [nucl-th/0207089].
  
\bibitem{Kaplan:1998tg} 
  D.~B.~Kaplan, M.~J.~Savage and M.~B.~Wise,
  Phys.\ Lett.\ B {\bf 424}, 390 (1998)
  [nucl-th/9801034].
  
\bibitem{Kaplan:1998we} 
  D.~B.~Kaplan, M.~J.~Savage and M.~B.~Wise,
  Nucl.\ Phys.\ B {\bf 534}, 329 (1998)
  [nucl-th/9802075].
  
\bibitem{Valderrama:2014vra} 
  M.~Pav{\'o}n Valderrama and D.~R.~Phillips,
  Phys.\ Rev.\ Lett.\  {\bf 114}, 082502 (2015)
  [arXiv:1407.0437 [nucl-th]].
  
\bibitem{Epelbaum:2014sza} 
  E.~Epelbaum, H.~Krebs and U.-G.~Mei{\ss}ner,
  Phys.\ Rev.\ Lett.\  {\bf 115}, 122301 (2015)
  [arXiv:1412.4623 [nucl-th]].
  
\bibitem{Entem:2017gor} 
  D.~R.~Entem, R.~Machleidt and Y.~Nosyk,
  Phys.\ Rev.\ C {\bf 96}, 024004 (2017)
  [arXiv:1703.05454 [nucl-th]].
  
\bibitem{Reinert:2017usi} 
  P.~Reinert, H.~Krebs and E.~Epelbaum,
  Eur.\ Phys.\ J.\ A {\bf 54}, 86 (2018)
  [arXiv:1711.08821 [nucl-th]].
  
\bibitem{Beane:2013kca} 
  S.~R.~Beane, S.~D.~Cohen, W.~Detmold, H.-W.~Lin and M.~J.~Savage,
  Phys.\ Rev.\ D {\bf 89}, 074505 (2014)
  [arXiv:1306.6939 [hep-ph]].
  
\bibitem{Chang:2017eiq} 
  E.~Chang {\it et al.} [NPLQCD Collaboration],
  Phys.\ Rev.\ Lett.\  {\bf 120}, 152002 (2018)
  [arXiv:1712.03221 [hep-lat]].
  
\bibitem{Donoghue:1991qv}
  J.~F.~Donoghue and H.~Leutwyler,
  Z.\ Phys.\ C {\bf 52}, 343 (1991).
  
\bibitem{Duflo:1995ep} 
  J.~Duflo and A.~P.~Zuker,
  Phys.\ Rev.\ C {\bf 52}, R23 (1995)
  [nucl-th/9505011].
  
\bibitem{Kaplan:1995yg} 
  D.~B.~Kaplan and M.~J.~Savage,
  Phys.\ Lett.\ B {\bf 365}, 244 (1996)
  [hep-ph/9509371].
  
\bibitem{Kaplan:1996rk} 
  D.~B.~Kaplan and A.~V.~Manohar,
  Phys.\ Rev.\ C {\bf 56}, 76 (1997)
  [nucl-th/9612021].
  
\bibitem{Mehen:1999qs} 
  T.~Mehen, I.~W.~Stewart and M.~B.~Wise,
  Phys.\ Rev.\ Lett.\  {\bf 83}, 931 (1999)
  [hep-ph/9902370].
  
\bibitem{Epelbaum:2004fk} 
  E.~Epelbaum, W.~Gl\"ockle and U.-G.~Mei{\ss}ner,
  Nucl.\ Phys.\ A {\bf 747}, 362 (2005)
  [nucl-th/0405048].
    
\bibitem{Hill:2013hoa} 
  R.~J.~Hill and M.~P.~Solon,
  Phys.\ Rev.\ Lett.\  {\bf 112}, 211602 (2014)
  [arXiv:1309.4092 [hep-ph]].
  
\bibitem{Arndt:2001ye} 
  D.~Arndt and M.~J.~Savage,
  Nucl.\ Phys.\ A {\bf 697}, 429 (2002)
  [nucl-th/0105045].
  
\bibitem{Chen:2001nb} 
  J.~W.~Chen and X.~d.~Ji,
  Phys.\ Rev.\ Lett.\  {\bf 87}, 152002 (2001)
  Erratum: [Phys.\ Rev.\ Lett.\  {\bf 88}, 249901 (2002)]
  [hep-ph/0107158].
  
\bibitem{Detmold:2005pt} 
  W.~Detmold and C.~J.~D.~Lin,
  Phys.\ Rev.\ D {\bf 71}, 054510 (2005)
  [hep-lat/0501007].
  
\bibitem{Diehl:2005rn}
  M.~Diehl, A.~Manashov and A.~Sch\"afer,
  Phys.\ Lett.\ B {\bf 622}, 69 (2005) 
  [hep-ph/0505269].
  
\bibitem{Ando:2006sk} 
  S.~i.~Ando, J.~W.~Chen and C.~W.~Kao,
  Phys.\ Rev.\ D {\bf 74}, 094013 (2006)
  [hep-ph/0602200].
  
\bibitem{Diehl:2006js} 
  M.~Diehl, A.~Manashov and A.~Sch\"afer,
  Eur.\ Phys.\ J.\ A {\bf 31}, 335 (2007)
  [hep-ph/0611101].
  
\bibitem{Wein:2014wma}
  P.~Wein, P.~C.~Bruns and A.~Sch\"afer,
  Phys.\ Rev.\ D {\bf 89}, 116002 (2014)  
  [arXiv:1402.4979 [hep-ph]].
  
\bibitem{Aubert:1983xm} 
  J.~J.~Aubert {\it et al.} [European Muon Collaboration],
  Phys.\ Lett.\  B {\bf 123}, 275 (1983).
  
\bibitem{Chen:2004zx} 
  J.~W.~Chen and W.~Detmold,
  Phys.\ Lett.\ B {\bf 625}, 165 (2005)
  [hep-ph/0412119].
  
\bibitem{Chen:2016bde} 
  J.~W.~Chen, W.~Detmold, J.~E.~Lynn and A.~Schwenk,
  Phys.\ Rev.\ Lett.\  {\bf 119}, 262502 (2017)
  [arXiv:1607.03065 [hep-ph]].
  
\bibitem{Hirai:2007sx} 
  M.~Hirai, S.~Kumano and T.-H.~Nagai,
  Phys.\ Rev.\ C {\bf 76}, 065207 (2007)
  [arXiv:0709.3038 [hep-ph]].
  
\bibitem{deFlorian:2011fp} 
  D.~de Florian, R.~Sassot, P.~Zurita and M.~Stratmann,
  Phys.\ Rev.\ D {\bf 85}, 074028 (2012)
  [arXiv:1112.6324 [hep-ph]].
  
\bibitem{Kovarik:2015cma} 
  K.~Kova\v r\'ik {\it et al.},
  Phys.\ Rev.\ D {\bf 93}, 085037 (2016)
  [arXiv:1509.00792 [hep-ph]].
  
\bibitem{Eskola:2016oht} 
  K.~J.~Eskola, P.~Paakkinen, H.~Paukkunen and C.~A.~Salgado,
  Eur.\ Phys.\ J.\ C {\bf 77}, 163 (2017)
  [arXiv:1612.05741 [hep-ph]].
  
\bibitem{Beane:2002wk} 
  S.~R.~Beane, V.~Bernard, E.~Epelbaum, U.-G.~Mei\ss ner and D.~R.~Phillips,
  Nucl.\ Phys.\ A {\bf 720}, 399 (2003)
  [hep-ph/0206219].

\bibitem{Hill:2014yxa}
  R.~J.~Hill and M.~P.~Solon,
  Phys.\ Rev.\ D {\bf 91}, 043505 (2015) 
  [arXiv:1409.8290 [hep-ph]].
  
\bibitem{Alexandrou:2018zdf} 
  C.~Alexandrou, M.~Constantinou, K.~Hadjiyiannakou, K.~Jansen, C.~Kallidonis, G.~Koutsou and A.~Vaquero Avil\'es-Casco,
  Phys.\ Rev.\ D {\bf 97}, 094504 (2018)
  [arXiv:1801.09581 [hep-lat]].

\bibitem{Crivellin:2013ipa} 
  A.~Crivellin, M.~Hoferichter and M.~Procura,
  Phys.\ Rev.\ D {\bf 89}, 054021 (2014)
  [arXiv:1312.4951 [hep-ph]].
  
\bibitem{Gasser:1974wd} 
  J.~Gasser and H.~Leutwyler,
  Nucl.\ Phys.\ B {\bf 94}, 269 (1975).
  
\bibitem{Borsanyi:2014jba} 
  S.~Borsanyi {\it et al.},
  Science {\bf 347}, 1452 (2015)
  [arXiv:1406.4088 [hep-lat]].
  
\bibitem{Gasser:2015dwa}
  J.~Gasser, M.~Hoferichter, H.~Leutwyler and A.~Rusetsky,
  Eur.\ Phys.\ J.\ C {\bf 75}, 375 (2015) 
  [arXiv:1506.06747 [hep-ph]].
  
\bibitem{Brantley:2016our} 
  D.~A.~Brantley, B.~Jo\'o, E.~V.~Mastropas, E.~Mereghetti, H.~Monge-Camacho, B.~C.~Tiburzi and A.~Walker-Loud,
  arXiv:1612.07733 [hep-lat].
  
\bibitem{Hoferichter:2015dsa} 
  M.~Hoferichter, J.~Ruiz de Elvira, B.~Kubis and U.-G.~Mei{\ss}ner,
  Phys.\ Rev.\ Lett.\  {\bf 115}, 092301 (2015)
  [arXiv:1506.04142 [hep-ph]].
  
\bibitem{Hoferichter:2015hva} 
  M.~Hoferichter, J.~Ruiz de Elvira, B.~Kubis and U.-G.~Mei{\ss}ner,
  Phys.\ Rept.\  {\bf 625}, 1 (2016)
  [arXiv:1510.06039 [hep-ph]].
  
\bibitem{Hoferichter:2016ocj} 
  M.~Hoferichter, J.~Ruiz de Elvira, B.~Kubis and U.-G.~Mei{\ss}ner,
  Phys.\ Lett.\ B {\bf 760}, 74 (2016)
  [arXiv:1602.07688 [hep-lat]].

\bibitem{Durr:2015dna}
  S.~D\"urr {\it et al.} [BMW Collaboration],
  Phys.\ Rev.\ Lett.\  {\bf 116}, 172001 (2016) 
  [arXiv:1510.08013 [hep-lat]].
  
\bibitem{Yang:2015uis}
  Y.~B.~Yang {\it et al.} [$\chi$QCD Collaboration],
  Phys.\ Rev.\ D {\bf 94}, 054503 (2016) 
  [arXiv:1511.09089 [hep-lat]].
  
\bibitem{Abdel-Rehim:2016won}
  A.~Abdel-Rehim {\it et al.} [ETM Collaboration],
  Phys.\ Rev.\ Lett.\  {\bf 116}, 252001 (2016)  
  [arXiv:1601.01624 [hep-lat]].
  
\bibitem{Bali:2016lvx}
  G.~S.~Bali {\it et al.} [RQCD Collaboration],
  Phys.\ Rev.\ D {\bf 93}, 094504 (2016) 
  [arXiv:1603.00827 [hep-lat]].
  
\bibitem{Gotta:2008zza} 
  D.~Gotta {\it et al.},
  Lect.\ Notes Phys.\  {\bf 745}, 165 (2008).
  
\bibitem{Strauch:2010vu} 
  T.~Strauch {\it et al.},
  Eur.\ Phys.\ J.\ A {\bf 47}, 88 (2011) 
  [arXiv:1011.2415 [nucl-ex]].
  
\bibitem{Hennebach:2014lsa} 
  M.~Hennebach {\it et al.},
  Eur.\ Phys.\ J.\ A {\bf 50}, 190 (2014) 
  [arXiv:1406.6525 [nucl-ex]].
  
\bibitem{Baru:2010xn} 
  V.~Baru, C.~Hanhart, M.~Hoferichter, B.~Kubis, A.~Nogga and D.~R.~Phillips,
  Phys.\ Lett.\ B {\bf 694}, 473 (2011)
  [arXiv:1003.4444 [nucl-th]].
  
\bibitem{Baru:2011bw} 
  V.~Baru, C.~Hanhart, M.~Hoferichter, B.~Kubis, A.~Nogga and D.~R.~Phillips,
  Nucl.\ Phys.\ A {\bf 872}, 69 (2011)
  [arXiv:1107.5509 [nucl-th]].
  
\bibitem{RuizdeElvira:2017stg} 
  J.~Ruiz de Elvira, M.~Hoferichter, B.~Kubis and U.-G.~Mei{\ss}ner,
  J.\ Phys.\ G {\bf 45}, 024001 (2018)
  [arXiv:1706.01465 [hep-ph]].
  
\bibitem{Weinberg:1958ut}
  S.~Weinberg,
  Phys.\ Rev.\  {\bf 112}, 1375 (1958).
    
\bibitem{Adler:1975he}
  S.~L.~Adler, E.~W.~Colglazier, Jr., J.~B.~Healy, I.~Karliner, J.~Lieberman, Y.~J.~Ng and H.~S.~Tsao,
  Phys.\ Rev.\ D {\bf 11}, 3309 (1975).
  
\bibitem{Gupta:2018lvp} 
  R.~Gupta, B.~Yoon, T.~Bhattacharya, V.~Cirigliano, Y.~C.~Jang and H.~W.~Lin,
  Phys.\ Rev.\ D {\bf 98}, 091501 (2018)
  [arXiv:1808.07597 [hep-lat]].
  
\bibitem{Gockeler:2006zu} 
  M.~G\"ockeler {\it et al.} [QCDSF and UKQCD Collaborations],
  Phys.\ Rev.\ Lett.\  {\bf 98}, 222001 (2007)
  [hep-lat/0612032].
 
\bibitem{Cirigliano:2017tqn} 
  V.~Cirigliano, A.~Crivellin and M.~Hoferichter,
  Phys.\ Rev.\ Lett.\  {\bf 120}, 141803 (2018)
  [arXiv:1712.06595 [hep-ph]].
 
 \bibitem{Baum:2011rm}
  I.~Baum, V.~Lubicz, G.~Martinelli, L.~Orifici and S.~Simula,
  Phys.\ Rev.\ D {\bf 84}, 074503 (2011) 
  [arXiv:1108.1021 [hep-lat]].
  
\bibitem{Hoferichter:2016duk} 
  M.~Hoferichter, B.~Kubis, J.~Ruiz de Elvira, H.-W.~Hammer and U.-G.~Mei{\ss}ner,
  Eur.\ Phys.\ J.\ A {\bf 52}, 331 (2016)
  [arXiv:1609.06722 [hep-ph]].
  
\bibitem{Hoferichter:2018zwu} 
  M.~Hoferichter, B.~Kubis, J.~Ruiz de Elvira and P.~Stoffer,
  Phys.\ Rev.\ Lett.\  {\bf 122}, 122001 (2019)
  [arXiv:1811.11181 [hep-ph]].
  
\bibitem{Martin:2009iq} 
  A.~D.~Martin, W.~J.~Stirling, R.~S.~Thorne and G.~Watt,
  Eur.\ Phys.\ J.\ C {\bf 63}, 189 (2009)
  [arXiv:0901.0002 [hep-ph]].
  
\bibitem{Buckley:2014ana} 
  A.~Buckley, J.~Ferrando, S.~Lloyd, K.~Nordstr\"om, B.~Page, M.~R\"ufenacht, M.~Sch\"onherr and G.~Watt,
  Eur.\ Phys.\ J.\ C {\bf 75}, 132 (2015)
  [arXiv:1412.7420 [hep-ph]].
  
\bibitem{Hahn:2004fe} 
  T.~Hahn,
  Comput.\ Phys.\ Commun.\  {\bf 168}, 78 (2005)
  [hep-ph/0404043].
  
\bibitem{Brown:2018dum} 
  A.~G.~A.~Brown {\it et al.} [Gaia Collaboration],
  Astron.\ Astrophys.\  {\bf 616}, A1 (2018)
  [arXiv:1804.09365 [astro-ph.GA]].
  
\bibitem{Necib:2018iwb} 
  L.~Necib, M.~Lisanti and V.~Belokurov,
  arXiv:1807.02519 [astro-ph.GA].
  
\bibitem{Eilers:2018}
 A.-C.~Eilers, D.~W.~Hogg, H.-W.~Rix and M.~Ness,
 Astrophys.\ J.\  {\bf 871}, 120 (2019)
 [arXiv:1810.09466 [astro-ph.GA]]. 
  
\bibitem{Evans:2018bqy} 
  N.~W.~Evans, C.~A.~J.~O'Hare and C.~McCabe,
  Phys.\ Rev.\ D {\bf 99}, 023012 (2019)
  [arXiv:1810.11468 [astro-ph.GA]].

\end{thebibliography}
\end{document}